\documentclass[3p]{elsarticle}
\usepackage{lineno}
\usepackage{cuted} 
\usepackage{hyperref}
\usepackage{color}
\usepackage{xcolor}
\usepackage{array, bigdelim}
\usepackage{arydshln} 
\usepackage{amsmath}
\usepackage{amssymb}
\usepackage{derivative}
\usepackage{cancel}
\usepackage{nomencl}
\usepackage{tabularx}
\usepackage{makecell}
\usepackage{booktabs}
\usepackage{multirow}
\usepackage{placeins}
\usepackage{etoolbox}
\usepackage{enumitem}
\usepackage{threeparttable}

\setlist[itemize]{leftmargin=*} 

\newcommand{\bfK}{\mathbf{K}}
\newcommand{\bfS}{\mathbf{S}}
\newcommand{\invKF}{\bfK_F^{-1}}
\newcommand{\bff}{\mathbf{f}}
\newcommand{\bfd}{\mathbf{d}}
\newcommand{\bfG}{\mathbf{G}}
\newcommand{\bfA}{\mathbf{A}}
\newcommand{\bfB}{\mathbf{B}}
\newcommand{\bfC}{\mathbf{C}}
\newcommand{\bfH}{\mathbf{H}}
\newcommand{\bfP}{\mathbf{P}}
\newcommand{\bfR}{\mathbf{R}}
\newcommand{\bfT}{\mathbf{T}}
\newcommand{\bfU}{\mathbf{U}}
\newcommand{\bfV}{\mathbf{V}}

\newcommand{\lri}{\langle i \rangle}

\newcommand{\dsrci}{\bfd_{src}^{\lri}}
\newcommand{\dtgti}{\bfd_{tgt}^{\lri}}

\newcommand{\AngBr}[1]{\langle #1 \rangle} 
\newcommand{\Lnorm}[1]{\left\| #1 \right\|}
\newcommand{\goal}[1]{ \bfd_{goal}^{\langle #1 \rangle}}
\newcommand{\combnk}[2]{\begin{pmatrix} #1 \\ #2 \end{pmatrix}} 
\newcommand{\NTo}[1]{\mathbb{N}_{\leq}^{#1}} 
\newcommand{\compo}[2]{\left.#1\right|_{#2}} 

\makenomenclature

\renewcommand\nomgroup[1]{%
  \item[\bfseries
  \ifstrequal{#1}{A}{Scalar}{%
  \ifstrequal{#1}{B}{Vector}{%
  \ifstrequal{#1}{C}{Tensor}{%
  \ifstrequal{#1}{D}{Set}{%
  \ifstrequal{#1}{E}{Notation and Operator} } }}}
  ]}

\modulolinenumbers[5]
\journal{TBD}
\begin{document}
\emergencystretch 4em 

\begin{frontmatter}
\title{A Substructure Perturbation Method for Systematic Design of Mechanical Metamaterials with Programmed Functionalities}

\author[meam]{Jiakun Liu}
\author[meam]{Adam Taylor}
\author[meam]{Sage Fulco}
\author[meam]{Sumukh S. Pande}
\author[meam]{Kevin T. Turner\corref{cor1}}\ead{kturner@seas.upenn.edu}
\address[meam]{Department of Mechanical Engineering and Applied Mechanics, University of Pennsylvania, Philadelphia, PA 19147, United States}
\cortext[cor1]{Corresponding author}

\begin{abstract}
Mechanical metamaterials utilize geometry to achieve exceptional mechanical properties, including those not typically possible for traditional materials. To achieve these properties, it is necessary to identify the proper structures and geometries, which is often a non-trivial and computationally expensive process. Here, we propose a Substructure Perturbation Method (SSPM) for systematic design and search of these materials with programmed deformation modes. We present the theoretical fundamentals and computational algorithms of the SSPM, along with four design problems to investigate the effect and performance of the SSPM. Results reveal the necessity of analyzing multiple substructures simultaneously in obtaining successful designs, and its effectiveness in speeding up numerical processes. In one design case, SSPM is shown to be effectively two orders of magnitude faster than another state-of-art approach while using less computational resources. We also show an experimental validation where the fabricated prototypes can grasp objects respectively by undergoing programmed deformations under corresponding inputs. The proposed SSPM provides new fundamentals and strategies for the design of mechanical metamaterials with advanced functionalities.
\end{abstract}

\begin{keyword}
Mechanical Metamaterials \sep
Disordered Cellular Lattices \sep
Computer-aided Structure Design \sep
Material and Structure Design
\end{keyword}

\end{frontmatter}

\tableofcontents

\nomenclature[A, 00]{$\alpha$}{An assumed magnitude of LSP for error analysis }
\nomenclature[A, 01]{$f_q^*$}{Attempted amount of stiffness change of the $q$th ligament when applying LSP}
\nomenclature[A, 02]{$f_q$}{Regularized stiffness change in the $q$th ligament}

\nomenclature[A, 10]{$O_{[j]}^{ss}$}{Order of SS~$[j]$ (number of its real ligaments)}
\nomenclature[A, 11]{$\Omega^{\lri}$}{Distance to goal ($L^2\text{-norm}$ of true errors) in mode $\lri$}
\nomenclature[A, 12]{$m_{tgt}^{\lri}$}{Number of DOF involved in target in mode $\lri$}
\nomenclature[A, 13]{$n_m$}{Number of functional modes to realize simultaneously}
\nomenclature[A, 14]{$\psi^{all}_{max}$}{Maximum absolute value of relative errors in all modes}
\nomenclature[A, 15]{$\widehat{\psi}$}{Threshold for magnitude of relative errors to reach}

\nomenclature[A, 20]{$n_{el}$}{Number of ligaments in the parent network}
\nomenclature[A, 21]{$n_{node}$}{Number of nodes in the parent network}
\nomenclature[A, 22]{$n^{E,min}_{el}$}{Number of ligaments whose stiffness are at allowed lower stiffness bound}

\nomenclature[A, 30]{$E_e$}{Elastic stiffness of the $e$th ligament}
\nomenclature[A, 31]{$\widehat{E}_{min}$}{Goal lower bound of ligament stiffness in final network}
\nomenclature[A, 32]{$\widehat{E}_{max}$}{Goal upper bound of ligament stiffness in final network}
\nomenclature[A, 33]{$E^*_{min}$}{Initial attainable lower bound of ligament stiffness}
\nomenclature[A, 40]{$A_e$}{Cross-sectional area of the $e$th ligament}

\nomenclature[B, 00]{$\beta^{\lri}_{p[j]}$}{Rate of change in relative errors at target DOF of mode $\lri$ due to an infinitesimal negative-LSP by SS~$[j]$}
\nomenclature[B, d00]{$\bfd$}{Responses (displacement components) at all DOF}
\nomenclature[B, d01]{$\bfd_E$}{Subvector of $\bfd$ consists of prescribed essential BCs}
\nomenclature[B, d02]{$\bfd_F$}{Subvector of $\bfd$ consists of displacement components at all free DOF}
\nomenclature[B, d03]{$\dsrci$}{Source input (essential BCs) in mode $\lri$}
\nomenclature[B, d04]{$\dtgti$}{Responses at target DOF in mode $\lri$}
\nomenclature[B, d05]{$\goal{i}$}{Goal values for responses at target DOF in mode $\lri$}
\nomenclature[B, 01]{$\Delta^{\lri}$}{True errors between responses at target DOF and corresponding goals in mode $\lri$}
\nomenclature[B, 02]{$\Psi^{\lri}$}{Relative errors between responses at target DOF and corresponding goals in mode $\lri$}
\nomenclature[B, f01]{$\bff$}{Force components at all DOF}
\nomenclature[B, f02]{$\bff_F$}{Subvector of $\bff$ consists of applied natural BCs}
\nomenclature[B, f03]{$\bfR_E$}{Subvector of $\bff$ consists of reaction forces at DOF prescribed with essential BCs}


\nomenclature[C, a01]{$\bfA^{\lri}$}{Actuation matrix mapping input at source to response at target DOF in mode $\lri$}
\nomenclature[C, a02a]{$\bfA^{\lri}_{p[j]}$}{Perturbed $\bfA^{\lri}$ due to LSP by SS~$[j]$ in mode $\lri$}
\nomenclature[C, a02b]{$\bfA^{\lri}_{p[j,k,l,\dots]}$}{Perturbed $\bfA^{\lri}$ due to LSP by multiple substructures in mode $\lri$}
\nomenclature[C, g]{$\bfG^{\lri}$}{Gather matrix slicing global DOF to target DOF in mode $\lri$ }
\nomenclature[C, h]{$\bfH_{e}$}{Gather matrix of the $e\text{th}$ element mapping its element-wise DOF to corresponding structure-wise global DOF}

\nomenclature[C, k01]{$\bfK$}{Global stiffness matrix of a mechanical system}
\nomenclature[C, k02]{$\bfK_{EF}$}{Partitioned submatrix of $\bfK$ based on essential BCs}
\nomenclature[C, k03]{$\bfK_{F}$}{Partitioned submatrix of $\bfK$ based on essential BCs}
\nomenclature[C, k04a]{$\bfK^{p[j]}$}{Perturbed $\bfK$ due to LSP by SS~$[j]$}
\nomenclature[C, k04b]{$\bfK^{p[j,k,l, \dots]}$}{Perturbed $\bfK$ due to LSP by multiple SS}

\nomenclature[C, k05]{$\bfK^{p[j]}_{EF}$}{Partitioned submatrix of $\bfK^{p[j]}$ based on essential BCs}
\nomenclature[C, k06]{$\bfK^{p[j]}_{F}$}{Partitioned submatrix of $\bfK^{p[j]}$ based on essential BCs}

\nomenclature[C, k07]{$\bfK_{TE2}^{ref}$}{Universal base element stiffness matrix of a two-node linear truss element in its reference CSYS}
\nomenclature[C, k08]{$\bfK_{e}^{glb}$}{Stiffness matrix of the $e\text{th}$ element in the global CSYS}
\nomenclature[C, k09]{$\bfK_{e}^{ss[j]}$}{Stiffness matrix of the $e\text{th}$ element in SS~$[j]$ in the global CSYS}
\nomenclature[C, k10]{$\bfK_{}^{ph}$}{Base stiffness matrix of phantom ligaments}
\nomenclature[C, k11]{$\bfK_{e,[j]}^{re}$}{Base stiffness matrix of real ligaments in SS $[j]$}

\nomenclature[C, p]{$\bfP^{\lri}_{[j]}$}{Perturbational effect matrix by SS~$[j]$ in mode $\lri$}
\nomenclature[C, r]{$\bfR_e$}{Rotation matrix of the $e\text{th}$ element transferring reference CSYS to global CSYS }
\nomenclature[C, s01]{$\bfS_{[j]}$}{Global stiffness matrix of SS $[j]$}
\nomenclature[C, s02]{$\bfS_{EF,[j]}$}{Partitioned sub-matrix of $\bfS_{[j]}$ based on essential BCs}
\nomenclature[C, s03]{$\bfS_{F,[j]}$}{Partitioned sub-matrix of $\bfS_{[j]}$ based on essential BCs}

\nomenclature[C, t01]{$\bfT^{\lri}$}{Transformation matrix mapping essential BCs to all free DOF in mode $\lri$}
\nomenclature[C, t02a]{$\bfT_{p[j]}^{\lri}$}{Perturbed $\bfT^{\lri}$ due to LSP by SS~$[j]$ in mode $\lri$}
\nomenclature[C, t02b]{$\bfT_{p[j,k,l,\dots]}^{\lri}$}{Perturbed $\bfT^{\lri}$ due to LSP by multiple substructures in mode $\lri$}

\nomenclature[D, 01]{$\mathbb{N}^{n}_{\leq}$}{All natural numbers from $1$ to $n$}
\nomenclature[D, 02]{$\mathbb{N}^{all}$}{All ligament labels in the parent network}
\nomenclature[D, 03]{$\mathbb{L}^{re}_{[j]}$}{Labels of real ligaments in SS~$[j]$} 
\nomenclature[D, 04]{$\mathbb{L}^{ph}_{[j]}$}{Labels of phantom ligaments in SS~$[j]$}

\nomenclature[E, n1]{${\lri}$}{Superscript in values related to mode $\lri$}
\nomenclature[E, n2]{${[j]}$}{Subscript in values related to a SS labeled by integer $j$}
\nomenclature[E, o1]{$u\odot v$}{Dot product between two vectors}
\nomenclature[E, o2]{$u\oslash v$}{Element-wise divide between two vectors}
\nomenclature[E, o3]{$\compo{u}{r}$}{The $r\text{-th}$ component of vector $u$}
\nomenclature[E, o4]{$\Lnorm{u}$}{The Euclidean norm ($L^2\text{-norm}$) of vector $u$}

\clearpage 

\printnomenclature

\newpage

\section{Introduction}\label{sec:Intro}

Mechanical metamaterials are an emerging class of nontraditional materials that demonstrate exceptional properties or distinctive responses, typically realized by specially-tailored structures at one or multiple length scales~\cite{Barchiesi2019.mechanical, Zaiser2023.disordered, Jiao2023.mechanical, Bertoldi2017.flexible}. 
Some of these materials demonstrate gradient or hierarchical structural patterns, others typically exhibit bio-mimetic or irregular/disordered geometries with non-deterministic features. These complicated structures commonly arise from design strategies and algorithms, making them difficult to interpret through human-led design.

For instance, in topology optimization, a structure is determined via iterative removal of pixels/voxels from an initially complete piece of continuum solid. The optimized result is typically an irregular structure with an improved property metric (e.g., strength-to-density ratio). 
However, classical topology optimization, when regarded as a design strategy, has inherent limitations in creating structures with desired responses, essentially because it undergoes gradient-based methods over a fixed reference domain within continuum elasticity regime. In contrast, many of the exceptional properties and exquisite behaviors in mechanical metamaterials require moderate to large structural deformation, rotation, and/or material non-linearity. In addition, representation of structures in form of pixels/voxels is computationally inefficient for structures with low relative densities (e.g., a sparse lattice).

Indeed, a broad range of alternative strategies have been presented in recent studies. 
For example: A combinatorial approach that determines patterns of constitutive unit cells (i.e., building blocks) in a cubic structure (assemblage) to realize textured deformations under associated boundary loads~\cite{Coulais2016.combinatorial}; generative approaches that selectively prune nodes (i.e., remove ligaments) in a lattice network to realize auxetic behavior~\cite{Reid2018.auxetic} or targeted input-output responses~\cite{Rocks2017.designing, Yan2017.architecture}; machine-learning frameworks via training of topologically distinct representative unit cells to perform inverse design of lattice-like metamaterials with, for instance, targeted anisotropic stiffness~\cite{Bastek2022.inverting}, or targeted uniaxial compressive stress‒strain curve~\cite{Ha2023.rapid}. Other representative studies may be found in the available reviews~\cite{Barchiesi2019.mechanical, Zaiser2023.disordered, Jiao2023.mechanical, Bertoldi2017.flexible, Gao2023.rational}.

This work is primarily focused on the spacial responses of mechanical metamaterials and their correlated textures and/or functionalities, with some of the most successful designs using lattice-based metamaterials that can realize prescribed input-output displacements. In this work, we will refer to input-output deformation responses as \textit{functional modes}, or simply, \textit{modes}. Metamaterials with complex functional modes have become interesting subjects with their potential usage in various fields such as actuators, mechanisms~\cite{Ion2016.metamaterial}, and mechanical computers~\cite{Yasuda2021.mechanical}.

More specifically, to obtain mechanical metamaterials with prescribed functional modes, different kinds of strategies and algorithms have been presented in the past few years. For example, Rocks et al.~\cite{Rocks2017.designing} showed a computational approach in which a portion of the ligaments in a network are removed in a step-by-step manner (one ligament at a time) to obtain a targeted extension between two nodes (output) under an applied extension on another pair of nodes (input). In each step, the ligament whose deletion will minimize the cost function is judged by calculating the discrete Green's function associated with the self-stress states and compatible stress states correlated to the equilibrium matrix of the network. Note that this kind of response is denoted by some as `allosteric' behavior, as it conceptually resembles allosteric proteins capable of varying their shapes in response to environmental stimuli (e.g., ligand binding) at a distant site. Pashine et al.~\cite{Pashine2023.reprogrammable} later showed a similar network but with variable stiffness ligaments, and used a local stress-based numerical algorithm (modified from Rocks et al.~\cite{Rocks2017.designing}) to repeatedly soften a portion of the ligaments to realize targeted allosteric modes. Yan et al.~\cite{Yan2017.architecture} presented a different approach to create allosteric networks, in which correlation analyses involving randomly generated sequences (i.e., pick some links in the network) are performed by Monte Carlo methods to help search for a suitable ligament to update in each step, gradually leading to the targeted allosteric displacements in a network.

Lee et al.~\cite{Lee2022.mechanical} used two types of optimization methods: genetic algorithms (GA) and partial pattern search (PPS), respectively, to find working combinations of ligament stiffness values in a truss network to realizes desired nodal outputs. GA works in a global way by exploring stiffness combinations for all ligaments and attempting to find one that would reduce mean squared error (MSE) between current and target response, whereas PPS works in a local way by randomly selecting one ligament at a time and checking to see if adjusting its stiffness would reduce MSE.

Bonfanti et al.~\cite{Bonfanti2020.automatic} applied a Monte Carlo method integrated with FIRE optimization~\cite{Bitzek2006.structural} and simulated annealing~\cite{Kirkpatrick1983.optimization} %
to realize input-output mode in a beam network. The structure is updated iteratively, during which one ligament is either removed or re-introduced in each step. They also demonstrated a machine learning framework and showed its effectiveness in assisting the structure updating processes in comparable design problems.

The above discussions reveal that current exemplary approaches typically utilize one or a combination of: established optimization methods, trial-error search (e.g, randomly remove one link and check if it works), data-driven analysis (e.g, correlation study, machining learning). Each of these methods, considering its inherent logistics, could still encounter challenges as design problems become more demanding. Iterative trial-and-error steps, especially if only involving one or a few structural features, may encounter difficulties in finding a path to working configurations, as will be shown in a later section. In optimization methods, new attempts typically depend on previously attempted values (e.g., stiffness combinations) and their resultant cost function value(s), whereas information like strain and stress fields may not be contained or informed during these processes. Consequently, in addition to their computational bottleneck in high dimension problems, optimization methods are likely incapable of incorporating mechanical and structural features. This is perhaps why correlation analyses or machine learning frameworks that include more underlying mechanisms have emerged recently. These techniques, however, commonly require intensive data collection and training, and hereby also largely depend on the scale and quality of data, and subsequently, may only work effectively if design problems are correlated to the scenarios covered in the trained data.

Therefore, in pursuit of designing advanced mechanical metamaterials with sophisticated functionalities over an immense design space, a robust and computationally efficient approach would be greatly beneficial. Under such a motivation, we propose a Substructure Perturbation Method (SSPM) that achieves optimized designs through consideration of possible substructures of a system, where each substructure can be perturbed relative to the reference system. These perturbed substructures are iteratively updated to realize optimized designs and concurrently considered to maintain computational efficiency.

In specific to this study, we implement this approach to design truss and beam networks with multiple functional modes. We first introduce the theoretical foundations, and then the associated computational algorithms. Thereafter, we include several numerical cases to help illustrate the essential procedures and also the effectiveness of the proposed approach. Moreover, we demonstrate a design and fabrication example of a mechanical metamaterials capable of grasping objects distinctively under different inputs.

\section{Theoretical framework}
In derivation of this approach, there are no specific restrictions on the topology of the initial network or preconceived final geometries. Nevertheless, a relatively `densely-connected' structure is preferred, for example, a triangular lattice as shown in Fig.~\ref{fig:network_and_SS}. Firstly, triangular lattices are stretch-dominated and relatively stiff~\cite{Calladine1978.buckminster, Gibson1997.cellular}. This ensures the prescribed essential boundary conditions (BCs) result in limited and localized deformations, allowing for iterative structural modification to gradually bring the structural responses to their goals. Secondly, densely connected network could provide more options for design strategies since there are more ligaments connected to a node as compared to other types of lattices (e.g., square lattices). These characters of a densely connected network thereby justify its usage as an optional initial configuration.

\subsection{Substructure of a network}
We define a concept called \textit{substructure} (abbreviated as `SS'), and use an integer inside square brackets (e.g., $[j]$) when referring to a particular SS. Variables denoted by subscript~$[j]$ are thereby properties or dependent values of SS~$[j]$. 

\begin{figure*}[th]
\centering
\includegraphics[width=0.75\textwidth]{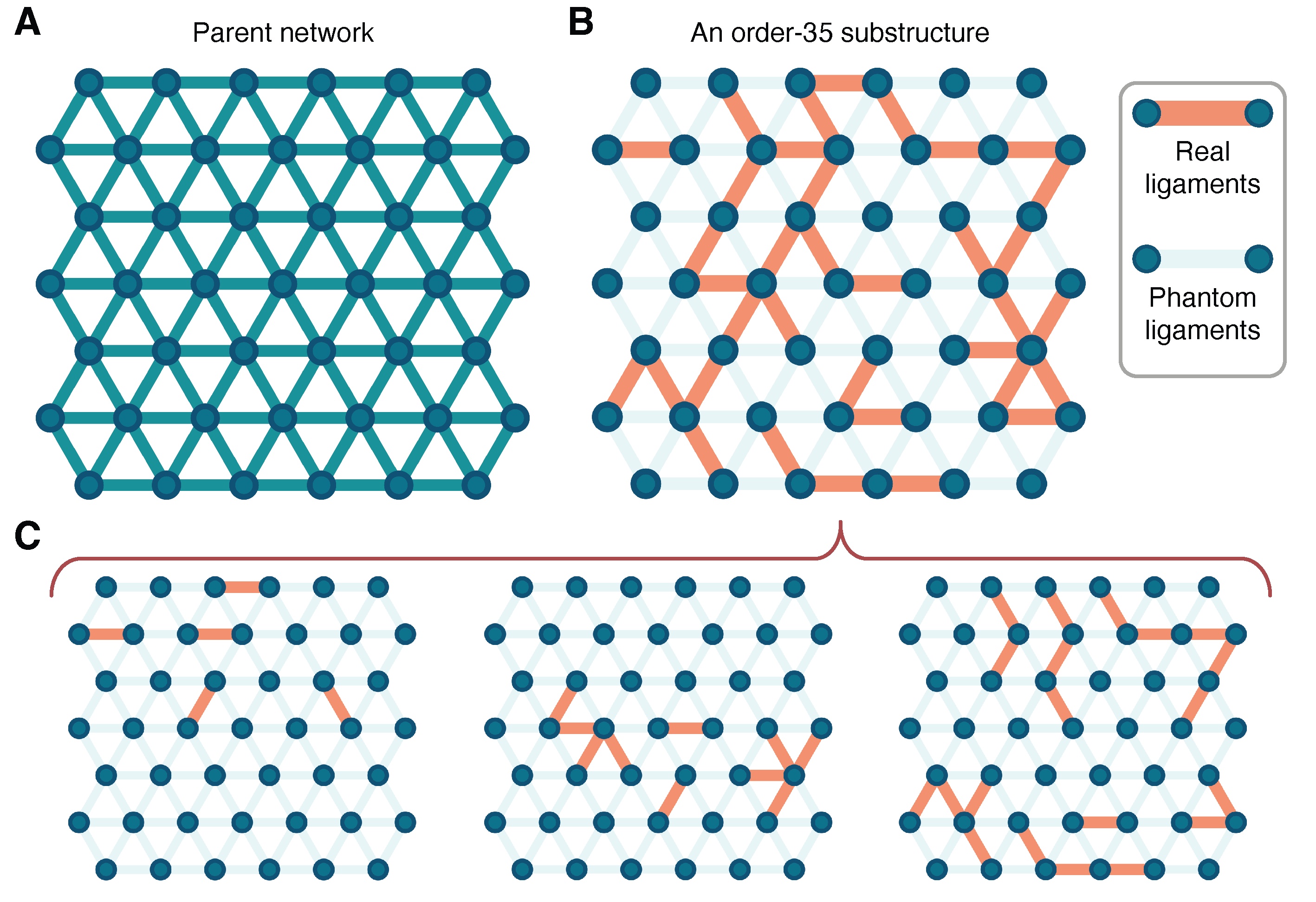}
\caption{Schematic of parent network, substructure (SS), and substructure decomposition.
(A)~Here a triangular lattice consists of 110 ligaments is used as the parent network. 
(B)~An example of an order-35 SS consists of 35 real ligaments (highlighted in orange) and 75 phantom ligaments (shown in light Teal blue). 
(C)~Three mutually disjoint SS having 5, 10, and 20 real ligaments, respectively. These three SS together forms one possible decomposition of the SS shown in panel~B.}
\label{fig:network_and_SS}
\end{figure*}

\textbf{Substructure.} The basic idea of substructure (SS) is to take a portion of the ligaments from their current parent network as the \textit{real ligaments}, and effectively removing the remaining ligaments by setting their stiffness to zero. Here we use \textit{phantom ligament} to describe such a topologically existing, but mechanically zero-stiffness ligament. This way, a SS still shares the same topology (nodes, ligaments, and connectivity) as its parent network, while consisting of real ligaments (positive stiffness) and phantom ligaments (zero stiffness). As an example, Fig.~\ref{fig:network_and_SS}B shows a SS with 35 real ligaments highlighted in orange, while the remaining are phantom ligaments shown in teal.

In a mathematical perspective, the real ligaments of a SS form a non-empty \textit{proper subset} of ligaments of the parent structure. We denote a set of continuous natural numbers from $1$ to $n$ as $\NTo{n}$, namely:
\begin{equation}\label{eq:set_N_leq_n}
\NTo{n} := \left\{ 1,2, \dots , n \right\}
\end{equation}
and assume a network has $n_{el}$ ligaments labeled by integers $1,2, \dots, n_{el}$. Then any \textit{proper subset} of $\NTo{n_{el}}$, excluding the empty set, will uniquely define a SS:
\begin{equation}\label{eq:set_N_ssj}
 \mathbb{L}^{re}_{[j]} := \left\{ x \big| x \in \mathbb{N}^{n_{el}}_{\leq} \right\}, 
 \,\, \mathbb{L}^{re}_{[j]} \subsetneqq \mathbb{N}^{n_{el}}_{\leq} := \mathbb{N}^{all}, 
 \,\, \mathbb{L}^{re}_{[j]} \neq \varnothing,
\end{equation}
where $\mathbb{N}^{all}$ is the set containing all ligament labels of the parent network, and $\mathbb{L}^{re}_{[j]}$ is the set containing labels of real ligament in a SS labeled by $[j]$. Also we define the \textit{order of a substructure}, $O_{[j]}^{ss}$, as the number of real ligaments it contains, hereby $ 1 \leq O_{[j]}^{ss} \leq n_{el}-1$.

Subsequently, the total number of all substructures (SS) with order-$k$ is determined by the combinatorial factor (also called the binomial coefficient) defined by:
\begin{equation}\label{eq:comb_cnk}
    C(n,k) = \combnk{n}{k} = \dfrac{n!}{k!(n-k)!} = \combnk{n}{n-k}\,, \quad k \in \mathbb{N}^n_{\leq},
\end{equation}
where $n$ is the sample size, and $k$ is the number of picked samples without repeat. The total number of all possible distinct SS of a parent network with $n_{el}$ ligaments is:
\begin{equation}\label{eq:total_Nss}
    \combnk{n_{el}}{1} + \combnk{n_{el}}{2} + \combnk{n_{el}}{3} + \dots + \combnk{n_{el}}{n_{el}-1} = 2^{n_{el}} - 2,
\end{equation}
which scales exponentially with $n_{el}$. It is thus typically impractical to analyze all possible SS for larger networks, and SS where $k$ approaches $n/2$. For example, the network shown in Fig.~\ref{fig:network_and_SS} has $110$ ligaments, so there are only 110 order-1 SS, and 5995 order-2 SS, but more than $4\times10^{21}$ order-20 SS.

\subsection{Response of a mechanical system}
Based on standard finite element analysis (FEA), the response of a linear elastic system with $n$ total degrees of freedoms (DOF) under static equilibrium is:
\begin{equation}\label{eq:Kd=f}
\underset{n \times n}{\bfK} \cdot \underset{ n \times 1}{\bfd} = \underset{ n \times 1}{\bff},
\end{equation}
where $\bfK$ is the stiffness matrix of the network, $\bfd$ is the nodal displacements associated with all DOF, and $\bff$ the corresponding nodal force components.

Assuming $r$ out of $n$ DOF are prescribed with essential boundary conditions with displacement values stored in a $r \times 1$ vector $\bfd_E$, then by moving all corresponding columns and rows to the top left corner in $\bfK$ and applying the same reordering to $\bfd$ and $\bff$, the system matrices can be reordered and partitioned to the following form (details can be found in Ref.~\cite{Fish2007.First}):

\begin{equation}\label{eq:partition_Kd=f}
\newcommand{\UP}[2]{\makebox[0pt]{\smash{\raisebox{1.5em}{$\phantom{#2}#1$}}}#2}
\newcommand{\LF}[1]{\makebox[0pt]{$#1$\hspace{4.6em}}}
\renewcommand{\arraystretch}{1.4}
\left[
\begin{array}{c @{}c : c }
    \LF{r}          & \UP{r}{\bfK_E} & \UP{n-r}{\bfK_{EF}} \\
    \hdashline
    \LF{n-r}        & \bfK_{EF}^T     & \bfK_F           \\
\end{array}\right] 
\cdot
\left[\begin{array}{c} \UP{1}{\bfd_E} \\ \hdashline \bfd_F \\ \end{array}\right]
=
\left[\begin{array}{c} \UP{1}{\bfR_E} \\ \hdashline \bff_F \\ \end{array}\right],
\end{equation}
where vector $\bfR_E$ is the reaction force components at element nodes prescribed with displacement loads (essential BC), $\bff_F$ is the external force applied on the free nodes (natural BC), and $\bfd_F$ is the response (nodal displacements) associated with all free DOF.

For the purpose of this study, body force, pre-defined fields, and external force-based loads are excluded, i.e., $\bff_F=\mathbf{0}$, hence the response in all free DOF can be determined entirely by solving:
\begin{equation}\label{eq:solution_dF}
\begin{aligned}
    \bfd_F & =  \invKF \left( \cancelto{\mathbf{0}}{\bff_F} - \bfK_{EF}^T \bfd_E \right) = \underbrace{-\invKF \cdot \bfK_{EF}^T}_{:=\bfT} \cdot \bfd_E \\ & = \bfT \cdot \bfd_E,
\end{aligned}
\end{equation}
where the defined term:
\begin{equation}\label{eq:def_T}
   \underset{(n-r) \times r}{\bfT} := -\invKF \bfK_{EF}^T
\end{equation}
is denoted as \textit{transform matrix}, which correlates prescribed displacements to the response at all free DOF.

Hereafter we shall use integers in angle brackets `$\lri$' to denote a functional mode. Without loss of generality, we assume an input in mode $\lri$ involves a total of $r$ prescribed DOF, and assume the desired output of this mode is defined over a total of $m$ DOF among all free DOF (i.e., $m <= n-r$). For the following, we use \textit{source} to refer to the $r$ DOF with prescribed essential BCs, use \textit{target}, or \textit{target DOF}, to refer to the $m$ DOF requested to realize desired displacements, and use \textit{goal} to indicate the goal/aimed values at a target. Following these terms, define vectors: $\dsrci$ as the prescribed (input) displacements at \textit{source}; $\dtgti$ as the nodal displacement components at \textit{target}; and $\goal{i}$ as the \textit{goal} displacement values at the $m$ target DOF. This way, $\dsrci$ and $\goal{i}$ are user-defined vectors associated with input and output in mode $\lri$, respectively. The core task is to bring the current response at \textit{target} $\dtgti$, under input at \textit{source} $\dsrci$, as close as possible to \textit{goal} $\goal{i}$.

The response vector at all $n-r$ free DOF in this mode, $\bfd_F^{\lri}$, can then be obtained by setting $\dsrci$ equal to $\bfd_E$ in Eq.~\ref{eq:partition_Kd=f} (with corresponding matrix partitions) and then solving it by Eq.~\ref{eq:solution_dF}. Let $\bfG^{\lri}$ be a \textit{gather matrix} associated with the target DOF in mode $\lri$, whose main usage is to store the mapping of entries (indices) of free DOF to the specific $m$ DOF in the target of this mode, namely:
\begin{equation}\label{eq:def_Gi}
\underset{m\times 1}{\dtgti} = \underset{m \times (n-r)}{\bfG^{\lri}} \cdot \underset{ (n-r) \times 1}{\bfd_F^{\lri}}.
\end{equation}
Given its usage, $\bfG$ is a matrix whose components can only take integers $0$ or $1$ (representing Boolean values). It strictly contains $m$ non-zero entries with exactly one unit value in each row. For instance, if a mechanical system has in total six free DOF ($\bfd_F$), in which the third and fifth entries are the two target DOF, then $\bfG$ can be obtained by `gathering' this mapping rule:
\begin{equation}\label{eq:example_of_G}
\begin{aligned}
&\text{if:} \,\ \underset{2\times 1}{\bfd_{tgt}}[1] = \underset{6 \times 1}{\bfd_F}[3], \,\ \underset{2\times 1}{\bfd_{tgt}}[2] = \underset{6 \times 1}{\bfd_F}[5] \\
&\text{then:} \,\ \bfG = \begin{bmatrix} 0 &  0 & 1 & 0 & 0 & 0 \\ 0 &  0 & 0 & 0 & 1 & 0 \end{bmatrix}.
\end{aligned}
\end{equation}
This way, by substituting Eq.~\ref{eq:solution_dF} into Eq.~\ref{eq:def_Gi}, the response at the target DOF in mode $\lri$ is: 
\begin{equation}\label{eq:dtgt=Adsrc}
\underset{m\times 1}{\bfd_{tgt}^{\lri}}
= \underbrace{\bfG^{\lri} \cdot \bfT^{\lri}}_{:=\bfA} \cdot \bfd_E^{\lri} 
= \underset{m\times r}{\bfA^{\lri}} \cdot \underset{r\times 1}\dsrci,
\end{equation}
where the defined matrix:
\begin{equation}\label{eq:def_A}
\underset{m\times r}{\bfA} := \bfG \bfT = -\bfG \bfK_{F}^{-1} \bfK_{EF}^T
\end{equation}
is named an \textit{actuation matrix} (of a specific mode), which correlates input at the source to response at the target. Eq.~\ref{eq:def_A} shows the general form of $\bfA$, where the superscripts labeling a specific mode, $\lri$, are skipped. Dimension and values of all matrices in Eq.~\ref{eq:def_A} are mode dependent, and hence they need to be generated respectively to each mode in multi-mode design problems.

\subsection{Stiffness matrix of a network and its substructures}
Assuming a network consists of truss ligaments with freely-rotating nodes (hinges), then its stiffness matrix can be obtained by meshing each ligament into one truss element and applying a stiffness assembly. From standard FEA, the universal stiffness matrix of a two-node, first-order, linear truss element is:
\begin{equation}\label{eq:def_K_TE2}
    \bfK_{TE2}^{ref} = 
        \begin{bmatrix} 1 & 0 & -1 & 0 \\ 0 & 0 & 0 & 0 \\ -1 & 0 & 1 & 0 \\ 0 & 0 & 0 & 0 \end{bmatrix},
\end{equation}
where the first two column/rows correspond to nodal DOF of one node of the truss element, and third and fourth column/rows correspond to DOF associated with the second node, respectively. The subscript TE2 is an abbreviation for a `truss element with two nodes', and the entries in $\bfK_{TE2}^{ref}$ are determined based on the correlation of DOF in its reference frame where the element is placed horizontally. 

\textbf{Stiffness matrix of parent network.} Subsequently, the global stiffness matrix of the entire network (with $n$ total DOF and $n_{el}$ ligaments) can be generated by stiffness assembly:
\begin{equation}\label{eq:asmb_glbK}
    \begin{aligned}
    \underset{n \times n}{\bfK} & = \sum_{e=1}^{n_{el}} \bfH_e^T \cdot \left[\bfR_e^T \cdot  \left( \frac{A_e E_e}{L_e} \cdot \bfK_{TE2}^{ref} \right) \cdot \bfR_e\right]  \cdot \bfH_e \\
    & = \sum_{e=1}^{n_{el}} \bfH_e^T \cdot \underbrace{\left[ \frac{A_e E_e}{L_e} \cdot \bfR_e^T \cdot \bfK_{TE2}^{ref} \cdot \bfR_e\right]}_{:=\bfK_{e}^{glb}}  \cdot \bfH_e \\
    & = \sum_{e=1}^{n_{el}} \underset{n \times 4}{\bfH_e^T} \cdot \underset{4 \times 4}{\bfK_{e}^{glb}} \cdot \underset{4 \times n}{\bfH_e}.
    \end{aligned}
\end{equation}
In the above Eq.~\ref{eq:asmb_glbK}: $E_e$, $A_e$, and $L_e$ are the material stiffness (Young's modulus), cross-section area, and length, respectively, $\bfR_e$ is the rotation matrix of this ligament based on its direction in the global coordinate system (CSYS), with $\bfK_{e}^{glb}$ defined by:
\begin{equation}\label{eq:def_Ke_glb}
    \bfK_{e}^{glb} :=\frac{A_e E_e}{L_e} \cdot \bfR_e^T \cdot \bfK_{TE2}^{ref} \cdot \bfR_e\,, \quad e \in \mathbb{N}^{all},
\end{equation}
called the \textit{standard stiffness matrix} of the $e\text{th}$ element (ligament) in the global CSYS of the network, and $\bfH_e$ is the gather matrix that maps components from the element-wise stiffness matrix to the structure-wise stiffness matrix in the same CSYS. Given its definition, $\bfH_e$ is also a Boolean-valued matrix (similar to $\bfG$, as shown in Eq.~\ref{eq:example_of_G}).

\textbf{Stiffness matrix of substructure.} The stiffness matrix of a SS, denoted as $\bfS_{[j]}$, is conceptually comparable to $\bfK$, and is constructed in the same manner as its parent network since they share the same topology. As a result, $\bfS_{[j]}$ has the same dimension as $\bfK$, and components from the same position (i.e., with the same column and/or row indices) in $\bfK$ and $\bfS_{[j]}$ correspond to the same nodal DOF because the gather matrices ($\bfH_e, \forall e \in \mathbb{N}^{all}$) are identical for all ligaments. The general form of $\bfS_{[j]}$ is thus:
\begin{equation}\label{eq:def_Sj_general}
     \underset{n \times n}{\bfS_{[j]}} = \sum_{e=1}^{n_{el}} \underset{n \times 4}{ \bfH_e^T} \cdot \underset{4 \times 4}{\bfK_{e}^{ss[j]}} \cdot \underset{4 \times n} {\bfH_e},
\end{equation}
where $\bfK_{e}^{ss[j]}$ is defined as the \textit{base stiffness matrix} of the $e\text{th}$ element (ligament) in SS~$[j]$, which takes a piece-wise form depending on whether this ligament is phantom or real in this SS:
\begin{equation}\label{eq:def_Ke_SSj_piecewise}
\bfK_{e}^{ss[j]} := 
    \begin{cases}
        \bfK_{}^{ph} & \text{if} \,\ e \notin \mathbb{L}^{re}_{[j]} \\
        \bfK_{e,[j]}^{re} & \text{if} \,\ e \in \mathbb{L}^{re}_{[j]}, \\
    \end{cases}
\end{equation}
where $\bfK_{}^{ph}$ is the \textit{base stiffness matrix of phantom ligaments}, $\bfK_{e,[j]}^{re}$ is the \textit{base stiffness matrix of real ligaments} in SS~$[j]$, and their expressions are explained below. 

The definition of phantom ligaments intuitively results in a matrix consisting of zeros in $\bfK^{ph}$. Specifically, by considering a case of material stiffness ($E$ in Eq.~\ref{eq:def_Ke_glb}) tending to zero, $\bfK^{ph}$ becomes a zero matrix regardless of its length and nodal positions:
\begin{equation}\label{eq:def_K_phan}
    \bfK_{}^{ph} := \lim_{E \to 0}\bfK_{e}^{glb} = \begin{bmatrix} 0 & 0 & 0 & 0 \\ 0 & 0 & 0 & 0 \\ 0 & 0 & 0 & 0 \\ 0 & 0 & 0 & 0 \end{bmatrix}
    := \mathbf{0}_{4\times4}.
\end{equation}
Thus, the phantom ligaments no longer contribute to the values of $\bfS_{[j]}$ in the stiffness assembly (Eq.~\ref{eq:def_Sj_general}), as if they are being skipped.

The form of the base stiffness matrix of a real ligament in a SS may be defined in many different ways, depending on what is most useful for the design strategy.
The general idea is to only retain unchanging terms in the standard ligament stiffness matrix of a ligament in its global CSYS during assembly ($\bfK_{e}^{glb}$), and replace adjustable parameters with reference constant values (e.g., unit). This way, $\bfS_{[j]}$ becomes a `base matrix' once created, without needing to be computed again during iterative design of a structure.

For instance, assuming ligament material stiffnesses $E$ are adjustable parameters during iterative structure design, then $\bfK_{e,[j]}^{re}$ can take the form:
\begin{equation}\label{eq:def_Ke_real}
\begin{aligned}
    \bfK_{e, [j]}^{re} &:= \frac{A_e}{L_e} \cdot \cancelto{1}{E_e} \cdot \bfR_e^T \cdot \bfK_{TE2}^{ref} \cdot \bfR_e \\
    & = \frac{A_e}{L_e} \cdot \bfR_e^T \cdot \bfK_{TE2}^{ref} \cdot \bfR_e \, , \quad e \in \mathbb{L}^{re}_{[j]},
\end{aligned}
\end{equation}
where the variable $E_e$ is replaced by one. 

Subsequently, if an order-1 SS~$[j]$ whose only real ligament has label $q$, then its stiffness matrix, denoted by $\bfS_{o1}^{\{q\}}$, takes form:
\begin{equation}\label{eq:def_Sj_o1}
    \begin{aligned}
    \text{if:} \,\ \mathbb{L}^{re}_{[j]} & = \left\{ q \right\} \\
    \text{then:} \,\ \underset{n \times n}{\bfS_{o1}^{\{q\}}} &= \sum_{e=1}^{n_{el}} \bfH_e^T \cdot \bfK_{e}^{ss[j]} \cdot \bfH_e = \bfH_q^T \cdot \bfK_{q,[j]}^{re} \cdot \bfH_q \\
    & = \frac{A_q}{L_q} \cdot \bfH_q^T \bfR_q^T \cdot \bfK_{TE2}^{ref} \cdot \bfR_q \bfH_q.
    \end{aligned}
\end{equation}
Note that $\bfK$ in general needs to be positive definite (i.e., non-singular) such that the network has valid and unique response, whereas $\bfS_{[j]}$ can be singular due to existing all-zero columns and rows associated with nodal DOF of phantom ligaments. For example, the stiffness matrices of the four SS shown in Fig.~\ref{fig:network_and_SS} are all singular.

\textbf{Decomposition of stiffness matrices.} 
The above derivations indicates that $\bfS_{[j]}$ follows superposition principals, whereby it can be decomposed into stiffness matrices of certain lower order substructures. For example, $\bfS_{[j]}$ can be expressed as a linear combination of the stiffness matrices of all order-1 SS each contains one of its real ligaments:
\begin{equation}\label{eq:Sj=sum_So1}
    \bfS_{[j]} = \sum_{q \in \mathbb{L}^{re}_{[j]}} \bfS_{o1}^{\{q\}}.
\end{equation}
Note that Eq.~\ref{eq:Sj=sum_So1} shows the \textit{ground-level decomposition}, or \textit{order-1 decomposition}, of $\bfS_{[j]}$. In fact, the terms in Eq.~\ref{eq:Sj=sum_So1} can be rearranged (into subgroups) to decompose $\bfS_{[j]}$ into stiffness matrices of some first and/or higher order substructures if their real ligaments are mutually distinct and together cover all the real ligaments in SS~$[j]$. More precisely:
\begin{equation}\label{eq:Sj_decomp_general}
\begin{aligned}
&\text{if:} \,\ \forall \ k,l \in \mathbb{N}^{n_d}_{\leq} \ \rightarrow \ \mathbb{L}^{re}_{[k]} \cap \mathbb{L}^{re}_{[l]} = \varnothing \ \text{and} \ \bigcup_{l=1}^{n_d} \mathbb{L}^{re}_{[l]} = \mathbb{L}^{re}_{[j]}\\
&\text{then:} \ \bfS_{[j]} = \sum_{l=1}^{n_d} \bfS_{[l]},
\end{aligned}
\end{equation}
where $n_d$ is the total number of substructures (in this decomposition) labeled by integers from set $\mathbb{N}^{n_d}_{\leq}$. As an example, Fig.~\ref{fig:network_and_SS}C shows three SS from one possible decomposition of the SS shown in Fig.~\ref{fig:network_and_SS}B.

Analogously, the stiffness matrix of the parent network, $\bfK$, can also be decomposed into its substructures in various ways. Based on Eq.~\ref{eq:asmb_glbK}, Eq.~\ref{eq:def_Sj_o1}, and Eq.~\ref{eq:Sj=sum_So1}, the order-1 decomposition of $\bfK$ takes form:
\begin{equation}\label{eq:K=sumSo1}
    \begin{aligned}
        \bfK &= \sum_{e=1}^{n_{el}} \bfH_e^T \cdot \bfK_{e}^{glb} \cdot \bfH_e = \sum_{e=1}^{n_{el}} E_e \cdot \bfS_{o1}^{\{e\}}.
    \end{aligned}
\end{equation}

\subsection{Substructure perturbation}\label{subsec:SSPM}
The overall concept of substructure perturbation is to change some structural features or properties in the parent network based on a collection of its substructures constructed in accordance with the types of features. In this study, we consider changing material stiffness of ligaments ($E_e$) to realize desired functional modes, and subsequently define the corresponding base stiffness matrix of real ligaments in substructures ($\bfK_{e,[j]}^{re}$ in Eq.~\ref{eq:def_Ke_real}). 

\textbf{Ligament stiffness perturbation by substructure(s)}
For a given SS~$[j]$, apply a small stiffness change, $-\alpha$, in each of the ligaments in the parent network whose label is in $\mathbb{L}^{re}_{[j]}$, (i.e., a real ligament in SS~$[j]$), referred to as a\textit{ligament stiffness perturbation} (LSP), which shall be considered as one subcategory of SSPM.

Then based on Eq.~\ref{eq:asmb_glbK}, Eq.~\ref{eq:Sj=sum_So1}, and Eq.~\ref{eq:def_Sj_o1}, the updated stiffness matrix of the parent network becomes:
\begin{equation}\label{eq:def_Kpj}
    \begin{aligned}
    \bfK^{p_{[j]}} &:= \sum_{q \notin \mathbb{L}^{re}_{[j]}} \frac{A_q E_q}{L_q} \bfH_q^T \bfR_q^T \cdot \bfK_{TE2}^{ref} \cdot \bfR_q \bfH_q + \sum_{q \in \mathbb{L}^{re}_{[j]}} \frac{A_q (E_q-\alpha)}{L_q} \bfH_q^T \bfR_q^T \cdot \bfK_{TE2}^{ref} \cdot \bfR_q \bfH_q \\  
    &= \sum_{q \in \mathbb{N}^{all}} \bfH_q^T  \bfK_{q}^{glb} \bfH_q - \sum_{q \in \mathbb{L}^{re}_{[j]}} \alpha \frac{A_q}{L_q} \bfH_q^T \bfR_q^T\bfK_{TE2}^{ref} \bfR_q \bfH_q \\
    & = \bfK - \alpha \sum_{q \in \mathbb{L}^{re}_{[j]}} \bfS_{o1}^{\{q\}} \\
    & = \bfK - \alpha \bfS_{[j]}, \\
    \end{aligned}
\end{equation}
where the defined term $\bfK^{p_{[j]}}$ is the resultant \textit{perturbed stiffness matrix} of the parent network due to a \textit{negative-LSP} by SS~$[j]$.

\textbf{Stiffness perturbation by multiple substructures.}
The relation in Eq.~\ref{eq:def_Kpj} can be extended to cases if applying LSP by multiple substructures sequentially, in which the perturbed stiffness matrix becomes:
\begin{equation}\label{eq:def_Kpjkl}
    \bfK^{p_{[j,k,l,\dots]}} = \bfK - \alpha_j \bfS_{[j]} - \alpha_k \bfS_{[k]} - \alpha_l \bfS_{[l]} - \dots,  
\end{equation}
where the perturbation coefficients, $\alpha_j, \alpha_k, \dots$, can be distinct, and can be positive or negative. The substructures do not need to be mutually disjoint, i.e., they can share common real ligaments.

\subsection{Change in structural response}\label{subsec:Kpj_and_approx}
A subsequent core interest is how LSP would affect the response of the network. Since $\bfS_{[j]}$ have the same dimension as $\bfK$, and they both are constructed with identical gather matrices $\bfH$ (i.e., same element-to-structure DOF mapping rule), the partitioned submatrices of $\bfS$ by Eq.~\ref{eq:partition_Kd=f}, i.e., $\bfS_{EF,[j]}$ and $\bfS_{F,[j]}$, also share the same dimension and DOF-mapping rule as $\bfK_{EF}$ and $\bfK_F$, respectively. Therefore, by applying the same matrix partitioning on both sides of Eq.~\ref{eq:def_Kpj}, the following two relations are obtained:
\begin{equation}\label{eq:Kpj_sub_matrices}
\begin{aligned}
    \bfK_{EF}^{p_{[j]}} &= \bfK_{EF} - \alpha \bfS_{EF,[j]} \\
    \bfK_{F}^{p_{[j]}} &= \bfK_{F} - \alpha \bfS_{F,[j]}.
\end{aligned}
\end{equation}

Then, by substituting Eq.~\ref{eq:Kpj_sub_matrices} into Eq.~\ref{eq:def_T}, the \textit{perturbed transformation matrix} due to a negative-LSP by SS~$[j]$ takes the form:
\begin{equation}\label{eq:def_Tpj(direct)}
    \bfT_{p_{[j]}} := - \left( \bfK^{}_{F} - \alpha \bfS_{F,[j]} \right)^{-1} \left( \bfK_{EF}^T - \alpha \bfS_{EF,[j]}^T \right).
\end{equation}

In the explicit form of $\bfT_{p[j]}$ in Eq.~\ref{eq:def_Tpj(direct)}, a term with significant computational cost is the matrix inverse. Here we apply Hua's identity~\cite{Cohn2002.further}, or so-called Shearman-Morrison-Woodbury (SMW) formula~\cite{Hager1989.updating}, to derive an approximation of $\bfT_{p_{[j]}}$. In a special case of the SMW formulas where $U,V$ are square matrices with the same dimension, it is shown that:
\begin{align}
     \left( \bfU - \bfV \right)^{-1} 
     & = \bfU^{-1} + \bfU^{-1} \bfV \left( \bfU - \bfV \right)^{-1} \label{eq:SMW_recursive} \\
     & = \sum_{k=0}^{\infty} \left(\bfU^{-1} \bfV\right)^{k} \bfU^{-1} \,\, \leftarrow \text{if} \,\ \bfU^{-1} \bfV < 1, \label{eq:SMW_inf_series}
\end{align}
where Eq.~\ref{eq:SMW_recursive} has a recursive structure and hence can be expanded to an infinite series form in Eq.~\ref{eq:SMW_inf_series} if the spectral radius of $\bfV$ is smaller than $\bfU$ (i.e., $\bfU^{-1} \bfV < 1$). 

By applying a small multiplier (scaling coefficient) on matrix $\bfV$ in Eq.~\ref{eq:SMW_inf_series}, the infinite series can then be approximated by neglecting higher order exponential terms:
\begin{equation}\label{eq:inv(U-V)_approx}
\begin{aligned}
     \left( \bfU - \alpha \bfV \right)^{-1} & = \sum_{k=0}^{\infty} \alpha^k  \left(\bfU^{-1} \bfV\right)^{k} \bfU^{-1} \\
     & = \bfU^{-1} + \alpha \bfU^{-1} \bfV \bfU^{-1} + \underbrace{\alpha^2 \left(\bfU^{-1} \bfV \right)^2 \bfU^{-1}}_{\mathcal{O}(\alpha^2)}
      + \underbrace{\alpha^3 \left(\bfU^{-1} \bfV \right)^3 \bfU^{-1}}_{\mathcal{O}(\alpha^3)} + \mathcal{O}(\alpha^4) + \dots  \\
     & \approx \bfU^{-1} + \alpha \bfU^{-1} \bfV \bfU^{-1}.
\end{aligned}
\end{equation}

Then by applying results in Eq.~\ref{eq:inv(U-V)_approx} onto Eq.~\ref{eq:def_T}, the \textit{perturbed transformation matrix} due to a negative-LSP by SS~$[j]$, $\bfT_{p_{[j]}}$, can be approximated by:
\begin{equation}\label{eq:Tpj_approx}
\begin{aligned}
    \bfT_{p_{[j]}} &= - \left( \bfK^{}_{F} - \alpha \bfS_{F,[j]} \right)^{-1} \left( \bfK_{EF}^T - \alpha \bfS_{EF,[j]}^T \right) \\
    &  \approx - \left(\invKF + \alpha \invKF \bfS_{F,[j]} \invKF \right) \left( \bfK_{EF}^T - \alpha \bfS_{EF,[j]}^T \right) \\
    &  = - \invKF \bfK_{EF}^T - \alpha\invKF\bfS_{F,[j]}\invKF\bfK_{EF}^T 
    + \alpha \invKF \bfS_{EF,[j]}^T + \underbrace{\alpha^2 \invKF \bfS_{F,[j]} \invKF \bfK_{EF}^T \bfS_{EF,[j]}^T}_{\mathcal{O}(\alpha^2)} \\
    & \approx  -\invKF\bfK_{EF}^T - \alpha \invKF \left( \bfS_{F,[j]} \invKF \bfK_{EF}^T - \bfS_{EF,[j]}^T \right)\\
    & = \bfT - \alpha \cdot \underbrace{\left( \invKF  \bfS_{F,[j]} \invKF \bfK_{EF}^T - \invKF \bfS_{EF,[j]}^T \right)}_{:= \mathbf{P}_{[j]}} \\
    &  = \bfT - \alpha  \mathbf{P}_{[j]}
\end{aligned}
\end{equation}
in which another higher order component $\mathcal{O}(\alpha^2)$ is neglected by assuming $\alpha$ is small. The defined term:
\begin{equation}\label{eq:Pj_def}
\mathbf{P}_{[j]} := \invKF  \bfS_{F,[j]} \invKF \bfK_{EF}^T - \invKF \bfS_{EF,[j]}^T
\end{equation}
is an important term denoted by \textit{perturbational effect matrix of SS~$[j]$}, which shows how effective is an infinitesimal LSP by SS~$[j]$ on the structural response (in a mode), and it does not depend on the assumed stiffness perturbation value $\alpha$. 

Note that Eq.~\ref{eq:Tpj_approx} is valid when $ \alpha\invKF\bfS_{F,[j]} < 1$ due to condition on spectral radius in Eq.~\ref{eq:SMW_inf_series}, which is true if $\alpha \to 0$. Higher order terms (i.e., $\mathcal{O}(\alpha^k)$ with $k\geq 2$) in Eq.~\ref{eq:inv(U-V)_approx}, and second order term $\mathcal{O}(\alpha^2)$ in Eq.~\ref{eq:Tpj_approx}, are neglected, which is justifiable when the magnitude of $\alpha$ is small. Also note that matrix $\bfU$ in Eq.~\ref{eq:SMW_recursive} needs to be positive definite (invertible), whereas square matrix $\bfV$ can be either invertible or singular. The stiffness matrices of the network and its substructures, $\bfK$ and $\bfS_{[j]}$, satisfy these conditions because $\bfK_F$ must be invertible to ensure non-trivial solution of structural response, while there is no requirement on singularity on $\bfS_{[j]}$.

The above derivation involving LSP by a single SS (Eq.~\ref{eq:def_Kpj}) can be directly extended to LSP by multiple SS (Eq.~\ref{eq:def_Kpjkl}). The resultant approximate \textit{perturbed transformation matrix} takes the form:
\begin{equation}\label{eq:Tpjkl_approx}
    \bfT_{p[j,k,l, \dots]} = \bfT - \alpha_j \mathbf{P}_{[j]} - \alpha_k \mathbf{P}_{[k]} - \alpha_l \mathbf{P}_{[l]} - \dots
\end{equation}
with more details on derivation of Eq.~\ref{eq:Tpjkl_approx} provided in the supplementary information (SI). 

Since the actuation matrix $\bfA$ is simply a submatrix of $\bfT$ (sliced by gather matrix $\bfG$ in Eq.~\ref{eq:def_A}), the approximate \textit{perturbed actuation stiffness matrix} of the parent network is derived following the same procedures as in Eq.~\ref{eq:Tpj_approx} and Eq.~\ref{eq:Tpjkl_approx}, which gives:
\begin{equation}\label{eq:Apj_approx}
\begin{aligned}
    \bfA_{p_{[j]}} & := -\bfG  \left( \bfK^{}_{F} - \alpha \bfS_{F,[j]} \right)^{-1} \left( \bfK_{EF}^T - \alpha \bfS_{EF,[j]}^T \right) \\
    & = -\bfG \cdot \bfT_{p[j]} \\
    & \approx -\bfG \left( \bfT - \alpha \bfP_{[j]} \right) \\
    & = \bfA - \alpha \bfG \bfP_{[j]},
\end{aligned}
\end{equation}
and similarly for LSP by multiple substructures:
\begin{equation}\label{eq:Apjkl_approx}
    \bfA_{p[j,k,l, \dots]} = \bfA - \alpha_j \mathbf{P}_{[j]} - \alpha_k \mathbf{P}_{[k]} - \alpha_l \mathbf{P}_{[l]} - \dots
\end{equation}

An important feature of $\bfT_{p[j,...]}$ and $\bfA_{p[j,...]}$  is that they only depend on the assumed stiffness perturbation values, $\alpha_j$, and perturbational effect matrix of each involved SS, $\bfP_{[j]}$. Calculation of $\bfP_{[j]}$ (Eq.~\ref{eq:Pj_def}) only requires solving for the current response(s) of the network by obtaining $\bfK^{-1}$. No extra matrix inverse besides $\bfK^{-1}$ is required, and $\bfS_{EF,[j]}, \bfS_{EF,[j]}$ are just characteristic properties of SS~$[j]$ that only need to be calculated once during their initiation. This provides significant advantages in computational efficiency in the proposed SSPM.

Another important term is the approximate \textit{rate of response at free DOF} due to negative-LSP by SS~$[j]$ in mode $\lri$, denoted by $\mathbf{v}_{F,p[j]}^{\lri}$. By combining Eq.~\ref{eq:solution_dF}, Eq.~\ref{eq:Tpj_approx}, and taking the derivative with respect to $\alpha$, it takes the form:
\begin{equation}\label{eq:def_vFpj}
    \mathbf{v}_{F,p[j]}^{\lri} := \pdv{}{\alpha} \left( \bfT^{\lri}_{p[j]} \dsrci - \bfT^{\lri}\dsrci \right) = -\mathbf{P}_{[j]}^{\lri} \dsrci,
\end{equation}
which is simply the perturbational effect matrix multiplied by the input.

\subsection{Errors analysis}\label{subsec:errors}
Here we introduce several measures to quantify the difference between current response at target DOF, $\dtgti$, and its desired goal, $\goal{i}$.
\begin{itemize}
\item \textbf{True error.} 
Define the \textit{true error} in a mode, vector $\Delta^{\lri}$, as the real difference between the current response at target DOF and its goal:
\begin{equation}\label{eq:def_Delta}
    \Delta^{\lri} = \dtgti - \goal{i}
\end{equation}

\item \textbf{Distance to goal.} Define the \textit{distance to goal}, scalar $\Omega^{\lri}$, as the $L^2\text{-norm}$ of the true error $\Delta^{\lri}$:
    \begin{equation}\label{eq:def_Omega}
        \begin{aligned}
            \Omega^{\lri} := & \Lnorm{\dtgti-\goal{i}} := \Lnorm{\Delta^{\lri}} \\
            = & \left[ \sum_{r=1}^{m}  \left( \dtgti - \goal{i} \right)_{r} \right]^{\frac{1}{2}} = \left[ \sum_{r=1}^{m}  \Delta^{\lri}_r \right]^{\frac{1}{2}}
        \end{aligned}
    \end{equation}

\item \textbf{Relative error.} Define the \textit{relative error}, vector $\Psi^{\lri}$, as the `normalized' true error at each DOF by its corresponding goal value, but skip normalization if the goal of a target DOF is zero:
\begin{equation}\label{eq:def_Psi}
\Psi^{\lri}_r = 
    \begin{cases}
        \dfrac{\compo{\dtgti}{r} - \compo{\goal{i}}{r}}{\compo{\goal{i}}{r}} & \text{if} \,\ \compo{\goal{i}}{r} \neq 0 \\
        \dtgti & \text{if } \compo{\goal{i}}{r} = 0.
    \end{cases}
\end{equation}
\end{itemize}

The piecewise form of relative error prevents division by zero and ensures that error is measured relatively rather than dependent on the magnitude of goal. The true error will be primarily used in the following theoretical analysis, whereas the relative error will be used as an objective to judge if a `solution' has been achieved in numeral algorithms.

To improve computational efficiency, the expected rate of change of errors are of primary focus. Firstly, given the complexity of the stiffness assembly (Eq.~\ref{eq:asmb_glbK}) and matrix inverse (Eq.~\ref{eq:solution_dF}), it is not efficient, especially for a large system (e.g., with more than $10^3$ DOF), to apply `trial-error' steps to frequently change the system and then recompute its responses to check whether the attempted configuration works. Secondly, given the immense number of possible substructures (Eq.~\ref{eq:total_Nss}), such a `trial-error' method could quickly become impractical even for a small system (e.g., with 50 total DOF) in scenarios where many higher order substructures may need to be inspected. 

Therefore, for a given SS, we analyze the tendency of errors if LSP by this SS were applied, rather than actually applying it in the network and then computing the errors. This helps to quickly determine the effectiveness of given substructure(s).

\textbf{Rate of change in distance to goal.} By taking the derivative of $\Delta^{\lri}$ to $\alpha$, and taking the limit as $\alpha \to 0$, we obtain the \textit{rate of change in distance} to goal under an infinitesimal negative-LSP by SS~$[j]$ (full derivations provided in SI):
\begin{equation}\label{eq:limit_of_dOmega}
\begin{aligned}
\lim_{\alpha \to 0} \pdv{\Omega^{\lri}_{p[j]}}{\alpha} &= \lim_{\alpha \to 0} \pdv*{ \left( \Lnorm{\bfA_{p_{[j]}}^{\lri}\dsrci-\goal{i}}-\Lnorm{\Delta^{\lri}} \right) }{\alpha} \\
& = \lim_{\alpha \to 0} \pdv*{\left( \Lnorm{\Delta^{\lri} - \alpha \bfG^{\lri} \bfP_{[j]}^{\lri} \dsrci} - \Lnorm{\Delta^{\lri}} \right)}{\alpha} \\
& = \dfrac{-\Delta^{\lri} \odot \bfG^{\lri} \bfP_{[j]}^{\lri}\dsrci }{\Lnorm{\Delta^{\lri}}},
\end{aligned}
\end{equation}
where the numerator is a dot product (denoted by $\odot$). This can seemingly be the condition for finding suitable substructures, since negative rate means decrease in distance to goal. However, since $\Omega^{\lri}$ is the norm of the vector containing true errors, a decrease in $\Omega^{\lri}$ does not guarantee that errors at all DOF are decreasing. Therefore, a more strict condition is introduced below, which ensures that errors at all DOF are expected to decrease.

\textbf{Rate of change in true errors.} Based on Eq.~\ref{eq:limit_of_dOmega} and Eq.~\ref{eq:def_vFpj}, we define the vector:
\begin{equation}\label{eq:def_beta_i}
\beta^{\lri}_{p[j]} := \left[ - \bfG^{\lri} \bfP_{[j]}^{\lri} \dsrci \right] \oslash \Delta^{\lri} = \left[ \bfG^{\lri} \mathbf{v}_{F,p[j]}^{\lri} \right] \oslash \Delta^{\lri},
\end{equation}
where $\oslash$ indicates element-wise division. The defined vector $\beta^{\lri}_{p[j]}$ is the \textit{rate of change in true error at target DOF} in mode $\lri$ due to an infinitesimal negative-LSP on the current parent network by SS~$[j]$. Note that $\beta$ represents a rate of change, and henceforth does not depend on $\alpha$. Its meaning can also be explained by observing that $-\alpha \bfP_{[j]}^{\lri} \dsrci$ is the \textit{change in response} in all free DOF (Eq.~\ref{eq:limit_of_dOmega}), and thus $-\bfG^{\lri}\bfP_{[j]}^{\lri}\dsrci$ is the \textit{rate of change in response} in target DOF. Subsequently the resultant $\beta^{\lri}_{p[j]}$ (normalized by current true error) is the rate of change in true error at the corresponding DOF. 

The meaning of vector $\beta$ leads to two useful non-negative, and dimensionless measures. The first is the \textit{relative standard deviation} (RSD) of $\beta$, i.e., its standard deviation $\sigma(\beta)$ normalized by the magnitude of their average $\mu(\beta)$:
    \begin{equation}\label{eq:def_RSD}
        RSD(\beta) := \dfrac{\sigma(\beta)}{|\mu(\beta)|} \geq 0,
    \end{equation}
which indicates the relative dispersion of the values in it. Here, it can be used to estimate how `evenly' the rates of decrease in true errors are. The second is the Euclidean norm ($L^2\text{-norm}$) of $\beta$, which shows overall how `rapidly' the true errors would decrease. A schematic showing the meaning of $\beta$ and its RSD for a target involving two DOF is given in Fig.~\ref{fig:schematic_of_beta} in SI. Ideally, if $\beta \to 0$, then the errors would decrease by the same percentage relative to their current magnitudes.

The above results are based on assumption of attempting a negative-LSP as defined in Eq.~\ref{eq:def_Kpj}. If swapping $-\alpha$ to $\alpha$ in Eq.~\ref{eq:def_Kpj}, hence applying a \textit{positive-LSP} (i.e., $\bfK + \alpha \bfS_{[j]}$), the derived approximate formula for perturbed matrices (Eqs.~\ref{eq:inv(U-V)_approx} and \ref{eq:Tpj_approx}) also work reversely by swapping the signs of $\alpha$ since higher order terms ($\alpha^2, \alpha^3, \dots$) are neglected during linear approximation in these equations.

\subsection{Beneficial ligament stiffness perturbation}

We can now determine if negative or positive-LSP by a substructure would simultaneously decrease errors in all target DOF in each mode. Here we define such a stiffness perturbation as a \textit{beneficial-LSP}, and the two constraints are:
\begin{subequations}
\begin{align}
& \textbf{Beneficial-LSP by SS~$[j]$ for mode} \lri: \nonumber \\
& \ \bfK - \alpha \bfS_{[j]} \quad \text{if} \,\ \forall \, r = 1,\dots, m_{tgt}^{\lri} \rightarrow \, \compo{\beta^{\lri}_{p[j]}}{r} < 0 \label{eq:def_EffSP_neg} \\
& \ \bfK + \alpha \bfS_{[j]} \quad \text{if} \,\ \forall \, r = 1,\dots, m_{tgt}^{\lri} \rightarrow \, \compo{\beta^{\lri}_{p[j]}}{r} > 0.\label{eq:def_EffSP_pos}
\end{align}
\end{subequations}
Namely, if all components of $\beta^{\lri}_{p[j]}$ are negative, then applying negative-LSP by SS~$[j]$ would reduce magnitudes of true errors at all target DOF in this mode, and reversely, if all components of $\beta^{\lri}_{p[j]}$ are positive, then positive-LSP by SS~$[j]$ (change $-\alpha$ to $+\alpha$ in Eq.~\ref{eq:def_Kpj}) would reduce all true errors. Moreover, if components of $\beta^{\lri}_{p[j]}$ are not sharing the same sign, LSP by this SS would not reduce errors at all DOF simultaneously. A special case is when $\beta$ contains zeros, indicating true errors at corresponding DOF would remain unchanged (see Fig.~\ref{fig:schematic_of_beta}), which may also be considered as a beneficial-LSP as long as the non-zero component(s) share the same sign.

When multiple modes need to be realized simultaneously by the network, a LSP by SS~$[j]$ is beneficial only if the positive or negative-LSP is beneficial for all modes concurrently. Mixed directions in LSP (e.g., negative for mode $\AngBr{1}$, positive for mode $\AngBr{2}$) are not allowed since multi-mode functions are realized by the same network. This requires that all components in $\beta$ of each mode must have the same sign, namely:
\begin{subequations}
\begin{align}
& \textbf{Beneficial-LSP by SS~$[j]$ for multiple modes}: \nonumber \\
& \ \bfK - \alpha \bfS_{[j]} \quad \text{if} \,\ \forall \,\, i \in \NTo{n_m}, \, r \in \NTo{m_{tgt}^{\lri}} \rightarrow \, \compo{\beta^{\lri}_{p[j]}}{r} < 0 \label{eq:def_EffSP_neg_multi} \\
& \ \bfK + \alpha \bfS_{[j]} \quad \text{if} \,\ \forall \,\, i \in \NTo{n_m}, \, r \in \NTo{m_{tgt}^{\lri}} \rightarrow \, \compo{\beta^{\lri}_{p[j]}}{r} > 0. \label{eq:def_EffSP_pos_multi}
\end{align}
\end{subequations}

The above condition can also be expressed in a simpler form by concatenating $\beta$ from all modes (and then checking its components):
\begin{equation}\label{eq:def_concate_beta}
\begin{aligned}
    \underset{ m_{\Sigma} \times 1}{\beta_{p[j]}^{\Sigma}} & := \left[ \left(\beta_{[j]}^{\AngBr{1}}\right)^T \,\, \left(\beta_{[j]}^{\AngBr{2}}\right)^T  \dots \,\, \left(\beta_{[j]}^{\AngBr{n_m}}\right)^T   \right]
    \\
    &= \left[
        \compo{\beta_{[j]}^{\AngBr{1}}}{1} \dots \compo{\beta_{[j]}^{\AngBr{1}}}{m_{tgt}^{\AngBr{1}}}
        ,
        \compo{\beta_{[j]}^{\AngBr{2}}}{1} \dots \compo{\beta_{[j]}^{\AngBr{2}}}{m_{tgt}^{\AngBr{2}}}
        ,
    \dots 
    \right]^{T},
\end{aligned}
\end{equation}
where $m_{\Sigma}= \sum_{i=1}^{n_m} m_{tgt}^{\AngBr{i}} $ is the summation of number of target DOF in each mode, and note that concatenated $\beta$ may contain multiple measures of one global DOF from different modes. Then the condition for beneficial-LSP by one substructure for one and/or multiple modes can be expressed alternatively as:
\begin{subequations}
\begin{align}
& \textbf{Beneficial-LSP by SS~$[j]$ for all modes}: \nonumber \\
& \ \bfK - \alpha \bfS_{[j]} \quad \text{if} \,\ \forall \,\, r \in \left\{1,2,\dots,m_{\Sigma} \right\} \rightarrow \, \compo{\beta^{\Sigma}_{p[j]}}{r} < 0 \label{eq:def_EffSP_neg_multi_concate} \\
& \ \bfK + \alpha \bfS_{[j]} \quad \text{if} \,\ \forall \,\, r \in \left\{1,2,\dots,m_{\Sigma} \right\} \rightarrow \, \compo{\beta^{\Sigma}_{p[j]}}{r} > 0. \label{eq:def_EffSP_pos_multi_concate}
\end{align}
\end{subequations}

\subsection{Other types of goals}\label{subsec:typeB-goal}
Note that the above error analyses are cases where the responses at target DOF have specific goal values $\dtgti$. In other situations the goal in a mode may be set to, for example, elongation of a ligament, or more generally, relative distance between any two nodes in the system. Furthermore, a goal may be set to a range, i.e., a lower and/or upper bound, instead of a specific value. In such cases, rate of responses and errors depends on the involved DOF and types of goals.

Here we show an example in which length of a ligament is of interest. Assume at a numerical step, the coordinates of two end nodes (labeled by $a,b$) of an element are:
\begin{equation}\label{eq:nodal_coords_two_nodes}
\mathbf{c}_{ele}^* := \left[ x_a^*, y_a^*, x_b^*, y_b^* \right]^T    
\end{equation}
Since the change of response in all free DOF is $-\alpha \bfP_{[j]}^{\lri} \dsrci$ (see Eq.~\ref{eq:limit_of_dOmega}), by constructing a gather matrix $\bfG_{ele}$ associated with the four DOF of this element, the change in nodal coordinates are:
\begin{equation}\label{eq:change_in_nodal_coords}
\begin{aligned}
    \underset{4 \times 1}{\delta \bfd_{ele}^*} & := \bfG_{ele} \cdot \left( -\alpha \bfP_{[j]}^{\lri} \dsrci \right) 
    =  -\alpha \bfG_{ele} \bfP_{[j]}^{\lri} \dsrci \\
    & :=  \alpha \cdot \left[ dx_a^*, dy_a^*, dx_b^*, dy_b^* \right]^T 
\end{aligned}
\end{equation}
Then by finding and adding up the incremental displacements along the current ligament's axial direction, the change in element length can be obtained by the dot product:
\begin{equation}\label{eq:change_in_len}
\begin{aligned}
    \delta L_{ele}^* &= \underbrace{\left[ -\cos{\theta^*}, -\sin{\theta^*}, \cos{\theta^*}, \sin{\theta^*} \right]}_{:=\Vec{\theta}_{ele}} \odot \delta \bfd_{ele}^* \\
    & = \Vec{\theta}_{ele} \odot \delta \bfd_{ele}^* = -\alpha \cdot \Vec{\theta}_{ele} \odot \bfG_{ele} \bfP_{[j]}^{\lri} \dsrci
\end{aligned}
\end{equation}
where 
\begin{equation}
    \theta^* = \arctan \left( \dfrac{y_b^* - y_a^*}{x_b^* - x_a^*} \right)
\end{equation}
is the angle between the current axial direction of this ligament and $x$-axis, assuming that axial direction is pointing from node-a to node-b. Then by taking its derivative with respect to $\alpha$, the \textit{rate of element length due to infinitesimal negative-LSP} is:
\begin{equation}\label{eq:rate_of_change_in_len}
    \pdv*{\delta L_{ele}^*}{\alpha} = - \Vec{\theta}_{ele} \odot \bfG_{ele} \bfP_{[j]}^{\lri} \dsrci
\end{equation}
which is a scalar independent of $\alpha$. Then based on the current length $L_{ele}^*$ and its goal value and type, conditions on whether LSP by a SS would be beneficial can be established accordingly. For instance, if goal is in terms of a lower bound and requires that $L_{ele} \geq L_{goal} = 3.0$ while currently $L_{ele}^*= 2.8$, then a beneficial LSP is one that would elongate this element, requiring $\pdv*{\delta L_{ele}^*}{\alpha} > 0$.

\FloatBarrier
\section{Computational algorithm}
\begin{figure*}[thb]
\centering
\includegraphics[width= 0.98\textwidth]{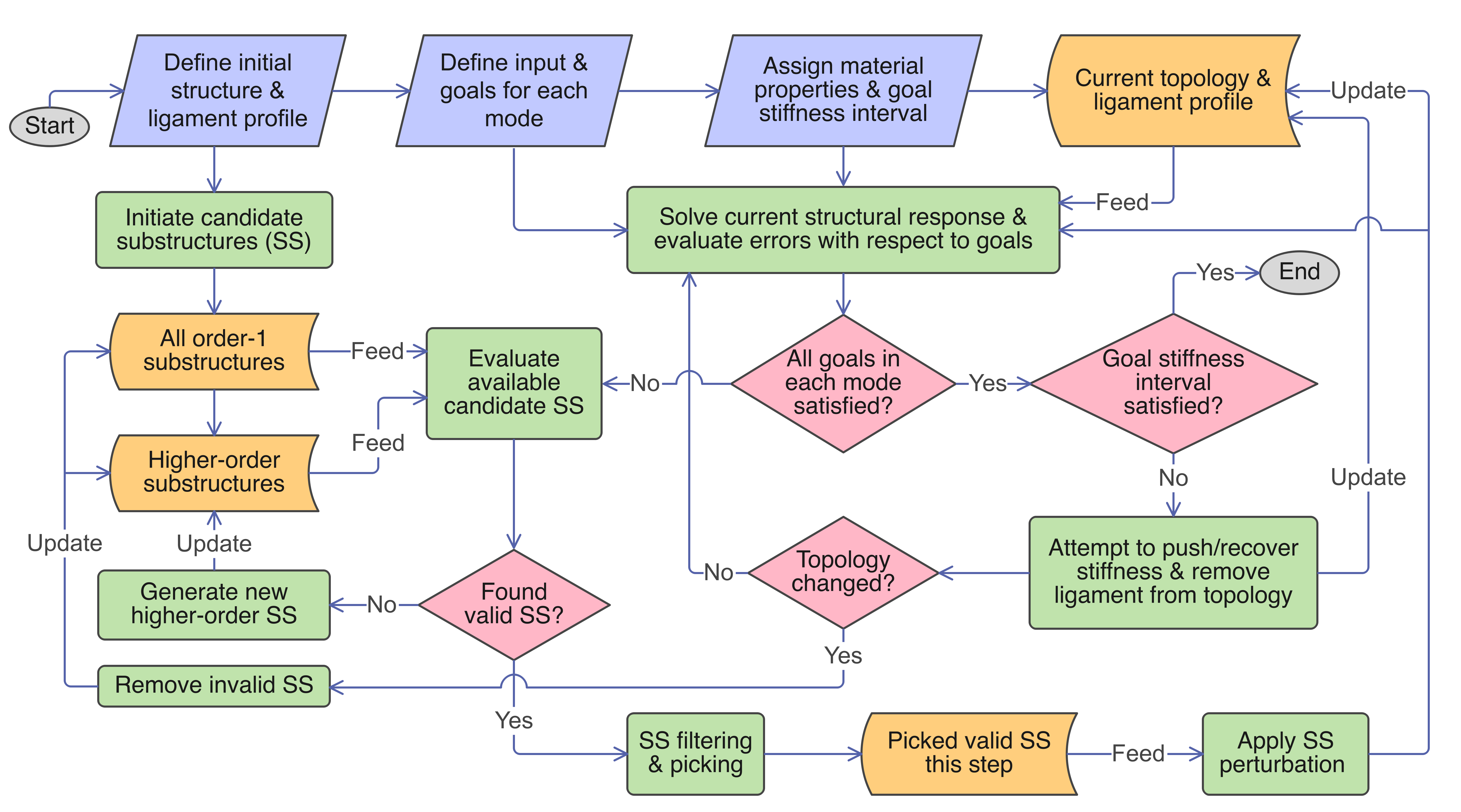}
\caption{Essential steps of the numerical algorithms implemented in this study. Note that examination and prevention of zero energy mode in a truss network, as explained in Section.~\ref{subsec:LSP-and-check-ZSEM}, are integrated in procedures involving ligament stiffness push/recover and topological removal of ligaments.
}\label{fig:flowchart}
\end{figure*}

The overall numerical framework is to start with a parent network, and then iteratively change a portion of its structural features in a stepwise manner until all target DOF at output reach the prescribed goals. At each step, based on the current structural responses and their difference to goals, scan or search through \textit{candidate substructures}, find \textit{valid substructures} among them (i.e., those providing beneficial LSP regarding all modes), then apply ligament stiffness changes in the parent network accordingly. A flowchart showing the essential computational procedures implemented in this study is given in Fig.~\ref{fig:flowchart}.

\subsection{Define input and output}\label{subsec:input-output}
Input of a mode is equivalent to essential boundary conditions. The output is flexible and depends on the desired functional modes. Two types of goals are implemented in this study to cover different senarios.
\begin{itemize}
    \item \textbf{Type-A goal.} Each type-A goal is to prescribe displacement component at a nodal DOF, which needs to be defined by a value and one of the three notations: 'EQ' (equals to), `HT' (higher than), or `LT ' (smaller than), to indicate the required relation for the DOF response to satisfy. For example, a type-A goal on DOF3 with ($0.2$,~EQ) means the response at DOF3 needs to be exactly $0.2$, while ($-0.1$,~LT) means the response needs to be lower than $-0.1$. 
    \item\textbf{Type-B goal.} This is for senarios where the change in distance between two nodes is of interest. It is defined by the same convention as above, i.e., a value and a condition notation. One type-B goal prescribe a desired change in distance between two nodes, and hence involves four nodal DOF in two-dimensional problems (Section.~\ref{subsec:typeB-goal}). By assigning two end nodes of a ligament as the nodes in this type, a goal involving elongation of a ligament can also be defined.
\end{itemize}

In each mode, the goals need to be defined on a unique set of DOF since each component in $\beta^{\lri}_{p[j]}$ (Eq.~\ref{eq:def_beta_i}) corresponds to one unique DOF. Therefore, if requiring the response of a DOF to lie between $(d_{0}, d_{1})$, then a type-A goal with ($d_0$,~HT) and another type-A goal with ($d_1$,~LT) can be defined on the same DOF respectively in two modes.

\subsection{Substructure generation and filtering}\label{subsec:SS-gen-and-filter}
\textbf{Candidate substructures.} Among all possible SS of a parent network, it is essential to always include all order-1 SS in the available candidate SS in each numerical step. The total number of all order-1 SS equals the number of ligaments, so analyzing all order-1 SS is computationally efficient. More importantly, this ensures that by default, each ligament in the network will be analyzed at least once (individually) in each numerical step.

For higher order ($O[j] \geq 2$) candidate SS, there can be various ways to generate them from the immense amount of possible configurations. For instance, randomly pick some ligament, pick all ligaments connected to a node, etc. In this work, we apply simple random sampling to generate higher-order substructures. Other modified methods to generate candidate SS need to rely on additional information. For instance, generate more SS from regions where ligaments have higher than average axial strains, i.e., by considering these regions as being more `mechanically relevant' such that structural changes in these regions may be more effective. Such kinds of methods are justifiable only if investigated in combination with specific structural and/or material mechanisms, which are subject to future studies.

\textbf{Valid substructures.} Define any SS that strictly satisfy either Eq.~\ref{eq:def_EffSP_neg_multi} or Eq.~\ref{eq:def_EffSP_pos_multi} as a \textit{valid SS}. This ensures that errors between responses and goals in all target DOF in all modes are expected to decrease. Finding valid SS from the candidate SS is a strict condition, extra candidate SS will be generated and evaluated if none of the available candidates are valid.

\textbf{Picking high-quality substructures.} Certain evaluation methods may be applied to pick `high-quality' SS from the valid SS to apply LSP on the current network. This is not a strictly necessary process, but can be an essential `filtering process' if the number of valid SS is large (e.g., more than 100). Firstly, selecting `higher quality' SS could improve performance of numerical algorithm. Secondly, limiting the amount of \textit{picked SS} may help to confine the range and total magnitude of features being adjusted, and hence could benefit numerical stability by avoiding abrupt structure change in one single step.

Here we consider two above mentioned measures (\textit{RSD} and \textit{$L^2$-Norm}) of the concatenated $\beta^{\Sigma}_{p[j]}$ of a valid SS (Eq.~\ref{eq:def_concate_beta}) to evaluate its quality. The first is \textit{RSD criterion}, which quantifies how evenly a LSP by this SS would decrease the errors, with lower RSD indicating higher quality. The second is \textit{$L^2$-norm criterion}, which indicates how rapidly a LSP would decrease the errors, with higher norms indicating higher quality. If we apply these criteria, a size limit on the picked SS ($N^{pick}_{ss}$) needs to be defined. Then, valid SS will be sorted according to the criterion and finally the first $N^{pick}_{ss}$ substructures will be picked. If the number of valid SS is lower than this limit, then the filtering criteria will be skipped and all valid SS will be picked.

\textbf{Exploitation vs. Exploration}
A concept in deep learning is the so-called exploitation and exploration strategies, and there is a fundamental trade-off between them. Exploitation prefers taking advantage of the existing known options/policies that lead to immediate reward, whereas exploration leans to new options (instead of short-term results) that can be potentially beneficial in the long run.

Under the substructure perturbation framework, when the number of candidate SS is large, validating all of them can be computational costly because concatenated $\beta$ has to be evaluated for each candidate SS. On the other side, since the amount of stiffness change in a step will be limited, it could take multiple consecutive steps to reduce its stiffness to close to zero, during which the corresponding candidate SS would remain as valid. Therefore, we implement \textit{SS validation priority}, inspired by the exploration-exploitation concept, and evaluate the associated SS accordingly. In each numerical step, valid SS from the previous step are of the highest priority and will be validated first (exploitation). If at least one of them is still valid, SS picking will be made directly. If none of them are valid, then all historically valid SS will be evaluated. If there is still no valid SS, all available candidate SS will be evaluated. Lastly, if none of the SS are valid, new SS will be generated (repeatedly if necessary) until at least one valid SS is found (exploration).

\subsection{Update structure while preventing zero mode}\label{subsec:LSP-and-check-ZSEM}
There are two important correlated aspects during iterative LSP procedures. First is the attainable range of ligament stiffness $E_e$. The second, specific to truss-based network, is whether a network contains artificial zero modes. These two are correlated because zero modes of a network depends on its topology (nodes and connectivity). In this study, when the structural responses are close to goals, the ligament stiffness profiles are used as indicators of whether a ligament should be contained or removed, and hence it affects the topology of the final design.

\textbf{Zero strain energy modes}
A \textit{zero strain energy mode} (ZSEM), or simply a \textit{zero mode}, is when at least one ligament can move and/or rotate without generating any stress (and hence strain energy) in the system. This is particularly common for truss networks since nodes are free to rotate (i.e., nodal rotational stiffness is zero). Note that for a two dimensional truss network, there are three inherent basic zero modes: rigid body translation in the i) horizontal and ii) vertical directions as well as iii) in-plane rigid body rotation. Therefore, the main task is to recognize and prevent any additional zero mode when changing the topology of a truss network. Specifically, an eigenvalue decomposition on the stiffness matrix of the current network, $\bfK$, is performed. Each zero eigenvalue represents one zero mode, with its corresponding eigenvector showing the relative nodal displacements in this zero mode. More details are provided in SI~\ref{subsec:detect_ZSEM}.

Theoretically, the total number of zero modes depends solely on the topology of a network, rather than its specific ligament stiffness magnitudes. For example, a ligament with just one end connected to other ligaments (while the other end isolated) will introduce one zero mode, whereas its mechanical parameters ($E, A, l$) does not affect the zero mode it introduces. Therefore, examination of zero modes is only necessary when changing the network topology.

\textbf{Stiffness bounds.} Since the input and output are both displacement-based, the relative values of ligament stiffness will determine the relation between input and output. This can be verified by applying a scaling factor $c$ on $\bfK$ and observing that the transformation matrix of the network (Eq.~\ref{eq:def_T}) remains the same:
\begin{equation}\label{eq:T_independent_of_c}
    -(c \cdot \bfK_F)^{-1} (c \cdot \bfK_{EF}^T) = -\invKF \bfK_{EF}^T = \bfT.
\end{equation}
Therefore, distributions of ligament stiffness can be used to predict structural responses, hereby $E_e$ can be considered as the relative stiffness in the network.

In a design problem, an interval $[\widehat{E}_{min}, \widehat{E}_{max}]$ indicating the \textit{goal stiffness interval} in the final network needs to be specified. Namely, the allowable ratio between maximum stiffness to minimum stiffness. Because the topology of the final design remains to be solved, this goal interval should not be enforced from the beginning. Instead, the strategy is to: i) initially allow a wide range of stiffness with an attainable lower bound close to zero, ii) start iterative ligament stiffness perturbation (LSP) by substructures, iii) once the structural responses have reached goals, attempt to topologically remove ligaments based on their relative stiffness, iv) repeat these processes until a final design has been achieved.

\textbf{I. Initial stiffness profile and interval}. The initial ligament stiffnesses are all set to $1.0$. The \textit{initial stiffness interval}, i.e., attainable range of $E_e$, is set to $[E^*_{min}, E^*_{max}] = [0.001, 1]$. Here denote ligaments whose stiffness are at the upper bound as \textit{full stiffness ligaments}, and those at the lower bound as \textit{small stiffness ligaments}. The advantages of setting a small but non-zero value as the initial lower bound are: i) Stiffness perturbation values (for involved ligaments) in each step can be either positive or negative. Therefore, retaining the small ligaments in the topology allows for their potential contribution to the structural responses if their stiffness are added back in later steps. If a ligament is topologically removed from the network, candidate SS would no longer contain that ligament. ii) Contribution by small stiffness ligaments to the nodal displacements are considered negligible, thereby change in structural responses are expected to be limited if they are removed topologically from the network. iii) For truss networks, implementation of small stiffness ligaments ensures that examination of zero modes only needs to be checked and ensured when attempting to change topology, and hence is not required during iterative LSP procedures.

\textbf{II. Iterative LSP by SS}. In each numerical step, after SS generation and filtering, LSP will be applied to the parent network based on the picked SS. Firstly, each picked SS is assigned with a perturbation coefficient ($\alpha_j, \alpha_k, \dots$ in Eq.~\ref{eq:def_Kpjkl}). Subsequently, the \textit{attempted amount of stiffness change} in each involved ligament, denoted by $f_q$ (with $q$ being the ligament label), can be obtained by:
\begin{equation}\label{eq:attempt_fq}
    f^*_q = \sum_{ j \in \mathbb{L}^{ss}_{pick} } \sum_{l \in \mathbb{L}^{re}_{[j]} } \delta_{ql} \cdot \alpha_j, \quad 
\end{equation}
where $\mathbb{L}^{ss}_{pick}$ contains labels of all picked SS, and $\delta_{ql}$ is the Kronecker delta operator. The sign of $\alpha_j$ depends on whether SS~$[j]$ corresponds to negative or positive-LSP (Eq.~\ref{eq:def_EffSP_neg_multi_concate} or \ref{eq:def_EffSP_pos_multi_concate}). Finally, the attempted stiffness change for each ligament $f_q$ will be adjusted if the change by $f_q$ would violate the currently assigned stiffness bounds:
\begin{equation}\label{eq:regularized_fq}
    f_q = \begin{cases}
        f_q^* &\text{if} \,\ E_{min} \leq E_q + f_q^* \leq E_{max} \\
        E_{min} - E_q < 0 &\text{if} \,\ E_q + f_q^* < E_{min} \\
        E_{max} - Eq > 0 &\text{if} \,\ E_q + f_q^* > E_{max}
    \end{cases}
\end{equation}
where $f_q$ is the final \textit{regularized stiffness change}. (Note that the stiffness bonds in Eq.~\ref{eq:regularized_fq} depend on the stages of numerical simulations, as will be explained below.) This way, stiffness of all ligaments will strictly stay in the assigned stiffness bounds in each numerical step. 

The relative error vector ($\Psi^{\lri}$ in Eq.~\ref{eq:def_Psi}) is used to determine if goals are reached in a mode. Specifically, a goal is reached if the magnitude (absolute value) of the corresponding relative error is smaller than a predefined threshold $\widehat{\psi}$. Denote $\psi^{all}_{max}$ as the \textit{maximum absolute value of relative errors in all modes}, then the structural responses will satisfy all goals when $\psi^{all}_{max} < \widehat{\psi}$, namely:
\begin{equation}\label{eq:criteria_all_goals}
\begin{aligned}
    &\text{if:} \quad \forall \,\, i \in \NTo{n_m}, \, r \in \NTo{m_{tgt}^{\lri}} \rightarrow \, |\Psi^{\lri}_r| < \widehat{\psi} \\
    &\text{then:} \quad \psi^{all}_{max} < \widehat{\psi} \quad \text{(all goals satisfied)}.
\end{aligned}
\end{equation}

\textbf{III. Topology update and stiffness recovery.} Once the above structural response criteria (Eq.~\ref{eq:criteria_all_goals}) is realized for the first time, attempts to make topological modification (i.e., removal of ligaments) will be activated. Firstly, all ligaments with stiffness smaller than a fraction of the goal stiffness lower bound, denoted as $f_{\downarrow} \times \widehat{E}_{min}$, will be `pushed' down to small stiffness ligaments (i.e., with stiffness of $0.001$), where $f_{\downarrow} \in (0, 1]$ is the \textit{stiffness push down coefficient}. Secondly, as many as possible small stiffness ligaments will be topologically removed from the network while ensuring there are no artificial zero modes in the network. Thirdly, after each topology change, all ligaments whose stiffness are below the lower bound of the goal stiffness interval, $\widehat{E}_{min}$, will be `recovered' to a value slightly above $\widehat{E}_{min}$. A predefined \textit{stiffness recover coefficient}, $f_{\uparrow}$, is used to determine the recover stiffness $E_{\uparrow}$:
\begin{equation}\label{eq:recovering_E}
E_{\uparrow} = \widehat{E}_{min} + f_{\uparrow} (\widehat{E}_{max} - \widehat{E}_{min}).
\end{equation}

Such modification on the network topology and stiffness profile will likely lead to non-negligible effects on the structural response so that goals are no longer satisfied. Ideally, a certain amount of LSP steps following one topology change and stiffness recover will lead to a final design where structural response satisfies all goals ($\psi^{all}_{max} < \widehat{\psi}$), all ligament stiffness are within $[\widehat{E}_{min}, \widehat{E}_{max}]$, and there are no zero modes in the network. In practice, it could take multiple such steps to realize a final design. Also, as will be shown in numerical examples, attempting to remove ligaments `on-the-fly' may improve computational speed and success rate. For example, a threshold on the number of existing small stiffness ligaments in the network, $n^{E,min}_{el}$, can be defined. If at the end of a step, $n^{E,min}_{el}$ is above the defined threshold (e.g., 5 ligaments), removal of ligaments from the network topology will be attempted. This helps to reduce the number of small stiffness ligaments when $\psi^{all}_{max} < \widehat{\psi}$, and hence can limit the amount of ligaments involved in stiffness push down and recovery. Note that for intermittent topology change when $\psi^{all}_{max} > \widehat{\psi}$, stiffness push down and recovery are not performed since the structural response are not yet close to the goals.

\textbf{IV. Configuration reset.} Occasionally, the above procedure involving topology and stiffness profile updates may result in significant change in structural responses. This is largely due to the uncertainty in ligament stiffness profile (i.e., distribution and values of $E_q, \, q \in \mathbb{N}^{all}$) and corresponding ligaments removal. To address this issue, once $\psi^{all}_{max}$ drops below a threshold percentage, $f^{\% \psi}_{\leftarrow}$, from the initial value, a \textit{configuration reset point} will be stored. If removal of small stiffness ligaments and subsequent stiffness recovery results in a $\psi^{all}_{max}$ larger than that of the stored reset point, the network configuration, including its topology and ligament stiffness values, will be reset to the stored configuration.

The above essential steps will be performed repeatedly until a final design is realized. The simulation will be terminated if the number of numerical steps has reached a limit (e.g., 1000 steps), or if the number of configuration resets has reached a limit (e.g., 5 times). More details on these procedures are exemplified in the demonstrative design examples in the next section.

\FloatBarrier
\section{Numerical design examples}
In this section, we show several design examples covering both truss and beam networks, and different types of goals and geometries. For each example, we also investigate the effect of different simulation parameters on the run time and success rate, and denote simulations with the same input parameters (for the same design problem) as a series. The success rate in a series is the percentage of simulations that successfully find working designs before reaching any termination criterion.

The first example, \textit{Pinch}, is explained in detail to illustrate the computational procedures. Another example, \textit{Actuator}, is established in specific to compare with another state-of-art method. All numerical results presented in this study are performed using an in-house developed custom finite element toolkit written in Python. Essential utility and functionality involving the presented SSPM algorithms in the above section (e.g., generation and evaluation of candidate SS) are integrated into the toolkit.

\subsection{Example: Pinch}
\begin{figure*}[th]
\centering
\includegraphics[width= 0.98\textwidth]{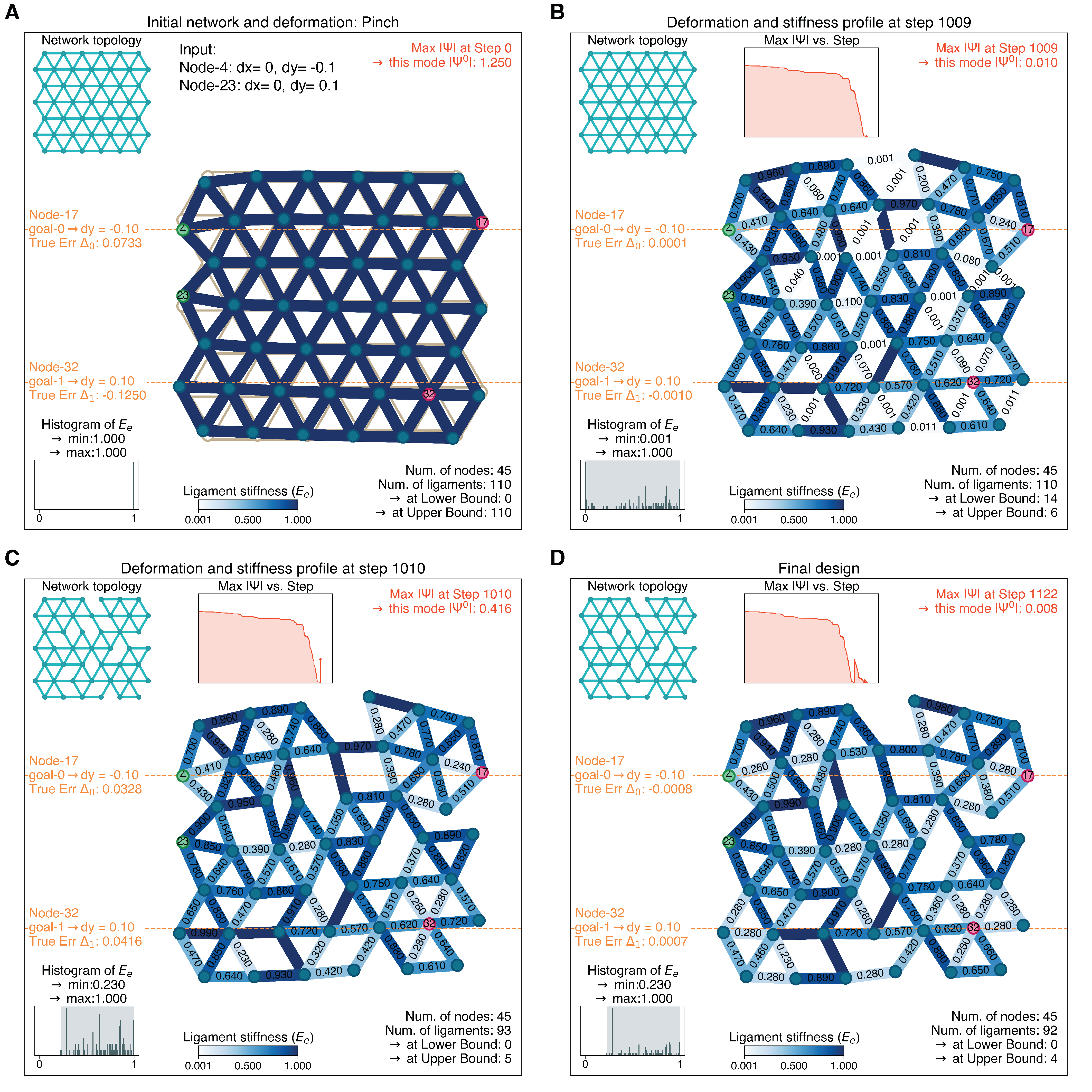}
\caption{An numerical example from Pinch (truss network), Series-2, to demonstrate the iterative structure update procedures.
(A)~Initially, at Step-0, all 110 ligaments in the network have stiffness ($E_e = 1.0, e \in \mathbb{N}^{all}$). The undeformed structure is shown in wire frame in light sand.
(B)~After iterative LSP, goal is realized (i.e., $\psi_{max}^{all} < \widehat{\psi} = 0.01$) for the first time at Step-1009 under the initial stiffness interval $[0.001, 1]$. The distribution of ligament stiffness is shown in the inset histogram, and annotated on each ligaments. 
(C)~Topology change and subsequent stiffness recovery is then attempted in Step-1010 (following panel~B), through which, 17 ligaments are topologically removed from the network, and $E_e$ of the leftover small stiffness ligaments are recovered to $E_{\uparrow}$. The network now has 93 ligaments, no zero modes, and stiffness profile satisfies the goal stiffness interval $[0.2, 1]$. However, due to the change in topology and stiffness profile, $\psi_{max}^{all}$ jumps up to 0.416, therefore iterative LSP continues.
Then, at Step-1101 (not shown here), goal is realized for the second time, and another attempt to change topology is made, in which, one ligament is removed, and consequently, $\psi_{max}^{all}$ jumps up to a value slightly above $\widehat{\psi} = 0.01$.
(D)~Eventually, at Step-1122, a final design that satisfies prescribed input-output mode, goal stiffness interval, and with no artificial zero mode, is obtained. An animation showing the iterative process of this design example is included in SI.}
\label{fig:example-Pinch-1}
\end{figure*}

The initial network is shown in Fig.~\ref{fig:example-Pinch-1}A. A truss network is assumed and all ligaments have length $0.5$ and initial stiffness $1.0$. One functional mode (denoted Mode-0) is considered. For input, a `pinch' along the vertical direction on two nodes on the left edge, node-4 and node-23, is applied as the essential BCs. The applied displacements along the y-axis on bottom and top nodes are $-0.1$ and $+0.1$, respectively, while the displacements along the x-axis are zero. The desired output is a `pinch' along the vertical direction on two randomly picked nodes on the right side, node-17 and node-32 in this case, which should have the same magnitude as the prescribed input pinch. Namely, one type-A goal with (+0.1, 'EQ') on y-displacement of node-32, and one type-A goal with (-0.1, 'EQ') on y-displacement of node-17, are prescribed for output (their horizontal displacements are not prescribed). These two goals can be visualized by the two `goal lines' as shown in orange dashed horizontal lines in Fig.\ref{fig:example-Pinch-1}.

In total, six numerical series with different simulation parameters are performed. To illustrate the necessity and effectiveness of higher-order SS (i.e., ability to consider and apply simultaneous stiffness change in multiple ligaments), we firstly run Series-1 by only generating all 110 order-1 SS as the candidate SS. Namely, at each step, the algorithm will investigate each one of the 110 ligaments in the network, and check if adjusting (increasing or reducing) its stiffness would reduce the errors to goal. Also, to prevent any dependency on SS picking and subsequent LSP, we run 110 simulations in Series-1 by changing the size limit of picked SS from 1 to 110. Interestingly, all simulations in Series-1 failed, indicating that, at some point, adjusting any one ligament in the network would not lead to a working design. Results from Series-1 will then be compared with other series where higher-order SS are considered. For the other five series (2 to 6), candidate SS with orders ranging from 1 to 20, with up to 1000 SS per order, are included. To investigate the effect of simulation parameters on the results and run time statistics, 500 simulations are performed for each series under the same input. Simulation parameters and statistics of results are summarized in Table~\ref{tab:Pinch_stats}.

As an example, three steps (Fig.~\ref{fig:example-Pinch-1}B to D) from one simulation in Series-2 are included to show the essential structure updating procedures (explained in Section~\ref{subsec:LSP-and-check-ZSEM}). From the initial network, 1009 numerical steps (iterative LSP) are taken to obtain a stiffness profile achieving the prescribed goals ($\psi_{max}^{all} < 0.01$) for the first time under the initial stiffness interval $[0.001, 1]$. Up to this point, the topology of the network has remained unchanged, but there are ligaments whose stiffness are at, or close to $0.001$ (see Fig.~\ref{fig:example-Pinch-1}B). Following this step, ligaments with $E_e < f_{\downarrow} \widehat{E}_{min}$ are pushed down to small stiffness ligaments (i.e., replace $E_e$ with $0.001$ if $E_e < f_{\downarrow} \widehat{E}_{min}$). Then 17 ligaments are topologically removed from the network, with stiffness of the leftover small stiffness ligaments recovered to $E_{\uparrow}$. The resultant network is as shown in Fig.\ref{fig:example-Pinch-1}C, where the network now has 93 ligaments and their stiffness are within the goal stiffness interval. However, due to these modifications, the structural response has diverted from the prescribed goals and $\psi_{max}^{all}$ jumps up to $0.416$, iterative LSP henceforth continues. Then at Step-1101 (not shown), $\psi_{max}^{all}$ drops back to below $0.01$ again, after which, one ligament is removed in another topology change attempt, thereby $\psi_{max}^{all}$ jumps up again such that iterative LSP continues. Then at Step-1122, a final design satisfying prescribed output and goal stiffness interval $[0.2, 1]$, and without any artificial zero energy mode, is achieved.

Starting with Series-2, higher order candidate SS are included and will be regenerated if none of the available candidate SS are valid. In both Series-2 and 3, only one valid SS (whose order can range from 1 to 20) is picked to apply LSP in each step, and different SS picking criteria (RSD and $L^2\text{-norm}$) are assigned. The success rate and run time are comparable because a fixed perturbation coefficient ($\alpha_j$ = 0.01) is defined in both cases. In Series-4, the topology change is attempted `on-the-fly' whenever the number of small stiffness ligaments is above 5. As a result, the success rate improved by about three times (from $\approx 21\%$ to $\approx 60\%$), and the median run time also reduces from about $48$ seconds to $33$ seconds. In Series-5, the size limit of picked SS is increased from 1 to 5. A further improvement in success rate (from $60\%$ to $71\%$) is observed, and the min and median run time also improved because more ligaments are expected to be adjusted in each step. Nevertheless, because exploitation is used during SS feeding, and also, because the number of valid SS can be lower than 5, the average number of picked SS (and hence the total amount of stiffness change) in all steps should be less than 5 times in Series-5 as compared to Series-2 to 4. In Series-6, the size limit of picked SS is increased from 5 to 10, but despite the reduction in minimum run time, the success rate decreased moderately and the median run time increased moderately. Whether extra time was spent in sorting valid SS, or in finding removable ligaments during topology change when $\psi_{max}^{all} < \widehat{\psi}$, requires further investigation. But a comparison between Series-5 and 6 indicates that applying a larger amount of stiffness change would not necessarily improve the results.

\begin{figure*}[th]
\centering
\includegraphics[width= 0.75\textwidth]{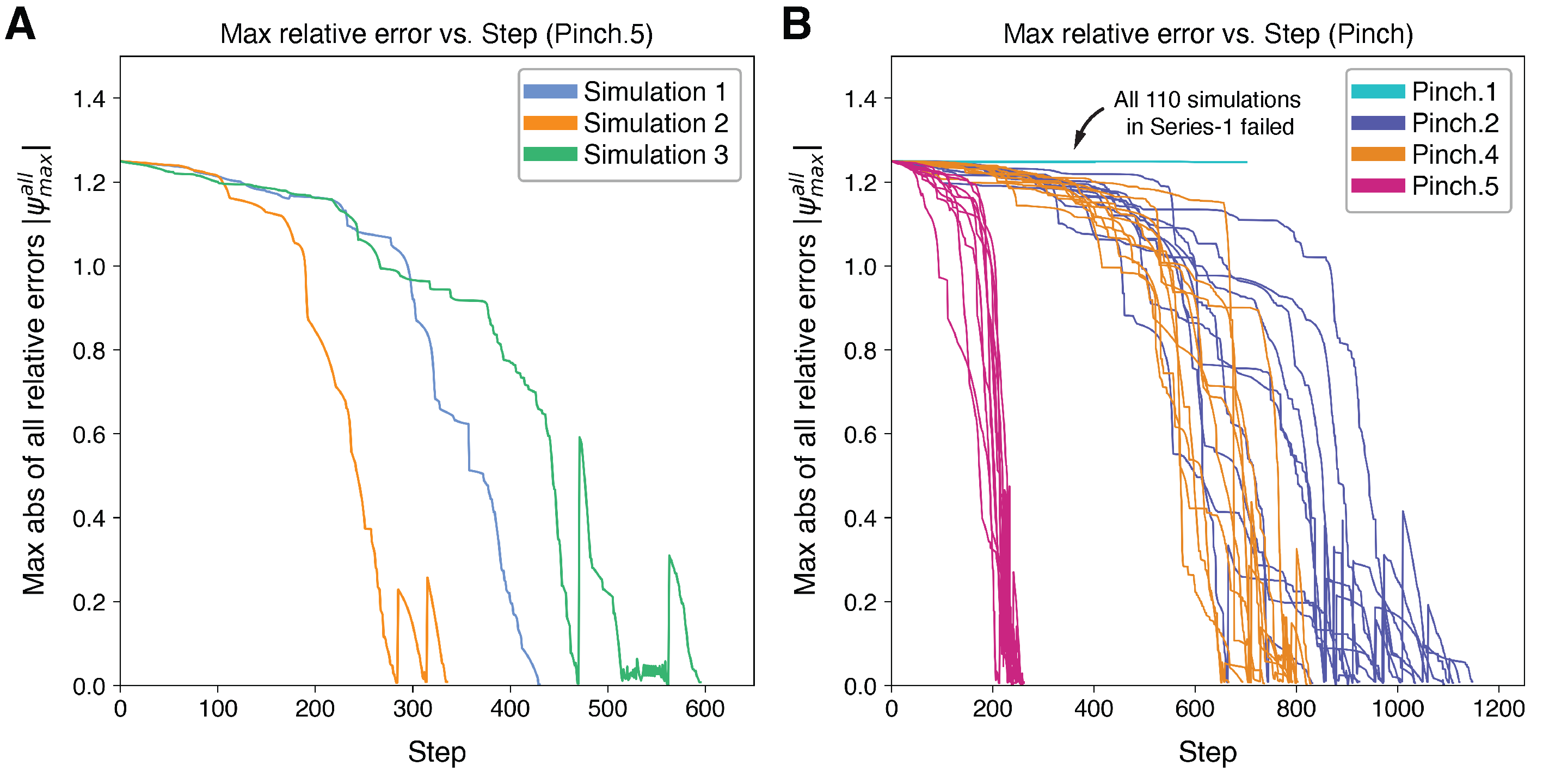}
\caption{Maximum magnitude of relative error vs numerical step: examples from Pinch.
(A)~Three representative successful simulations from Pinch.5 are included to show $\psi_{max}^{all}$ vs step. In Simulation-1, upon first realization of goal, one topology change attempt directly results in a working design. In Simulation-2, multiple rounds of topology change are performed to reach the final design. In Simulation-3, one topology reconfiguration (to a value close to $50\%$ of the initial $\psi_{max}^{all}$) is made, while some moderate oscillations occurred in later stages.
(B)~Comparison between different series to show the effect of higher-order SS. For Series-1, all 110 simulations are shown, but because only one ligament is adjusted in each step, the decreasing rates of $\psi_{max}^{all}$ are significantly lower than the other series, thus the 110 curves almost overlap with each other, and together look like one flat line. All simulations in Series-1 failed after about 50 to 650 steps. For each of the other series, ten simulations that took the smallest number of steps to complete are shown. In Series-2, higher-order SS are considered, and one valid SS is picked for LSP in each step. In Series-4, ligament removal is attempted during the simulation if the number of small stiffness ligaments surpassed five. Compared to Series-2, run time statistics and success rate are improved. In Series-5, the size limit of picked SS to apply LSP is increased to 5 (whereas in Series-2 and Series-4, only one valid SS can be picked). As shown, the decreasing rates of $\psi_{max}^{all}$, and thereby the run time statistics, are further improved. This comparison study demonstrates the necessity of higher-order SS not only in providing paths to working designs, but also in enhancing the design efficiency.}
\label{fig:Pinch-Psi-vs-step}
\end{figure*}

Fig.~\ref{fig:Pinch-Psi-vs-step}A, as an example, shows three typical $\psi_{max}^{all}$ curves to visualize the effect of stiffness push and recovery, topology change, and reconfiguration on the historical values of $\psi_{max}^{all}$. Fig.~\ref{fig:Pinch-Psi-vs-step}B demonstrates the effect of considering higher-order SS in not only finding a `path' to final designs, but also in improving the `convergence rates' of simulations. Details are explained in the figure caption.

It should be noted that success rate depends on the input parameters and also the termination criteria (e.g., allow a maximum of 2000 numerical steps). If, for example, allowing more total steps, the success rate will increase, but at the cost of increasing average run time. Also, an investigation on the failed cases reveals that the main obstacle is the condition on zero energy mode (specific to truss networks), not due to the LSP procedures. A working configuration under the initial stiffness interval ($[0.001, 1]$) can always be achieved, whereas simultaneously ensuring goal stiffness interval, prescribed outputs, and no zero modes, is the main challenge.

In brief, through this design example (Pinch), we demonstrate the essential numerical procedures, study the effect of several input parameters (representing different strategies), and highlight the important role of higher-order SS (i.e., considering and applying multiple substructural features simultaneously) not only in finding paths to successful designs, but also in enhancing the speed and efficiency of the iterative structure update process.

\begin{table*}[thb]
\small
\centering
\caption{Input parameters and run time for design example: Pinch (truss network)} \label{tab:Pinch_stats}
\begin{threeparttable}
\begin{tabular}{l c c c c c c c c c}
    \toprule
    \multicolumn{2}{c}{ \multirow{7}{*}{ \makecell{Common \\ parameters} } } &
    \multicolumn{8}{l}{Initial network topology: $n_{el}=110$, $n_{node}=45$} \\
    & & \multicolumn{8}{l}{Goal ligament stiffness interval in final design: $\widehat{E}_{min}= 0.20$, $\widehat{E}_{min} = 1$} \\ 
    & & \multicolumn{8}{l}{Allowed total number of numerical steps (before termination): 2000} \\
    & & \multicolumn{8}{l}{Perturbation coefficient for each picked SS: $\alpha_j= 0.01$} \\
    & & \multicolumn{8}{l}{Stiffness push down coefficient: $f_{\downarrow} = 0.25$} \\
    & & \multicolumn{8}{l}{Stiffness recover coefficient: $f_{\uparrow}= 0.1$} \\
    & & \multicolumn{8}{l}{Allowed maximum relative error in all modes in final design: $\psi^{all}_{max} < \widehat{\psi} = 0.01$} 
    \\
    [1ex]
    \hline \\[-2.5ex]
    \multirow{3}{*}{Series} & \multirow{3}{*}{\makecell{Threshold of \\ $n^{E,min}_{el}$ to \\ change topology}} &
    \multicolumn{2}{c}{\makecell{Order \& limit \\ of candidate SS\tnote{1}} } & \multicolumn{2}{c}{\makecell{Criterion \& size \\ limit of picked SS\tnote{2}} } & 
    \multicolumn{2}{c}{\makecell{Numerical \\ Results\tnote{3}}} &
    \multicolumn{2}{c}{ \makecell{Run time\tnote{4} \\(seconds)} } \\ [1ex] 
    \cmidrule(lr){3-4} 
    \cmidrule(lr){5-6}
    \cmidrule(lr){7-8}
    \cmidrule(lr){9-10}
     & & $O_{[j]}^{ss}$ & limit/order & Criterion & limit & Success (\%) & Total & Min & Median \\[1ex]
    \hline
    1 & \textit{None} & 1 & \textit{NA} & \textit{Any} & 1 to 110 & \textit{All failed} & 110 & \textit{NA} & \textit{NA}\\
    2 & \textit{None} & 1 to 20 & 1000 & RSD & 1 & 21.2\% & 500 & 15.34 & 47.70 \\
    3 & \textit{None} & 1 to 20 & 1000 & L2-Norm & 1 & 21.4\% & 500 & 17.04 & 47.84 \\
    4 & 5 & 1 to 20 & 1000 & RSD & 1 & 59.8\% & 500 & 10.98 & 33.25 \\
    5 & 5 & 1 to 20 & 1000 & RSD & 5 & 71.2\% & 500 & 6.55 & 27.81 \\
    6 & 5 & 1 to 20 & 1000 & RSD & 10 & 66.0\% & 500 & 6.02 & 35.21 \\
    \bottomrule
\end{tabular}
\begin{tablenotes}\footnotesize
\item[*] \textit{NA} indicates `not applicable'.
\item[1] Number of distinct SS to generate for each order. If $C(n,k)$ is lower than limit, all order-$k$ SS will be included.
\item[2] Criteria to pick SS from valid SS to apply LSP. If number of valid SS is smaller than the size limit, filtering criteria will be skipped and all valid SS (if any) will be picked. 
\item[3] Successfully found a design before reaching any termination criterion (e.g., 2000 allowed numerical steps).
\item[4] Statistics for successful results. All simulations were performed on a laptop with an Apple M1 Pro chip and 16 GB of RAM.
\end{tablenotes}
\end{threeparttable}
\end{table*}

\FloatBarrier
\subsection{Example: Auxetic}
\begin{figure*}[th]
\centering
\includegraphics[width= 0.98\textwidth]{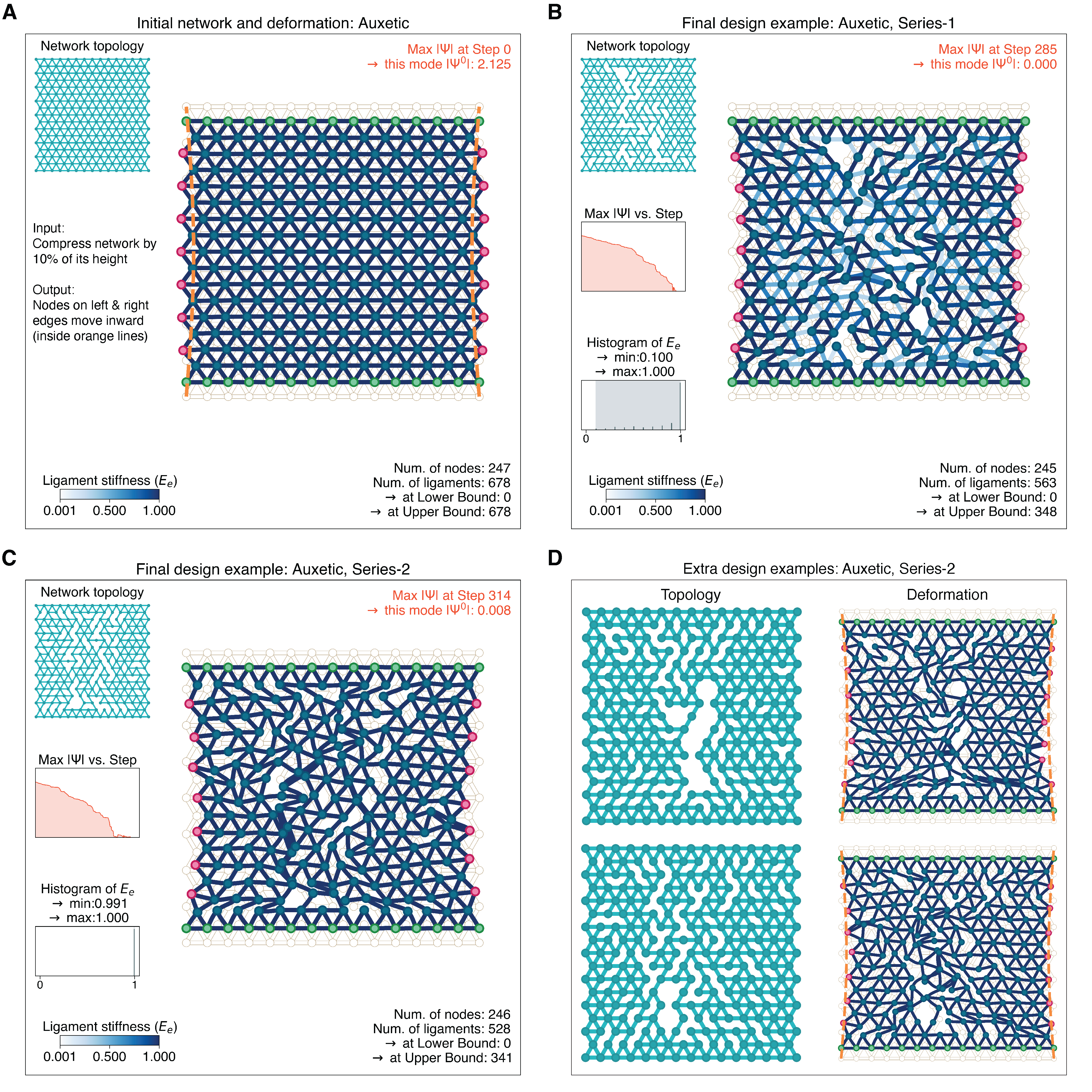}
\caption{Numerical example: Auxetic (beam network).
(A)~Initial topology and deformation.
(B)~One design example from Series-1, where the goal stiffness interval of ligaments is set to $[0.1, 1]$.
(C)~One design example from Series-2. In this series, the goal stiffness interval is set to $[0.99, 1]$ to obtain networks consist of ligaments with effectively uniform stiffness, such that the prescribed outputs are realized solely by the tailored network topology.
(D)~Two extra design examples from Series-2.
}
\label{fig:example-Auxetic}
\end{figure*}

Auxetic behavior is a well-known functionality realizable through specially designed geometrical patterns in architected materials~\cite{Bertoldi2017.flexible, Lakes1987.foam, Saxena2016.three}. Here we start from a regular triangular lattice as shown in Fig.~\ref{fig:example-Auxetic}A, and apply SSPM to realize negative Poisson's ratio under compression. The initial network contains 678 ligaments (Euler beams) and 247 nodes. For input, a $10\%$ compressive strain along the vertical direction is applied via nodal displacement BCs on the top and bottom edges. As shown, the left and right edges of the structure protrude out under compression, exhibiting typical positive Poisson's ratio. The prescribed output is for the nodes on the edges (shown in pink) to move inward. More specifically, an effective Poisson's ratio $\nu= -0.5$ is prescribed at the mid-section, i.e., the two nodes on the horizontal symmetric axis (one from the left edge, one from the right edge) need to move in by at least $2.5\%$ of the structure's width, respectively. The required displacements for the remaining nodes are then determined based on sin functions spanning over the structures' height, as marked by the orange dashed lines in Fig.~\ref{fig:example-Auxetic}A. Then, type-A goal on the horizontal nodal DOF of these nodes are prescribed accordingly.

Four series are performed for this example, the input parameters and results are summarized in Table~\ref{tab:Auxetic_stats}. For Series-1, we allow variation in ligament stiffness in the final design with goal stiffness interval $[0.1, 1]$, one design example is given in Fig.~\ref{fig:example-Auxetic}B. As expected, the nodes involved with target DOF move into the prescribed dashed lines, thereby realizing auxetic behavior under compression. For Series-2, we set the goal stiffness interval to $[0.99, 1]$, i.e., request that the ligaments in the final design have effectively uniform stiffness. An example is given in Fig.~\ref{fig:example-Auxetic}C, in which the auxetic behavior is realized solely by the specially tailored geometries. Two additional design examples are given in Fig.~\ref{fig:example-Auxetic}D.

Interestingly, the deformation mechanisms and geometrical characters of the designed structures conceptually resemble a classic design strategy of auxetic structures, where collapsing and closure of inner voids/cutouts are utilized to realize structural shrinkage under compressive load. The fundamental difference is that the structures in this example are designed systematically via SSPM procedures without any prior experience or knowledge of auxetic structures. This indicates the versatility of the SSPM method and its potential implementation in solving challenging problems.

For run time, one design typically takes around 140 to 200 seconds to complete. As mentioned above (in example Pinch), the main obstacle to obtaining a working configuration in truss networks is to ensure no zero energy modes. In contrast, checking and preventing zero modes are no longer necessary here (in example Auxetic) since beam ligaments are assigned. As a result, the success rates are almost $100\%$ for all four series, with only 2 of the 400 simulations terminated due to reaching allowed number of steps. 

Also note that there are cases where ligaments start to contact and penetrate each other as the compressive strain increases. In such scenarios, Poisson's ratio may become a strain dependent parameter wherein the prescribed value ($\nu=-0.5$) may not retain all the way to the applied $10\%$ compressive strain. Nevertheless, contact mechanisms and structural non-linearity are beyond the focus of this study.

\begin{table*}[thb]
\small
\centering
\caption{Input parameters and run time for design example: Auxetic (beam network)} \label{tab:Auxetic_stats}
\begin{threeparttable}
\begin{tabular}{l c c c c c c c c c}
    \toprule
    \multicolumn{3}{c}{ \multirow{9}{*}{\makecell{Common \\ parameters}} } &
    \multicolumn{7}{l}{Initial network topology: $n_{el}=678$, $n_{node}=247$} \\
    & & & \multicolumn{7}{l}{Criterion for picking valid SS: RSD} \\ 
    & & & \multicolumn{7}{l}{Allowed total number of numerical steps (before termination): 1000} \\
    & & & \multicolumn{7}{l}{Perturbation coefficient for each picked SS: $\alpha_j= 0.1$} \\
    & & & \multicolumn{7}{l}{Threshold of $n^{E,min}_{el}$ to change topology (remove ligaments): 5} \\
    & & & \multicolumn{7}{l}{Stiffness push down coefficient: $f_{\downarrow} = 0.25$} \\
    & & & \multicolumn{7}{l}{Stiffness recover coefficient: $f_{\uparrow} = 0.1$} \\
    & & & \multicolumn{7}{l}{Allowed maximum relative error in all modes in final design: $\psi^{all}_{max} < \widehat{\psi} = 0.01$} \\
    & & & \multicolumn{7}{l}{Threshold of relative error at a target DOF to skip concatenation in $\beta$: 0.01}
    \\
    [1ex]
    \hline \\[-2.5ex]
    \multirow{3}{*}{Series} & \multicolumn{2}{c}{\makecell{Goal $E_e$ interval \\ in final design} } &
    \multicolumn{2}{c}{\makecell{Order \& limit \\ of candidate SS} } & \makecell{Size limit of\\ picked SS} & 
    \multicolumn{2}{c}{\makecell{Numerical \\ Results}} &
    \multicolumn{2}{c}{ \makecell{Run time\tnote{1} \\(seconds)} } \\ [1ex] 
    \cmidrule(lr){2-3}
    \cmidrule(lr){4-5} 
    \cmidrule(lr){6-6}
    \cmidrule(lr){7-8}
    \cmidrule(lr){9-10}
     & $\widehat{E}_{min}$ & $\widehat{E}_{max}$ & $O_{[j]}^{ss}$ & limit/order & limit & Success (\%) & Total & Min & Median \\[1ex]
    \hline
    1 & 0.10 & 1.0 & 1 to 5 & 3000 & 5 & 100\% & 100 & 76.71 & 143.19 \\
    2 & 0.99 & 1.0 & 1 to 5 & 3000 & 5 & 98\% & 100 & 70.31 & 184.62 \\
    3 & 0.99 & 1.0 & 1 to 5 & 3000 & 10 & 100\% & 100 & 67.20 & 155.69 \\
    4 & 0.99 & 1.0 & 1 to 20 & 750 & 10 & 100\% & 100 & 81.31 & 209.64 \\
    \bottomrule
\end{tabular}
\begin{tablenotes}\footnotesize
\item[1] Statistics for successful results. All simulations were performed on a laptop with an Apple M1 Pro chip and 16 GB of RAM.
\end{tablenotes}
\end{threeparttable}
\end{table*}

\FloatBarrier
\subsection{Example: Actuator}
This design example is investigated specifically to compare with a state-of-art approach by Bonfanti et al.~\cite{Bonfanti2020.automatic} as mentioned in Section~\ref{sec:Intro}. In their work, beam networks are utilized to transform vertical displacement on the top left (input) to horizontal nodal displacement on the top right (output). Different network sizes realizing comparable functional modes are also studied to obtain the scaling law of their algorithms (i.e., simulation time versus initial network size).

\begin{figure*}[th]
\centering
\includegraphics[width= 0.98\textwidth]{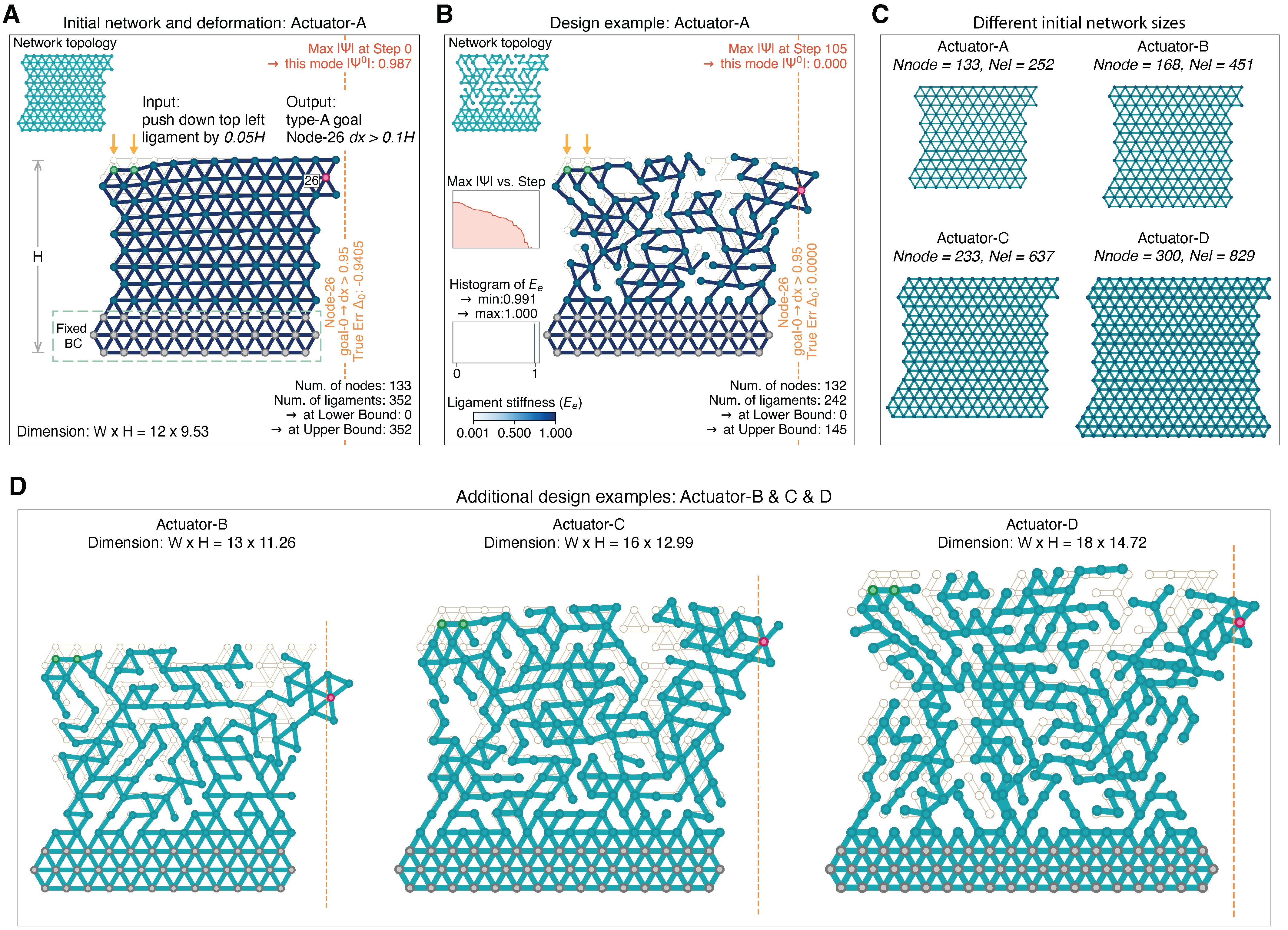}
\caption{Numerical example: Actuator (beam network).
(A)~Initial network, deformation under input, and prescribed output (Actuator-A).
(B)~One design example (Actuator-A).
(C)~Different sizes of actuators (labeled by letters A to D) with increasing number of ligaments are included for investigation of computation times.
(D)~Additional design examples from Actuator-B to Actuator-D.
}
\label{fig:example-Actuator}
\end{figure*}

As shown in Fig.~\ref{fig:example-Actuator}A, we start by generating a beam network (denoted as Actuator-A) with identical geometry and prescribed input-output to one of the demonstrated actuator examples by Bonfanti et al.~\cite{Bonfanti2020.automatic}. This initial network consists of 352 ligaments and 133 nodes. A term named `efficiency' is also introduced in their study, which is the ratio between magnitude of the output displacement to that of the input displacement. Since the structure searching and updating algorithms in their study are based on adding/removing ligaments under a Monte Carlo approach integrated with optimization and data-driven methods, resultant efficiency values in their final design are non-deterministic and vary roughly from 2.0 to 13.0, depending on the structure sizes~\cite{Bonfanti2020.automatic}. In this study, since the output has to be defined explicitly through type-A and type-B goals, an efficiency of at least 2.0 is prescribed. Specifically, the input is to fix the bottom three layers of nodes, and push down a ligament on the top left by a magnitude equivalent to $5\%$ of the structure's height $H$, and the output is to horizontally displace a node on the top right (shown in pink) at least twice the input magnitude (defined by Type-A goal). Fig.~\ref{fig:example-Actuator}B shows a design example from Actuator-A.

To investigate the scaling law of our algorithms in this design problem, three additional actuator examples with different sizes, labeled by letters B, C, and D, are studied. The inputs and prescribed functional modes are comparable, i.e., push down top left ligament by $5\%$ of the network's height, and obtain a nodal horizontal displacement on the top right by at least twice the input magnitude. For each actuator size, series with different control parameters, labeled by numbers, are also included. For each series, we continue the simulations until 100 successful designs are collected. Details are summarized in Table~\ref{tab:Actuator_stats}. An animation showing the iterative design process in a simulation example (Actuator-A) is given SI.

\begin{table*}[thb]
\small
\centering
\caption{Input parameters and run time for design example: Actuator (beam network)} \label{tab:Actuator_stats}
\begin{threeparttable}
\begin{tabular}{l c c c c c c c c c c}
    \toprule
    \multicolumn{3}{c}{ \multirow{7}{*}{\makecell{Common \\ parameters}} } & 
    \multicolumn{8}{l}{Goal ligament stiffness interval in final design: $\widehat{E}_{min}= 0.99$, $\widehat{E}_{min} = 1$} \\
    & & & \multicolumn{8}{l}{Criterion for picking valid SS: RSD} \\ 
    & & & \multicolumn{8}{l}{Allowed total number of numerical steps (before termination): 2000} \\
    & & & \multicolumn{8}{l}{Orders of candidate SS to generate: 1 to 10} \\
    & & & \multicolumn{8}{l}{Size limit of candidate SS each order: 2000} \\
    & & & \multicolumn{8}{l}{Size limit of picked valid SS for LSP: 10} \\
    & & & \multicolumn{8}{l}{Perturbation coefficient for each picked SS: $\alpha_j= 0.1$} \\
    & & & \multicolumn{8}{l}{Stiffness push down coefficient: $f_{\downarrow} = 0.25$} \\
    & & & \multicolumn{8}{l}{Stiffness recover coefficient: $f_{\uparrow}= 0.1$} \\
    & & & \multicolumn{8}{l}{Allowed maximum relative error in all modes in final design: $\psi^{all}_{max} < \widehat{\psi} = 0.01$} \\
    [1ex]
    \hline \\[-2.5ex]
    \multirow{3}{*}{Series\tnote{*}} & \multicolumn{2}{c}{\makecell{Initial \\ topology}} & \multirow{3}{*}{\makecell{Threshold of \\ $n^{E,min}_{el}$ to \\ change topology}} &
    \multicolumn{2}{c}{\makecell{Numerical \\ Results}} &
    \multicolumn{3}{c}{\makecell{Run time of successful \\ results in seconds (sec)}} &
    \multicolumn{2}{c}{\makecell{Run time of \\ all simulations (sec)}} \\ [1ex]
    \cmidrule(lr){2-3} 
    \cmidrule(lr){5-6}
    \cmidrule(lr){7-9}
    \cmidrule(lr){10-11}
     & $n_{el}$ & $n_{node}$ & & Success & Total & Min & Median & Average & Median & Average \\[1ex]
    \hline
    A.1 & 352 & 133 & 10 & 100 & 147 & 10.16 & 88.45 & 109.02 & 116.01 & 149.79 \\
    A.2 & 352 & 133 & 20 & 100 & 166 & 10.29 & 55.83 & 98.47 & 91.14 & 133.07 \\
    A.3 & 352 & 133 & 30 & 100 & 193 & 10.77 & 50.83 & 77.77 & 84.07 & 117.43 \\
    \hline 
    B.1 & 451 & 168 & 10 & 100 & 144 & 16.03 & 140.26 & 203.07 & 163.82 & 279.02 \\
    B.2 & 451 & 168 & 20 & 100 & 173 & 16.32 & 88.47 & 148.20 & 128.53 & 180.68 \\
    B.3 & 451 & 168 & 30 & 100 & 195 & 17.33 & 104.50 & 159.62 & 138.03 & 218.44 \\
    \hline
    C.1 & 637 & 233 & 10 & 100 & 129 & 34.50 & 315.67 & 641.15 & 354.22 & 687.87 \\
    C.2 & 637 & 233 & 20 & 100 & 175 & 36.81 & 203.48 & 382.52 & 301.66 & 634.12 \\
    \hline
    D.1 & 829 & 300 & 10 & 100 & 140 & 86.08 & 607.16 & 1249.15 & 738.11 & 1547.88 \\
    D.2 & 829 & 300 & 20 & 100 & 162 & 93.38 & 491.43 & 1073.00 & 507.92 & 1146.29 \\
    \bottomrule
\end{tabular}
\begin{tablenotes}\footnotesize
\item[*] This example contains different initial network size and geometries, labeled by letters A, B, C and D. For each initial network, different series are labeled by Arabic numbers.
\item[*] In each series, simulation continues until 100 successful designs are collected.
\item[*] All simulations were performed on a laptop with an Apple M1 Pro chip and 16 GB of RAM.
\end{tablenotes}
\end{threeparttable}
\end{table*}

Our approach demonstrates significant advantages in both computational speed and required hardware resources. For example, in Actuator-A, Series-2, it took 166 simulations to obtain 100 successful designs. The minimum, median, and average run times of the 100 successful simulations are only 10.29, 55.83, and 98.47 seconds, respectively. 
For the identical problem (with 352 ligaments in the initial network) in their study, the execution times (to collect 100 designs) range from about 1 to 16 hours, with a median of around 4 hours~\cite{Bonfanti2020.automatic} (exact value of median and average run time are not listed). The median run time of the successful results by SSPM is therefore about 258x faster. 
The median run time of all 166 simulations is 91.14 seconds, so if accounting for success rate, the effective median to obtain 100 designs is 91.14*1.66 = 151.3 seconds, still about 95x faster than 4 hours. The run time statistics of the other two series (with different input values of $n^{E,min}_{el}$) in Actuator-A are also in comparable range.

Moreover, the simulations in their study were executed on a 16-core desktop CPU, with corresponding data training (required as a prerequisite) performed on a high end GPU. In contrast, our simulations are performed on a laptop with a 10-core Apple M1 Pro CPU with 16GB unified RAM. Therefore, the proposed SSPM is effectively about two orders of magnitude faster (in this design example) while using less computational resources, and without the need to train or prepare any data.

\begin{figure*}[thb]
\centering
\includegraphics[width= 0.98\textwidth]{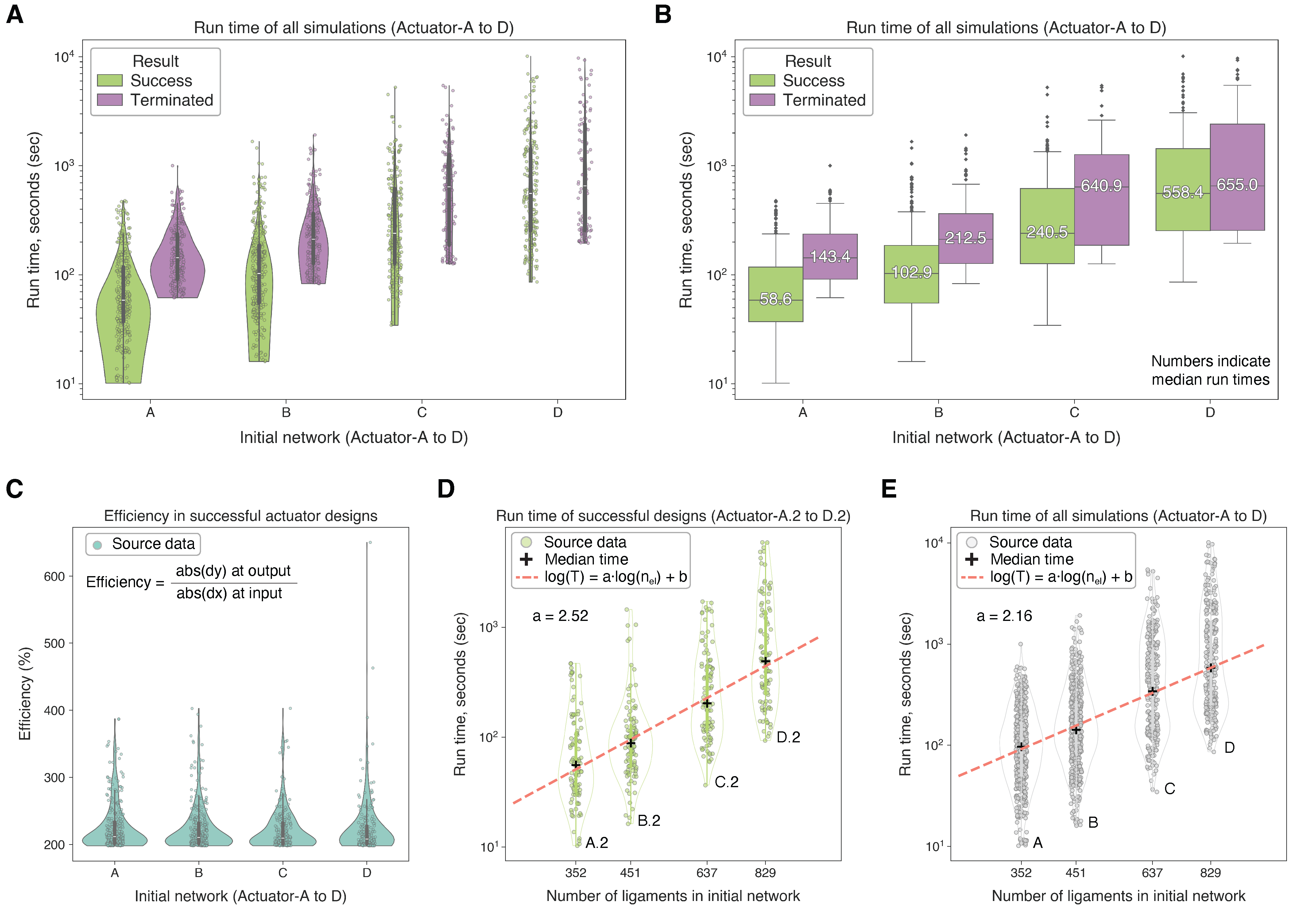}
\caption{Statistics of run time and efficiency of actuator design examples. Vertical axes that represent run time are in logarithmic scale. Data are also summarized in Table~\ref{tab:Actuator_stats}.
(A)~Violin plot of run times of both successful and terminated simulations in all examples (Actuator-A to D).
(B)~Box plot of run times of all simulations. The labeled numbers indicate median run times.
(C)~Violin plot of efficiency ratio in successful designs (the prescribed goal is efficiency $\geq 2.0$). 
(D)~Scaling law of computation time versus total number of ligaments, based on median run times of successful simulations from series A.2 to D.2.
(E)~Scaling law based on median run times of both successful and terminated simulations from all series.
}
\label{fig:runtTimes-Actuator}
\end{figure*}

Fig.~\ref{fig:runtTimes-Actuator}A and Fig.~\ref{fig:runtTimes-Actuator}B show violin plots of all simulations (both successful and terminated) from Actuator-A to Actuator-D. As the network dimension increases, the range and distribution of run times also expand. This is likely due to higher level of uncertainties in both candidate SS generation and ligament removal procedures when dealing with larger network dimensions. Also as expected, the terminated simulations generally take longer to run because more attempts were made to search for valid SS before reaching termination criteria.
Fig.~\ref{fig:runtTimes-Actuator}C shows the statistics of the efficiency (magnitude of horizontal displacement at output divided by magnitude of vertical input displacement). The goal value requires the efficiency to be at least 2.0, and the resulting efficiency values typically range from 2.0 to 2.5.

Fig.~\ref{fig:runtTimes-Actuator}D shows the approximate scaling law of computation time (versus the number of ligaments in the initial network) based on series A.2 to D.2. These four series share the same simulation control parameters, with the only difference being the actuator size. Both the horizontal and vertical axes are plotted in logarithmic scale. The orange dashed line shows the linear polynomial fit of the four median run times. As shown, the median values lie closely to the fitted line. The slope of the fitted line (i.e., coefficient indicating the exponential growth rate) is 2.52. 
Fig.~\ref{fig:example-Actuator}E shows approximate scaling law based on all series from Actuator-A to D (both successful and terminated simulations). The slope of the fitted line is 2.16. In contrast, the exponential coefficient in their study is 2.97~\cite{Bonfanti2020.automatic}. It should also be noted that the scaling law applies for this design example, and it may not represent the inherent computational complexity of the SSPM algorithms because the run time depends on various simulation parameters (e.g., amount of candidate SS generated when necessary, limit of total steps) and computational resources. Nevertheless, the essential parameters (e.g., number and order of candidate SS) of the included series are consistent through the four different actuator sizes to ensure the amount of work needed are scaling up proportionally to their sizes.

\FloatBarrier
\subsection{Example: Lobster}
So far, all the above demonstrated examples have only one functional mode. In the following two design examples, \textit{Lobster-A} and \textit{Lobster-B}, two prescribed functional modes are realized simultaneously. Truss networks are assumed in both examples.

\begin{figure*}[thb]
\centering
\includegraphics[width= 0.98\textwidth]{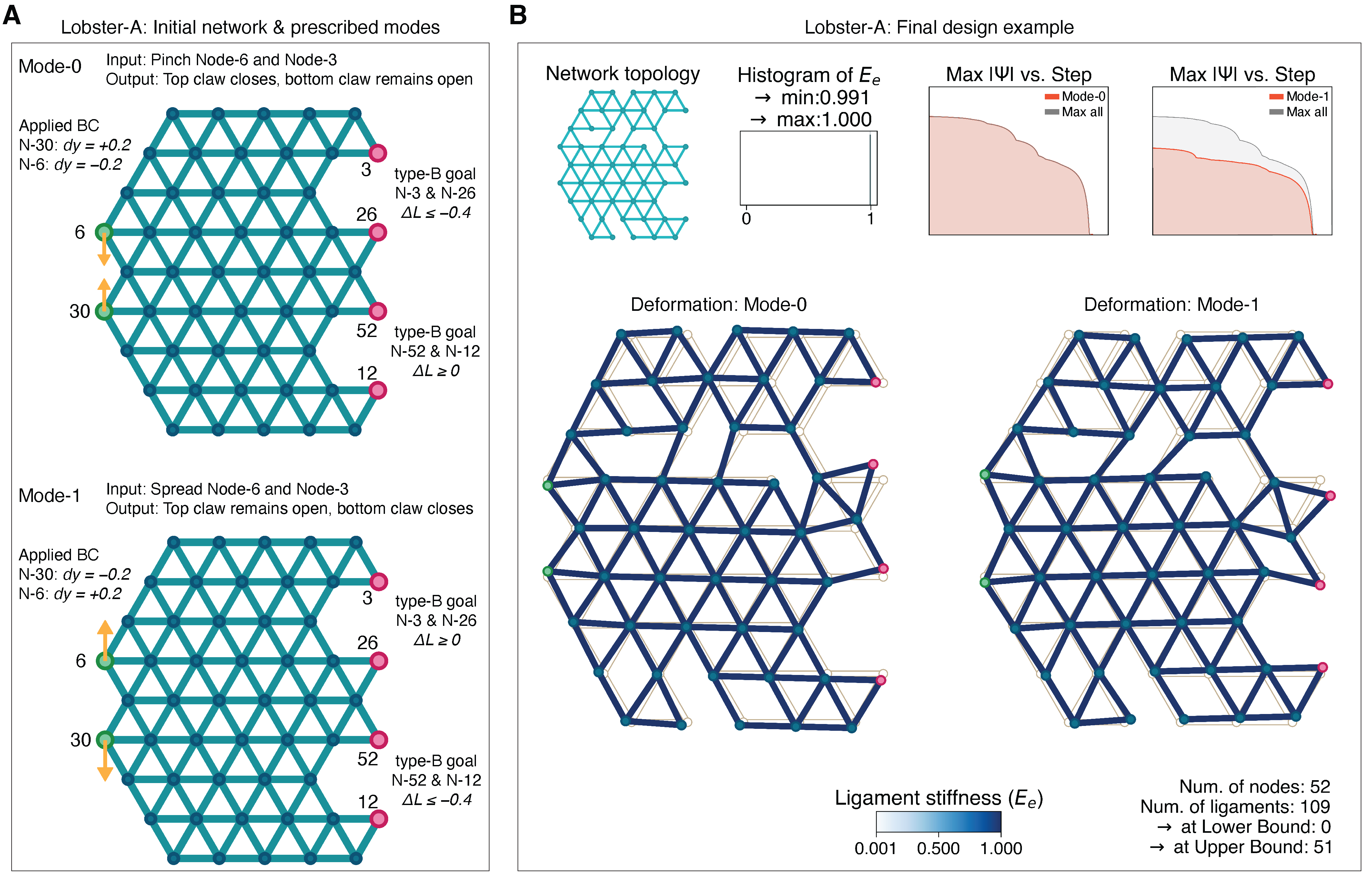}
\caption{Design example: Lobster-A (truss network).
(A)~The initial network consists of 125 ligaments and 52 nodes. The prescribed input and output in two modes are as denoted.
(B)~One design example and its deformations in the two prescribed modes. As expected, if pinching the tail (Mode-0), the top claw squeezes while the bottom claw remains open, if spreading the tail (Mode-1), the bottom claw squeezes while the top claw remains open. The goal stiffness interval is set to [0.99, 1] so the final design consists of ligaments with effectively uniform stiffness.
}
\label{fig:LobA-example}
\end{figure*}

For Lobster-A, the initial geometry is as shown in Fig.~\ref{fig:LobA-example}A. The two pockets on the right edge conceptually resemble two `claws', while the two end nodes on the left edge resemble its `tail'. The initial network consists of 125 ligaments and 52 nodes. For the first prescribed functional mode (denoted as Mode-0), the input is to pinch the two nodes on the left along the vertical direction by a magnitude of 0.2 ($20\%$ of the ligament length), and the output is for the top claw to close by at least twice the input magnitude while the bottom claw remains open. These responses are defined by two type-B goals on the two end nodes of each claw (shown in pink). For output in Mode-0, the first goal requires change in distance between node-3 and node-26 to be lower than -0.4 (i.e., the top claw closeup by at least 0.4). The second goal requires change in distance between node-52 and node-12 to be higher than zero (i.e., the bottom claw remains open). For the second prescribed mode (denoted as Mode-1), the input is to spread the input nodes (reverse to pinch) by 0.2, and the output is for the bottom claw to closeup by at least 0.4 while the top claw remains open. Two type-B goals for prescribed output in Mode-1 can hereby be defined similarly. Fig.~\ref{fig:LobA-example}B shows one design example. As expected, the top claw squeezes if pinching the tail (Mode-0), whereas the bottom claw squeezes if spreading the tail (Mode-1).

\begin{figure*}[tb]
\centering
\includegraphics[width= 0.75\textwidth]{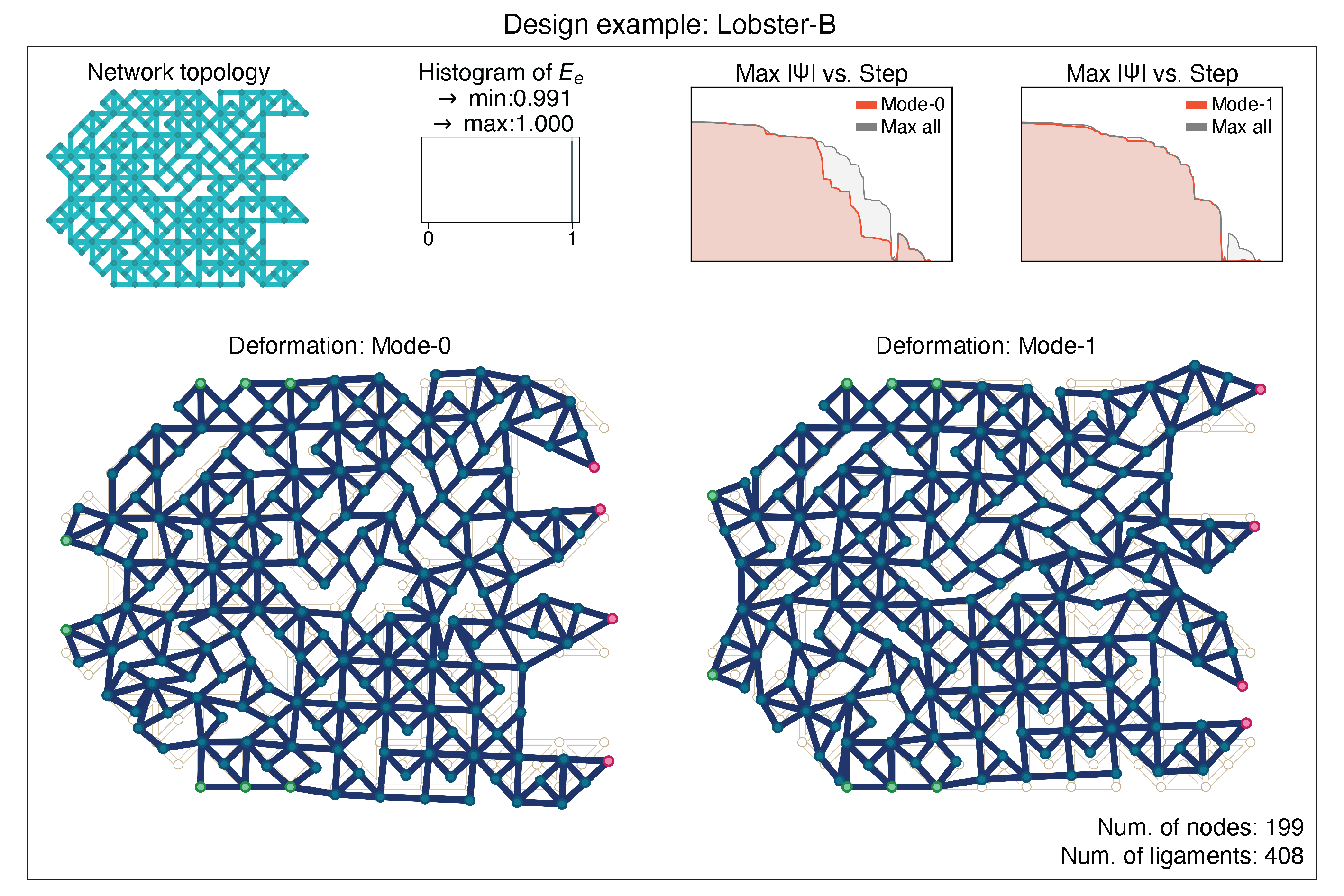}
\caption{Design example: Lobster-B (truss network). The initial network consists of 538 ligaments and 199 nodes. Comparable input-output functional modes to Lobster-A are prescribed.
}
\label{fig:LobB-example}
\end{figure*}

For Lobster-B, a square lattice filled with diagonal ligaments is used as the initial network. The input-output modes are comparable to Lobster-A. One design example is given in Fig.~\ref{fig:LobB-example}. Two series are performed for each example, and the results are summarized in Table~\ref{tab:Lobster_stats}. In Series-1, topology change is only attempted upon reaching the goal while Series-2 attempts to remove ligaments on-the-fly if $n^{E,min}_{el} \geq 5$. Improved success rates and decreased run times are then observed for Series-2 over Series-1 in both design examples. For Lobster-A, the median and average run times are only around 10 and 12 seconds, respectively. For Lobster-B, the median and average run times are all less than 60 seconds.

\begin{table*}[hb]
\small
\centering
\caption{Input parameters and run time for design example: Lobster (truss network)}\label{tab:Lobster_stats}
\begin{threeparttable}
\begin{tabular}{c c c c c c c c c c c}
    \toprule
    \multicolumn{3}{c}{ \multirow{7}{*}{\makecell{Common \\ parameters}} } & 
    \multicolumn{8}{l}{Goal ligament stiffness interval in final design: $\widehat{E}_{min}= 0.99$, $\widehat{E}_{min} = 1$} \\
    & & & \multicolumn{8}{l}{Criterion for picking valid SS: RSD} \\
    & & & \multicolumn{8}{l}{Allowed total number of numerical steps (before termination): 1000} \\
    & & & \multicolumn{8}{l}{Size limit of picked valid SS: $5$} \\
    & & & \multicolumn{8}{l}{Perturbation coefficient for each picked SS: $\alpha_j= 0.01$} \\
    & & & \multicolumn{8}{l}{Stiffness push down coefficient: $f_{\downarrow} = 0.25$} \\
    & & & \multicolumn{8}{l}{Stiffness recover coefficient: $f_{\uparrow}= 0.1$} \\
    & & & \multicolumn{8}{l}{Allowed maximum relative error in all modes in final design: $\psi^{all}_{max} < \widehat{\psi} = 0.01$} \\
    [1ex]
    \hline \\[-2.5ex]
    \multirow{3}{*}{Series\tnote{*}} & \multicolumn{2}{c}{\makecell{Initial \\ topology}} & \multirow{3}{*}{\makecell{Threshold of \\ $n^{E,min}_{el}$ to \\ change topology}} &
    \multicolumn{2}{c}{\makecell{Order \& limit \\ of candidate SS} } & 
    \multicolumn{2}{c}{\makecell{Numerical \\ Results}} &
    \multicolumn{3}{c}{Run time\tnote{1} \ (seconds)} \\ [1ex]
    \cmidrule(lr){2-3} 
    \cmidrule(lr){5-6}
    \cmidrule(lr){7-8}
    \cmidrule(lr){9-11}
     & $n_{el}$ & $n_{node}$ & & $O_{[j]}^{ss}$ & limit & Success (\%) & Total & Min & Median & Average \\[1ex]
    \hline
    A.1 & 125 & 52 & \textit{None} & 1 to 10 & 1000 & 83.8\% & 500 & 6.46 & 11.03 & 14.38 \\
    A.2 & 125 & 52 & 5 & 1 to 10 & 1000 & 91.8\% & 500 & 6.40 & 10.21 & 12.06 \\
    \hline
    B.1 & 538 & 199 & \textit{None} & 1 to 10 & 1000 & 51.4\% & 500 & 18.87 & 48.88 & 59.67 \\
    B.2 & 538 & 199 & 5 & 1 to 10 & 1000 & 72.6\% & 500 & 20.37 & 41.93 & 54.25 \\
    \bottomrule
\end{tabular}
\begin{tablenotes}\footnotesize
\item[*] This example contains different initial network size and geometries, labeled by letters A and B. Different series for the same network are labeled by Arabic numbers.
\item[1] Statistics for successful results. All simulations were performed on a laptop with an Apple M1 Pro chip and 16 GB of RAM.
\end{tablenotes}
\end{threeparttable}
\end{table*}

\FloatBarrier
\section{Prototyping example}
Here we demonstrate prototyping and experimental studies based on the Lobster-A design example (Fig.~\ref{fig:LobA-example}). In the numerical simulations, the axial stresses in a ligament generally scales linearly with its relative elongation (positive or negative), wherein no restrictions on their magnitudes are implied. In reality, large positive stress may result in tensile failure in ligaments or joints, while large compressive stress may lead to buckling or bending in ligaments. Therefore, for prototyping purposes here, it is preferred that a lattice structure realizes prescribed modes with less strain energy. In a truss network, this is when the ligaments tend to rotate, rather than stretch, under the input displacements. Therefore, among all the successful designs, we pick two with minimum overall magnitude of ligament compressive strains (in both two modes).

Each prototype is fabricated by laser-cutting a silicone rubber (polysiloxane) sheet with out-of-plane thickness of 1.6~mm and Shore hardness value of A90. The joint-to-joint ligament length in the prototypes is 25~mm. Given the fabrication processes, the joints in the prototypes have non-negligible rotation stiffness, whereas truss networks are assumed in the Lobster-A design example. To address this issue, the in-plane cross-sectional width of the ligaments is designed to be 2.0~mm, but are tapered to 1.25~mm around their ends to alleviate ligament bending. 

\begin{figure*}[tb]
\centering
\includegraphics[width= 0.98\textwidth]{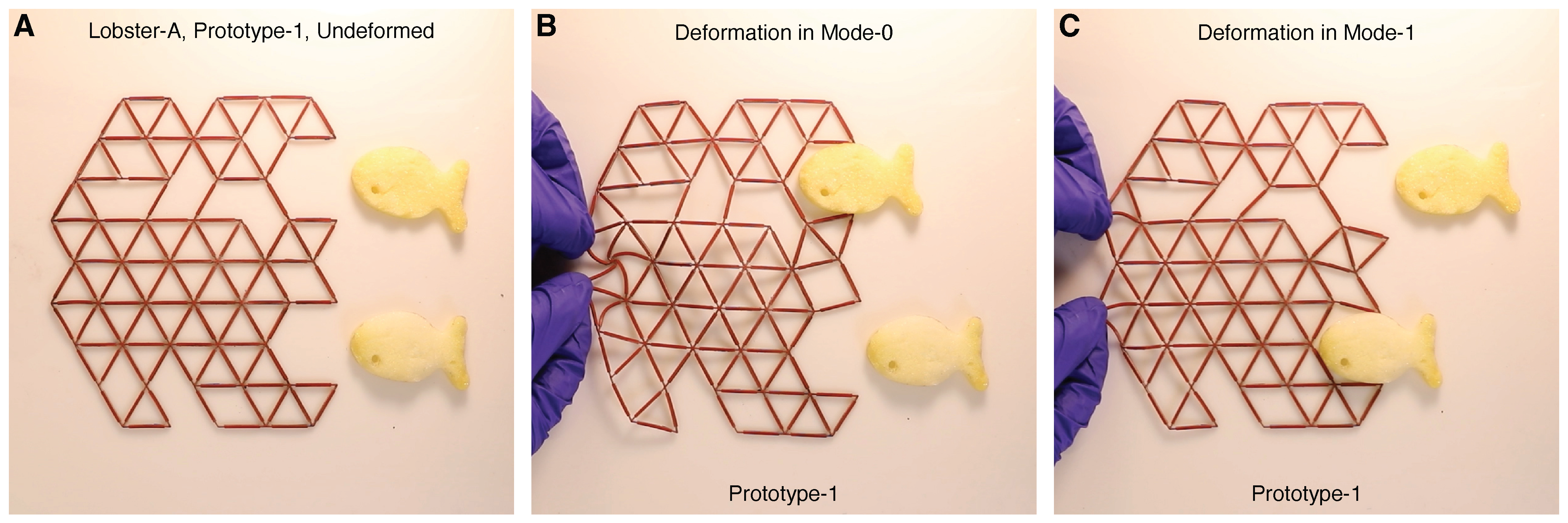}
\caption{Fabricated Prototype-1 from Lobster-A (same topology as in Fig.~\ref{fig:LobA-example}).
(A)~Two foam objects are placed near the claws to demonstrate the prescribed functionalities (selective seizing of objects).
(B)~In Mode-0, the top claw seizes one object, while the bottom claw remains open.
(C)~In Mode-1, the bottom claw seizes the other object, while the top claw remains open.
}
\label{fig:LobA-P1-snaps}
\end{figure*}

Fig.~\ref{fig:LobA-P1-snaps}A shows the fabricated sample of the first design, Prototype-1, whose topology is identical to the network shown in Fig.~\ref{fig:LobA-example}B. To illustrate the output functionalities, we place two objects (made by laser-cutting foam sheets) near the two `claws' of the prototype and actuate the structure in both modes via 'pinch' and 'spread' to grasp either object. As shown in Fig.~\ref{fig:LobA-P1-snaps}B and Fig.~\ref{fig:LobA-P1-snaps}C, the deformation mechanisms in both Mode-0 and Mode-1 are in close agreement with numerical predictions, and hereby realize the prescribed multi-mode functionalities.

Due to the inherent material characteristics and compressive instabilities of the ligaments, noticeable in-plane bending of ligaments is observed near the two input nodes, especially in Mode-0 when these nodes are being squeezed. There are also moderate out-of-plane deformations in a portion of the network, likely caused by the non-planar inputs (due to the manually applied load) and by friction between the network and the base platform (an acrylic plate). Placing a small plate to cover the prototype near its tail (where the input displacements are applied) would alleviate such out-of-plane deformations, but the overall structural responses would not differ by much. As a result of these factors, the magnitudes of change in distance between the end nodes of claws (i.e., claw opening or closure displacements) are smaller than the numerical predictions. We measure the opening and closure magnitude of each claw in each mode, and compare the results with numerical predictions, details are listed in SI. In addition, prototype of another design is fabricated, and animations showing the functionalities of the two prototypes are provided in SI.

\FloatBarrier
\section{Discussion and Conclusion}
In this study, we present a novel Substructure Perturbation Method (SSPM) for the systematic design and search of mechanical metamaterials with prescribed input-output functionalities. We firstly explain the theoretical fundamentals and derivations of the developed SSPM, and then, describe the concept of substructures (SS) corresponding to the features that can be potentially adjusted (i.e., ligament stiffness in this study). Subsequently, to illustrate the logistics and potential implementation of SSPM, we present the developed computational algorithms for the purpose of this study, and introduce four design problems covering truss and beam networks with different kinds of prescribed functionalities (i.e., in form of different types of goals). Results demonstrate the effectiveness of the SSPM framework and its advantages in computational performance. In the Actuator design example, we compared the current  numerical algorithms of SSPM to another state-of-art approach relying on iterative addition/removal of ligaments and data-driven methods, and found that SSPM is effectively about two orders of magnitude faster while using significantly less computing resources.

Unlike existing methods that typically rely on trial-and-error search, global optimizations, or machine learning frameworks, the proposed SSPM explicitly analyzes the effect of infinitesimal and simultaneous change on multiple substructural features, and thereby, prompts significant computational efficiency and versatility in its prospective applications. The methodology of SSPM in this work can be generalized and extended to support multiple structural features in different types of structures, such as concurrent nodal positional and cross-sectional area change in a three-dimensional network consists of beam or shell elements. SSPM may also be extended to design materials with prescribed elastic properties if, for instance, adapted to work in corporation with periodical boundary conditions and computational homogenization methods. Therefore, the proposed SSPM  provides new fundamentals for the design and optimization of mechanical metamaterials with advanced functional modes and properties.

We anticipate potential integration of advanced data-driven (machining learning) techniques with the proposed SSPM to further enhance its numerical performance. In fact, the required computational times for SS generation and validation (in a complete simulation) are both noticeably higher than the other procedures. This is not only because generation of each SS requires stiffness assembly while SS validation requires multiplication of several matrices (both are relatively computationally intensive), but also, more critically, time required for SS generation and validation scales with the amount of candidate SS being processed in a simulation. Under these circumstances, the `quality' of candidate SS (e.g., ratio of valid SS in available candidate SS) can play an important role in the computational speed and success rate in finding a design (given a step limit). In the numerical algorithms in this study, generation of SS is solely based on random sampling, but as discussed earlier, there can be various ways and considerations when generating candidate SS. Hence, we plan to investigate methods for improvement in the efficiency of SS generation and validation procedures, which would enhance the overall numerical performance of SSPM, especially for large networks or complicated multi-mode design problems.

Following the above discussions, future work can be categorized into two directions. The first involves further development of the proposed SSPM. For instance, SSPM may be extended to support higher dimensional finite elements (e.g., shell elements), and in accordance, may be generalized to support more substructural features (e.g., positional perturbation of nodes/joints in a network). The second direction is to use SSPM to design advanced mechanical metamaterials with programmed functionalities and/or properties, such as robotic mechanisms, actuators, and auxetic structures.

\section{Acknowledgement}
This research was funded primarily by the National Science Foundation (NSF) MRSEC program (award DMR-2309043).

\FloatBarrier
\newpage
\bibliography{mybibfile}

\begin{thebibliography}{10}
\expandafter\ifx\csname url\endcsname\relax
  \def\url#1{\texttt{#1}}\fi
\expandafter\ifx\csname urlprefix\endcsname\relax\def\urlprefix{URL }\fi
\expandafter\ifx\csname href\endcsname\relax
  \def\href#1#2{#2} \def\path#1{#1}\fi

\bibitem{Barchiesi2019.mechanical}
E.~Barchiesi, M.~Spagnuolo, L.~Placidi, Mechanical metamaterials: a state of
  the art, Mathematics and Mechanics of Solids 24~(1) (2019) 212--234.
\newblock \href {https://doi.org/10.1177/1081286517735695}
  {\path{doi:10.1177/1081286517735695}}.

\bibitem{Zaiser2023.disordered}
M.~Zaiser, S.~Zapperi, Disordered mechanical metamaterials, Nature Reviews
  Physics (2023) 1--10\href {https://doi.org/10.1038/s42254-023-00639-3}
  {\path{doi:10.1038/s42254-023-00639-3}}.

\bibitem{Jiao2023.mechanical}
P.~Jiao, J.~Mueller, J.~R. Raney, X.~Zheng, A.~H. Alavi, Mechanical
  metamaterials and beyond, Nature Communications 14~(1) (2023) 6004.
\newblock \href {https://doi.org/10.1038/s41467-023-41679-8}
  {\path{doi:10.1038/s41467-023-41679-8}}.

\bibitem{Bertoldi2017.flexible}
K.~Bertoldi, V.~Vitelli, J.~Christensen, M.~Van~Hecke, Flexible mechanical
  metamaterials, Nature Reviews Materials 2~(11) (2017) 1--11.
\newblock \href {https://doi.org/10.1038/natrevmats.2017.66}
  {\path{doi:10.1038/natrevmats.2017.66}}.

\bibitem{Coulais2016.combinatorial}
C.~Coulais, E.~Teomy, K.~De~Reus, Y.~Shokef, M.~Van~Hecke, Combinatorial design
  of textured mechanical metamaterials, Nature 535~(7613) (2016) 529--532.
\newblock \href {https://doi.org/https://doi:10.1038/nature18960}
  {\path{doi:https://doi:10.1038/nature18960}}.

\bibitem{Reid2018.auxetic}
D.~R. Reid, N.~Pashine, J.~M. Wozniak, H.~M. Jaeger, A.~J. Liu, S.~R. Nagel,
  J.~J. de~Pablo, Auxetic metamaterials from disordered networks, Proceedings
  of the National Academy of Sciences 115~(7) (2018) E1384--E1390.
\newblock \href {https://doi.org/10.1073/pnas.1717442115}
  {\path{doi:10.1073/pnas.1717442115}}.

\bibitem{Rocks2017.designing}
J.~W. Rocks, N.~Pashine, I.~Bischofberger, C.~P. Goodrich, A.~J. Liu, S.~R.
  Nagel, Designing allostery-inspired response in mechanical networks,
  Proceedings of the National Academy of Sciences 114~(10) (2017) 2520--2525.
\newblock \href {https://doi.org/10.1073/pnas.1612139114}
  {\path{doi:10.1073/pnas.1612139114}}.

\bibitem{Yan2017.architecture}
L.~Yan, R.~Ravasio, C.~Brito, M.~Wyart, Architecture and coevolution of
  allosteric materials, Proceedings of the National Academy of Sciences
  114~(10) (2017) 2526--2531.
\newblock \href {https://doi.org/10.1073/pnas.1615536114}
  {\path{doi:10.1073/pnas.1615536114}}.

\bibitem{Bastek2022.inverting}
J.-H. Bastek, S.~Kumar, B.~Telgen, R.~N. Glaesener, D.~M. Kochmann, Inverting
  the structure--property map of truss metamaterials by deep learning,
  Proceedings of the National Academy of Sciences 119~(1) (2022) e2111505119.
\newblock \href {https://doi.org/10.1073/pnas.2111505119}
  {\path{doi:10.1073/pnas.2111505119}}.

\bibitem{Ha2023.rapid}
C.~S. Ha, D.~Yao, Z.~Xu, C.~Liu, H.~Liu, D.~Elkins, M.~Kile, V.~Deshpande,
  Z.~Kong, M.~Bauchy, et~al., Rapid inverse design of metamaterials based on
  prescribed mechanical behavior through machine learning, Nature
  Communications 14~(1) (2023) 5765.
\newblock \href {https://doi.org/10.1038/s41467-023-40854-1}
  {\path{doi:10.1038/s41467-023-40854-1}}.

\bibitem{Gao2023.rational}
J.~Gao, X.~Cao, M.~Xiao, Z.~Yang, X.~Zhou, Y.~Li, L.~Gao, W.~Yan, T.~Rabczuk,
  Y.-W. Mai, Rational designs of mechanical metamaterials: Formulations,
  architectures, tessellations and prospects, Materials Science and
  Engineering: R: Reports 156 (2023) 100755.

\bibitem{Ion2016.metamaterial}
A.~Ion, J.~Frohnhofen, L.~Wall, R.~Kovacs, M.~Alistar, J.~Lindsay, P.~Lopes,
  H.-T. Chen, P.~Baudisch, Metamaterial mechanisms, in: Proceedings of the 29th
  annual symposium on user interface software and technology, 2016, pp.
  529--539.
\newblock \href {https://doi.org/10.1145/2984511.2984540}
  {\path{doi:10.1145/2984511.2984540}}.

\bibitem{Yasuda2021.mechanical}
H.~Yasuda, P.~R. Buskohl, A.~Gillman, T.~D. Murphey, S.~Stepney, R.~A. Vaia,
  J.~R. Raney, Mechanical computing, Nature 598~(7879) (2021) 39--48.
\newblock \href {https://doi.org/10.1038/s41586-021-03623-y}
  {\path{doi:10.1038/s41586-021-03623-y}}.

\bibitem{Pashine2023.reprogrammable}
N.~Pashine, A.~M. Nasab, R.~Kramer-Bottiglio, Reprogrammable allosteric
  metamaterials from disordered networks, Soft Matter 19~(8) (2023) 1617--1623.
\newblock \href {https://doi.org/10.1039/D2SM01284G}
  {\path{doi:10.1039/D2SM01284G}}.

\bibitem{Lee2022.mechanical}
R.~H. Lee, E.~A. Mulder, J.~B. Hopkins, Mechanical neural networks: Architected
  materials that learn behaviors, Science Robotics 7~(71) (2022) eabq7278.
\newblock \href {https://doi.org/10.1126/scirobotics.abq7278}
  {\path{doi:10.1126/scirobotics.abq7278}}.

\bibitem{Bonfanti2020.automatic}
S.~Bonfanti, R.~Guerra, F.~Font-Clos, D.~Rayneau-Kirkhope, S.~Zapperi,
  Automatic design of mechanical metamaterial actuators, Nature communications
  11~(1) (2020) 4162.
\newblock \href {https://doi.org/https://10.1038/s41467-020-17947-2}
  {\path{doi:https://10.1038/s41467-020-17947-2}}.

\bibitem{Bitzek2006.structural}
E.~Bitzek, P.~Koskinen, F.~G{\"a}hler, M.~Moseler, P.~Gumbsch, Structural
  relaxation made simple, Physical review letters 97~(17) (2006) 170201.
\newblock \href {https://doi.org/10.1103/physrevlett.97.170201}
  {\path{doi:10.1103/physrevlett.97.170201}}.

\bibitem{Kirkpatrick1983.optimization}
S.~Kirkpatrick, C.~D. Gelatt~Jr, M.~P. Vecchi, Optimization by simulated
  annealing, science 220~(4598) (1983) 671--680.
\newblock \href {https://doi.org/10.1126/science.220.4598.671}
  {\path{doi:10.1126/science.220.4598.671}}.

\bibitem{Calladine1978.buckminster}
C.~R. Calladine, Buckminster fuller's “tensegrity” structures and clerk
  maxwell's rules for the construction of stiff frames, International journal
  of solids and structures 14~(2) (1978) 161--172.
\newblock \href {https://doi.org/10.1016/0020-7683(78)90052-5}
  {\path{doi:10.1016/0020-7683(78)90052-5}}.

\bibitem{Gibson1997.cellular}
L.~J. Gibson, M.~F. Ashby, Cellular Solids: Structure and Properties, 2nd
  Edition, Cambridge Solid State Science Series, Cambridge University Press,
  1997.
\newblock \href {https://doi.org/10.1017/CBO9781139878326}
  {\path{doi:10.1017/CBO9781139878326}}.

\bibitem{Fish2007.First}
J.~Fish, T.~Belytschko, A First Course in Finite Elements, John Wiley \& Sons,
  Inc., 2007.

\bibitem{Cohn2002.further}
P.~M. Cohn, Further algebra and applications, Springer Science \& Business
  Media, 2002.
\newblock \href {https://doi.org/10.1007/978-1-4471-0039-3}
  {\path{doi:10.1007/978-1-4471-0039-3}}.

\bibitem{Hager1989.updating}
W.~W. Hager, Updating the inverse of a matrix, SIAM review 31~(2) (1989)
  221--239.
\newblock \href {https://doi.org/10.1137/1031049} {\path{doi:10.1137/1031049}}.

\bibitem{Lakes1987.foam}
R.~Lakes, Foam structures with a negative poisson's ratio, Science 235~(4792)
  (1987) 1038--1040.
\newblock \href {https://doi.org/10.1126/science.235.4792.1038}
  {\path{doi:10.1126/science.235.4792.1038}}.

\bibitem{Saxena2016.three}
K.~K. Saxena, R.~Das, E.~P. Calius, Three decades of auxetics research -
  materials with negative poisson's ratio: A review, Advanced Engineering
  Materials 18~(11) (2016) 1847--1870.
\newblock \href {https://doi.org/https://doi.org/10.1002/adem.201600053}
  {\path{doi:https://doi.org/10.1002/adem.201600053}}.

\end{thebibliography}

\FloatBarrier
\newpage
\appendix
\onecolumn

\section{Supporting information for theoretical framework}
\subsection{Approximation of perturbed transformation matrix by multiple substructures}
In the main text, we derived an estimation for the perturbed transformation matrix by a single substructure ($T_{p[j]}$ in Eq.~\ref{eq:Tpj_approx}), and showed the estimation formula for the perturbed transformation matrix by multiple substructures ($T_{p[j,k,l...]}$ in Eq.~\ref{eq:Tpjkl_approx}). This is derived based on the recursive rule in SMW formulas when introducing extra perturbational matrices. 

Specifically, in Eq.~\ref{eq:inv(U-V)_approx} we obtained:
\begin{equation}\label{eqSI:inv(A-B)_approx}
    ( \bfA - \alpha \bfB )^{-1}  = \bfA^{-1} + \alpha \bfA^{-1} \bfB \bfA^{-1} + \mathcal{O}(\alpha) \approx \bfA^{-1} + \alpha \bfA^{-1} \bfB \bfA^{-1}.
\end{equation}
Now apply another infinitesimal perturbation on $\bfA$, i.e., subtract it by another matrix $\bfC$ scaled by a small coefficient $\rho$. Then by treating $\bfA - \alpha \bfB$ as the base matrix and $\rho \bfC$ as the perturbation, we apply the same relationship shown in Eq.~\ref{eqSI:inv(A-B)_approx} and obtain the approximation formula of $(\bfA-\alpha\bfB - \rho\bfC)^{-1}$ by neglecting higher order terms:
\begin{equation}\label{eqSI:inv(A-B-C)_approx}
\begin{aligned}
    (\bfA-\alpha\bfB - \rho\bfC)^{-1} & = \left[(\bfA-\alpha\bfB)-\rho\bfC\right]^{-1} \\
    & =(\bfA - \alpha \bfB)^{-1} + \rho (\bfA-\alpha\bfB)^{-1} \bfC (\bfA-\alpha\bfB)^{-1} + \mathcal{O}(\rho) \\
    & \approx (\bfA - \alpha \bfB)^{-1} + \rho \left(\bfA^{-1} + \alpha \bfA^{-1} \bfB \bfA^{-1} \right) \bfC \left(\bfA^{-1} + \alpha \bfA^{-1} \bfB \bfA^{-1} \right)  \\
    & = (\bfA - \alpha \bfB)^{-1} + \left( \rho\bfA^{-1}+\alpha\rho \bfA^{-1}\bfB\bfA^{-1} \right) \bfC \left( \bfA^{-1}+\alpha \bfA^{-1}\bfB\bfA^{-1} \right) \\
    & = (\bfA - \alpha \bfB)^{-1} + \left[ \rho\bfA^{-1}+ \mathcal{O}(\alpha\rho) \right] \bfC \left( \bfA^{-1}+\alpha \bfA^{-1}\bfB\bfA^{-1} \right) \\
    & \approx \bfA^{-1} + \alpha \bfA^{-1} \bfB \bfA^{-1} + \beta \bfA^{-1} C \left( \bfA^{-1}+\alpha \bfA^{-1}\bfB\bfA^{-1} \right) \\
    & = \bfA^{-1} + \alpha \bfA^{-1} \bfB \bfA^{-1} + \beta \bfA^{-1} C \bfA^{-1} +\alpha \rho \bfA^{-1} \bfA^{-1} \bfB\bfA^{-1} \\
     & = \bfA^{-1} + \alpha \bfA^{-1} \bfB \bfA^{-1} + \beta \bfA^{-1} C \bfA^{-1} + \mathcal{O}(\alpha\rho) \\
     & \approx \bfA^{-1} + \alpha \bfA^{-1} \bfB \bfA^{-1} + \beta \bfA^{-1} C \bfA^{-1}. 
\end{aligned}
\end{equation}
Namely, if the amount of perturbations on the base matrix are sufficiently small such that the higher order terms are considered negligible, then the approximation formula for single perturbation can be extended in a recursive and extendable form for multiple perturbations. Approximation of $T_{p[j,k,l...]}$ is henceforth derived by taking advantage of these properties. Since the actuation matrix $\bfA$ is a sub-matrix of $\bfT$, these approximation formulas apply directly to $\bfA$ as well (Eq.~\ref{eq:Apj_approx} and Eq.~\ref{eq:Apjkl_approx}).

\subsection{Derivation of rate of change in distance to goal}
In Section~\ref{subsec:Kpj_and_approx}, we showed the rate of change in distance ($L^2\text{-norm}$ of true errors) due to infinitesimal LSP by a substructure. The full derivation of Eq.~\ref{eq:limit_of_dOmega} is as follows. For simplicity of notations, here replace the notations of two vectors:
\begin{equation}\label{eqSI:uv_replacement}
    \Delta{\lri} \rightarrow u, \quad \bfG^{\lri} \bfP_{[j]}^{\lri} \dsrci \rightarrow v.
\end{equation}
Then Eq.~\ref{eq:limit_of_dOmega} is equivalent to:
\begin{equation}\label{eqSI:uv_form_of_limit_of_dOmega}
    \lim_{\alpha \to 0} \pdv{\Omega_{p[j]}^{\lri}}{\alpha} = \lim_{\alpha \to 0} \pdv*{\left( \Lnorm{u - \alpha v } - \Lnorm{v} \right) }{\alpha}. \\
\end{equation}
Firstly, note that the second vector $v$ is independent of $\alpha$, the derivative term (before taking limit) can be simplified as below:
\begin{equation}\label{eqSI:simplify_dOmega}
\begin{aligned}
    \pdv*{\left( \Lnorm{u - \alpha v } - \Lnorm{v} \right) }{\alpha} 
    & = \pdv*{\Lnorm{u-\alpha v}}{\alpha} - 0  = \pdv*{\sqrt{\sum_1^r \left( u_r - \alpha v_r \right)}}{\alpha}
    = \dfrac{\sum_1^r \pdv*{\left( u_r - \alpha v_r \right)^2}{\alpha}  }{2\sqrt{\sum_1^r \left( u_r - \alpha v_r \right)}} \\
    & = \dfrac{ - \sum_1^r (u_r - \alpha v_r) v }{ \Lnorm{u-\alpha v_r}} 
    = \dfrac{-\sum_1^r(u_r v_r - \alpha v_r^2) }{ \Lnorm{u-\alpha v}} 
    = \dfrac{-\sum_1^r u_r v_r + \alpha \sum_1^r v_r^2 }{\Lnorm{ u_r - \alpha v_r}} \\
    & = \dfrac{- u \odot v + \alpha ( v \odot v) }{\Lnorm{ u - \alpha v}}.
\end{aligned}
\end{equation}
Secondly, substitute Eq.~\ref{eqSI:simplify_dOmega} into Eq.~\ref{eqSI:uv_form_of_limit_of_dOmega} and then change back the variable names, the derived result in the main text (Eq.~\ref{eq:limit_of_dOmega}) can be obtained:
\begin{equation}
    \lim_{\alpha \to 0} \dfrac{- u \odot v + \alpha ( v \odot v) }{\Lnorm{ u_r - \alpha v_r}} =  \dfrac{- u \odot v}{\Lnorm{ u_r }} = \dfrac{-\Delta^{\lri} \odot \bfG^{\lri} \bfP_{[j]}^{\lri}\dsrci }{\Lnorm{\Delta^{\lri}}}.
\end{equation}

\subsection{Schematic illustration of rate of change in relative errors}
Here we show a schematic illustration of $\beta$, which is the main metric to determine the validity and quality of a candidate substructure. Assume a target only involves two degrees of freedom (DOF1, DOF2), then each component in $beta$ shows the expected rate of change in true error in its related DOF if applying infinitesimal substructure perturbation. The dispersion of all values in $\beta$, quantified by relative standard deviation (RSD), indicates how evenly the attempted perturbation would reduce all errors.

\begin{figure*}[th]
\centering
\includegraphics[width= 0.5\textwidth]{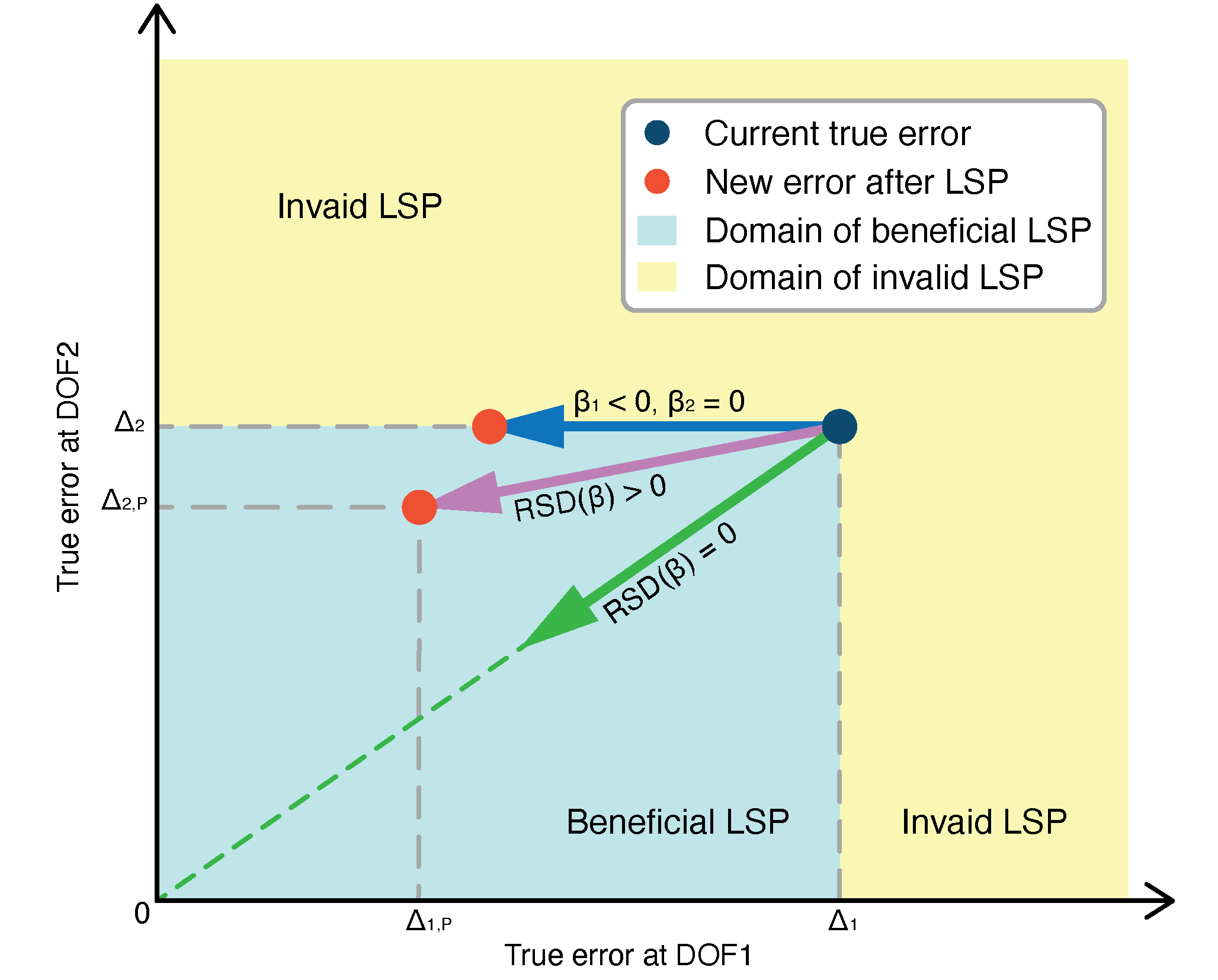}
\caption{Schematic illustration of $\beta_{p[j]}$. Here assume the target associated with this $\beta$ only involves two DOF so $\beta$ is a $2 \times 1$ vector (i.e., $\beta = [\beta_1, \beta_2]^T$). Valid perturbation is when at least one of the true errors would decrease while the remaining error is non-increasing, which can be categorized into three cases: i) Purple arrow shows a general case where both errors would decrease; ii) Yellow arrow shows a special case where the component in $\beta$ associated with one target is zero, indicating that true error would remain unchanged in this DOF (DOF2 in this example). iii) Green arrow shows an ideal case where all of the errors are expected to reduce evenly, this is unlikely to happen since zero is the absolute lower bond of RSD of a vector (Eq.~\ref{eq:def_RSD}). Domain of valid perturbation is highlighted in shaded blue, whereas domain of invalid perturbation is shown in shaded yellow.}\label{fig:schematic_of_beta}
\end{figure*}

\subsection{Finite element formulation of beam elements}
In two design examples (Auxetic, Actuator), we demonstrated network consists of two-node beam elements. The formulas are based on Euler-Bernoulli beam theory with extra consideration of axial displacement. Each end node has three DOF: horizontal displacement, vertical displacement, rotation angle. The stiffness matrix of such a two-node beam element, in its local coordinate, has form:
\begin{equation}
\bfK_{BE2} = E \cdot \bfK_{BE2}^{ref} =
\begin{bmatrix} 
    a & 0 & 0 & -a & 0 & 0 \\ 
    0 & 12b & 6bL & 0 & -12b & 6bL \\ 
    0 & 6bL & 4bL^2 & 0 & -6bL & 2bL^2 \\ 
    -a & 0 & 0 & a & 0 & 0 \\
    0 & -12b & -6bL & 0 & 12b & -6bL \\ 
    0 & 6bL & 2bL^2 & 0 & -6bL & 4bL^2 
\end{bmatrix},
\quad a = \frac{E A}{L}, \,\, b = \frac{E I}{L^3}.
\end{equation}
Therefore, the stiffness matrix of a beam element in the global CSYS has form:
\begin{equation}\label{eq:Ke_glb_beam}
\bfK_{e}^{glb} =E_e \cdot \bfR_e^T \cdot \bfK_{BE2}^{ref} \cdot \bfR_e\,
\end{equation}
where $\bfK_{BE2}^{ref}$ depends on the beam element's length, cross-sectional area, and inertia ($L$, $A$, and $I$). This indicates that for stiffness change in beam networks, we can apply the same formulas as those derived for ligament stiffness perturbation by SS in truss networks, and thereby implement the same error analysis and SS evaluation criteria. Note that this only applies to stiffness perturbation because $E$ is a common coefficient of the element stiffness matrix $\bfK_{BE2}^{ref}$. For change of other features such as cross-sectional area $A$, different strategies and derivations for approximated formulas of perturbed stiffness matrix of a network are required, which are subject to future studies.

\FloatBarrier
\section{Supporting information for numerical procedures}

\subsection{Speed and accuracy of using approximate formulas to evaluate change in structural response}
In Section~\ref{subsec:errors}, we derived expressions for nodal responses at targeted DOF under prescribed input, then derived approximate formulas to evaluate the change in nodal response at targeted DOF (here denote as $\delta \bfd_{tgt}$) if applying LSP by a substructure. The exact and approximate formulas are:
\begin{equation}\label{eqSI:exact_vs_approx_Delta}
    \delta \bfd_{tgt} = \bfG \left( \bfT_{p[j]} - \bfT \right) \bfd_E 
    \begin{cases}
    = \bfG \left[ \left( \bfK^{}_{F} - \alpha_j \bfS_{F,[j]} \right)^{-1} \left( \bfK_{EF}^T - \alpha_j \bfS_{EF,[j]}^T \right) - \bfT \right] \bfd_{src} & \, \text{exact} \\
    \approx - \alpha_j \bfG \bfP_{[j]} \bfd_{src} = - \alpha_j \bfG \left( \invKF  \bfS_{F,[j]} \invKF \bfK_{EF}^T - \invKF \bfS_{EF,[j]}^T \right) \bfd_{src} & \, \text{approx}
    \end{cases}
\end{equation}

As shown in Eq.~\ref{eqSI:exact_vs_approx_Delta}, if using the exact formula to compute change in response, inverse of the perturbed stiffness matrix of the updated network has to be computed first. As discussed, this is not efficient because matrix inverse is relatively time consuming while it has to be performed for each candidate SS. But in general, a large amount of candidate SS need to be validated to obtain a successful design in a simulation.

In contrast, the approximate formula avoids the need to compute any additional matrix inverse when analyzing candidate SS, and instead, takes advantage of the perturbational effect matrix $\bfP_{[j]}$, in which the stiffness matrices of a SS (i.e., $\bfS_{F,[j]}$, $\bfS_{EF,[j]}^T$) are its attributes that only need to be computed once during its creation, while the remaining terms (i.e., $\bfK_F^{-1}$, $\bfK_{EF}^T$) are dependent matrices of the current network, which only need to be evaluated once at the beginning of each numerical step when solving for structural response.

Here we investigate the computational speed and error between the exact and approximate formulas. We firstly pick the stiffness matrix $\bfK_F$ of the initial network of Lobster-A, which has dimensions $100 \times 100$ (50 free nodes in a 2D truss network). The number of ligaments in the initial network is 125, thus the order of a candidate SS can range from 1 to 124. 

\begin{figure*}[!th]
\centering
\includegraphics[width= 0.98\textwidth]{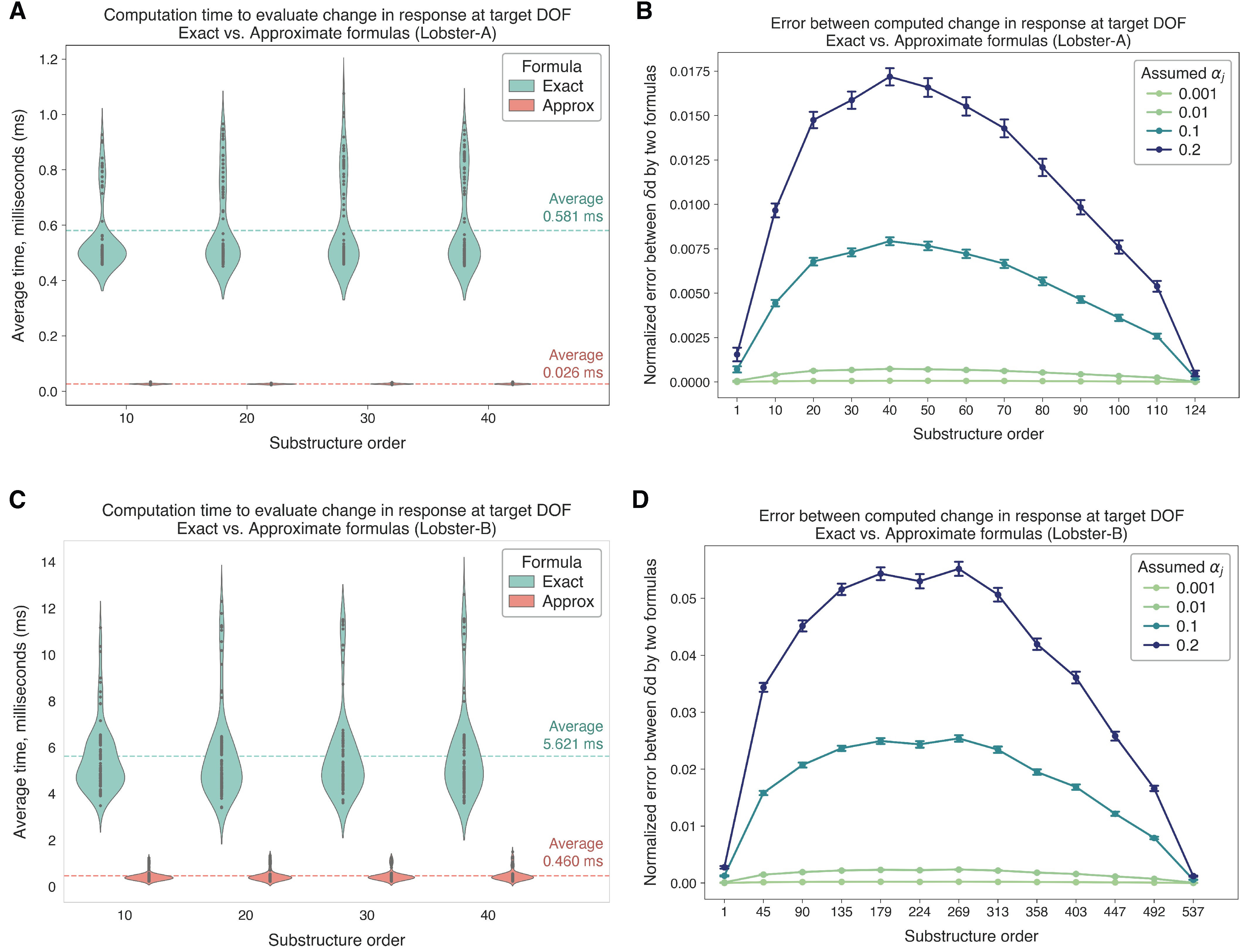}
\caption{Comparison of speed and accuracy between exact and approximate formulas to compute change in nodal response at target DOF due to LSP by one SS ($\delta\bfd_{tgt}$). 
(A)~Statistics of average run time to compute $\delta\bfd_{tgt}$ by the two formulas (Lobster-A). 
(B)~Statistics of normalized error between evaluated $\delta\bfd_{tgt}$ by the two formulas (Lobster-A). 
(C)~Run times by the two formulas (Lobster-B). 
(D)~Normalized errors between the two formulas (Lobster-B). In the time comparisons (panel~A and panel~C), 400 SS are generated and each data point represents the average value of 10000 runs of identical code snippets. In the error comparisons (panel~C and panel~D), up to 500 SS are generated for each picked SS order ranging from 1 to $n_{el}-1$.
}
\label{fig:delta-d-exact-vs-approx}
\end{figure*}

For comparison of computation times, we randomly generate 100 SS of order 10, 20, 30, and 40, respectively (there are 400 SS in total). For each SS, we measure the average run time to compute $\delta\bfd_{tgt}$ by the exact and approximate formulas, respectively. A perturbation coefficient $\alpha_j = 0.01$ is assigned in time comparison study. For each SS and formula, we run the identical code snippets for 10000 times and take the average (mean) value. The results are summarized by violin plots in Fig.~\ref{fig:delta-d-exact-vs-approx}A. Each data point (dot) in the violin plot is thereby an average over 10000 runs, and there are 800 data points. The run times of the exact formula demonstrate noticeable variations, ranging from about 0.45 to 1.07~milliseconds~(ms). In contrast, the run times of approximate formula show less dispersion, ranging from 0.023 to 0.035 ms. As annotated by the dashed lines in Fig.~\ref{fig:delta-d-exact-vs-approx}A, the average value of average run times for all 400 SS by exact formula and approximate formula is 0.026~ms and 0.581~ms, respectively. Computations based on the approximate formula is hence on average about 22 times faster than the exact formula (in this example). Also, the statistical measures are comparable between SS of different orders, which is expected because the dimensions of stiffness matrices of SS are uniform (the same as $\bfK$) and do not depend on the orders of SS.

For comparison of errors, We randomly generate up to 500 SS with orders 1, 10, 20, ..., 110, 124. In total, 5750 SS are generated since only 125 SS are available for order 1 and 124, respectively. Since the magnitudes of components of $\delta\bfd_{tgt}$ depend on the assigned perturbation coefficient $\alpha_j$, we compute $\delta\bfd_{tgt}$ for each SS by the two formulas with different values of $\alpha_j$: 0.001, 0.01, 0.1, and 0.2. Then we quantify the normalized error by finding the $L^2\text{-norm}$ of the difference between $\delta \bfd$ by the two formulas and then dividing it by the assigned $\alpha_j$:
\begin{equation}
    \delta\bfd^{err}_{tgt} = \frac{\Lnorm{\delta\bfd^{approx}_{tgt} - \delta\bfd^{exact}_{tgt}}}{\alpha_j}, \quad \alpha_j= 0.001, 0.01, 0.1, 0.2.
\end{equation}

Fig.~\ref{fig:delta-d-exact-vs-approx}B shows the statistics of errors categorized by the assigned perturbation coefficients. As expected, the normalized error is smaller when $\alpha_j$ decreases, because the approximation of perturbed matrix inverse (Eq.~\ref{eq:SMW_inf_series}) converges to the exact value when $\alpha \to 0$. The errors are in the magnitude of $10^{-5}, 10^{-4}$ and $10^{-3}$ for $\alpha_j=0.001, 0.01$ and $0.1$, respectively. This indicates the effectiveness and accuracy of the derived approximate formulas.

Interestingly, the errors are larger when the SS orders are around the middle region (e.g., from 30 to 60), and decrease when the SS orders are close 1 or the upper bound ($n_el-1$ in this study). This can be explained by the definition of SS and the fact that transformation matrix (Eq.~\ref{eq:def_T}) is independent of the ligaments stiffness magnitudes, but instead, their relative stiffness values (Eq.~\ref{eq:T_independent_of_c}). Which is exactly why relative stiffness are used in this study (characterized by interval $[0, 1]$). In specific, applying LSP by higher order SS is equivalent to applying a reverse LSP by its complementary lower order SS (assuming stiffness bounds are not encountered). For instance, if there are 100 ligaments with uniform stiffness $E_e = 0.5$ in a network, then reducing the stiffness of one ligament from 0.5 to 0.4 (a negative-LSP by an order-1 SS) is equivalent to increasing the stiffness of the remaining 99 ligaments from 0.5 to 0.625 (a positive-LSP by the order-99 SS complementary to that ligament). Therefore, SS orders around the middle regions would result in more dispersed relative stiffness changes in the network.

In addition to comparison based on the Lobster-A, we pick the initial network of Lobster-B and apply analogous comparative studies. The dimensions of $\bfK_F$ is $382 \times 382$ and the number of ligaments in the initial network is 538. As shown in Fig.~\ref{fig:delta-d-exact-vs-approx}C and Fig.~\ref{fig:delta-d-exact-vs-approx}D, similar relative differences between the two formulas are observed. Regarding the computation times, the approximate formula is on average 12 times faster than the exact formula. The reduced speed up factor (as compared to Lobster-A) is likely due to limited number of computing units (all simulations in this study are run on a laptop CPU with 10 cores). As dimensions of the matrices increases, more processing units may be utilized to enhance performance of matrix multiplications.

In brief, the comparative studies in this subsection indicates the efficiency and accuracy of the derived approximate formulas to evaluate expected change in nodal responses at targeted DOF.

\subsection{Identification of zero strain energy mode (ZSEM)}\label{subsec:detect_ZSEM}
Zero modes of a truss network can be identified via eigenvalue decomposition of its global stiffness matrix $\bfK$ (Eq.~\ref{eq:Kd=f}). The eigenvectors associated with zero eigenvalues represent infinitesimal displacements (in all nodal DOF) that would result in zero strain energy. Note that due to floating point errors, the solved `zero' eigenvalues are not exactly zero, but typically at least eight orders of magnitude smaller than the other `non-zero' values. Henceforth, all eigenvalues are firstly normalized by the maximum value, then a small threshold (e.g., $10^{-10}$) is applied to count the number of `zero' eigenvalues. In numerical simulation of truss networks, number of zero modes is strictly limited to three when attempting to change network topology. This ensures that there would be no artificial zero modes.

\begin{figure*}[th]
\centering
\includegraphics[width= 0.75\textwidth]{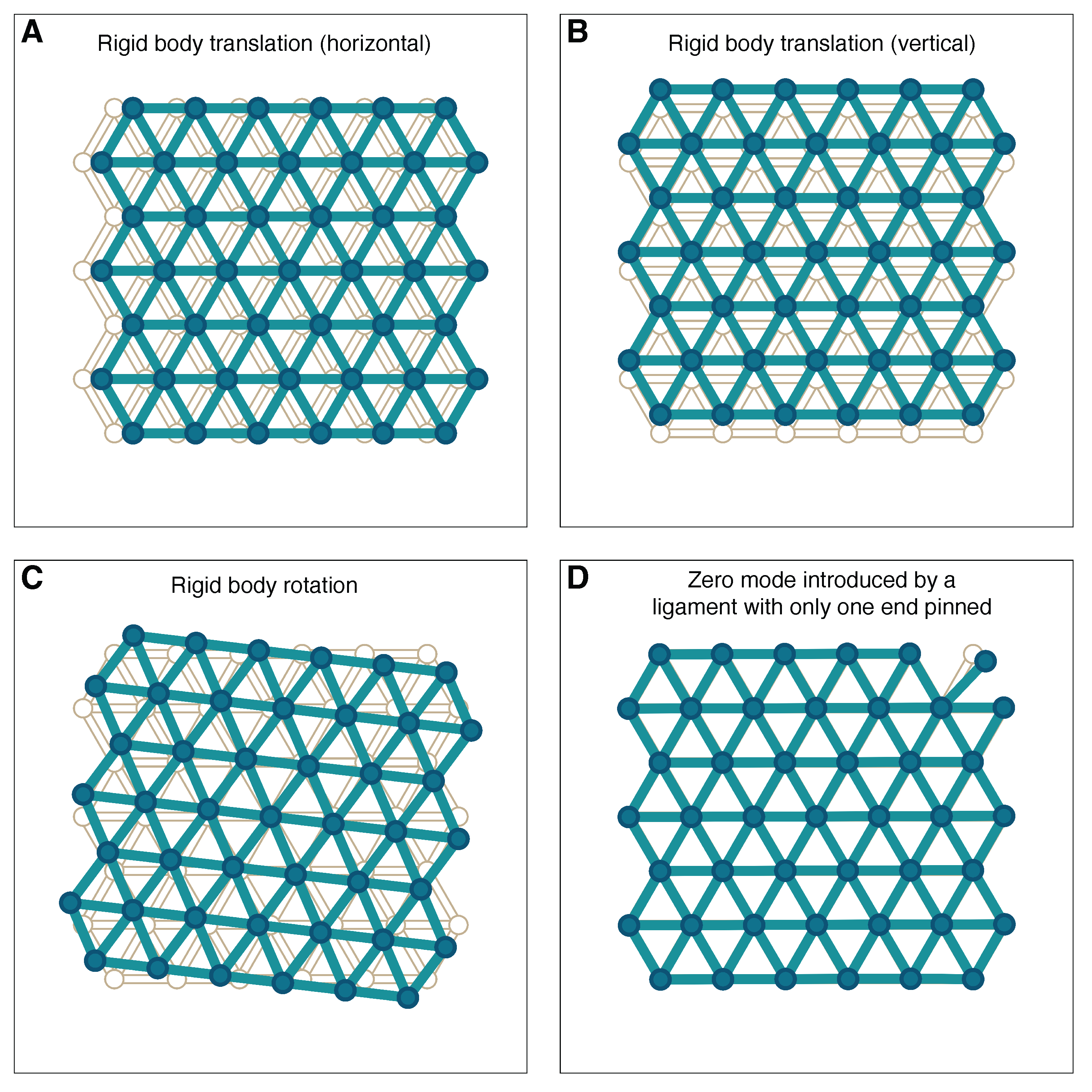}
\caption{Visualization of zero strain energy modes for a 2D truss network. The reference nodal positions of the network is shown in wire frames in light sand. Nodal displacements associated with each zero mode is obtained from the eigenvector corresponding to each zero eigenvalue. In specific, the nodal coordinates in the reference frame are offset accordingly by the scaled components of the eigenvector of a zero mode to visualize its representative nodal displacements. 
The first three figures show, respectively, three inherent zero modes associated with rigid body motions: (A) horizontal translation, (B) vertical translation, and (C) rotation. (D) Two ligaments on the top right are removed from the network, resulting in a ligament with only one end connected. The newly introduced zero mode represents a free rotation of this ligament about its connected end.}
\label{fig:detecting-zero-modes}
\end{figure*}

Take the network in Fig.~\ref{fig:network_and_SS} as an example. This network is a triangular lattice and there is no extra zero modes beyond the three inherent rigid body modes. Eigenvalue decomposition of its stiffness matrix results in exactly three zero eigenvalues. The three zero modes are visualized in Fig.~\ref{fig:detecting-zero-modes}A to C, where the solid lines reveal the representative `nodal displacements' that would result in zero strain energy in the network. In Fig.~\ref{fig:detecting-zero-modes}D, the network is intentionally modified such that there is one ligament that can freely rotate about its connected end. As expected, eigenvalue decomposition of the stiffness matrix $\bfK$ of this network gives exactly four zeros, and the nodal displacements associated with the extra artificial zero mode is visualized accordingly based on the corresponding eigenvector.

\FloatBarrier
\section{Supporting information for prototyping}

In the Lobster-A design example (see Fig.~\ref{fig:LobA-example}), inputs are in form of pinching or spreading the two controlling nodes along the vertical direction, and outputs (goals) are in form of relative change in distance between the two end nodes of the top or bottom `claw' of the network (i.e., claw opening or closure displacement). Here we take the Prototype-1 (Fig.~\ref{fig:LobA-P1-snaps}) and investigate the quantitative differences in these metrics between the prototype and numerical predictions. 

\begin{figure*}[htb]
\centering
\includegraphics[width= 0.75\textwidth]{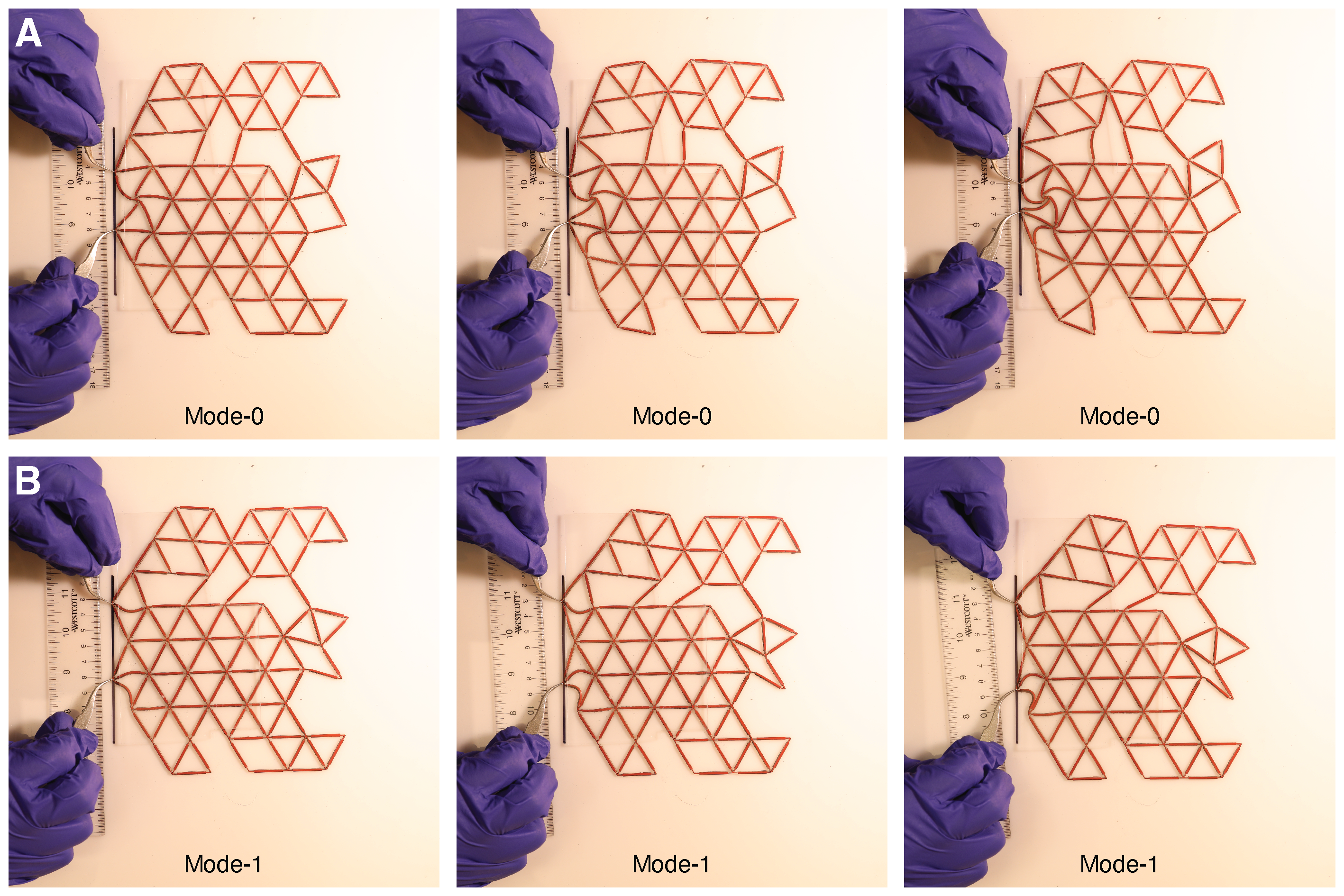}
\caption{Measurement of input and output magnitudes in each mode (Prototype-1).
(A) Applied inputs (i.e., three different displacement load magnitudes) and corresponding structural responses in Mode-0.
(B) Applied inputs and corresponding structural responses in Mode-1. All relevant data are listed in Table~\ref{tab:LobA-P1-measurement}.
}
\label{fig:LobA-P1-measures}
\end{figure*}

As shown in Fig.~\ref{fig:LobA-P1-measures}, we use two tweezers to apply displacement controlled load on the two input nodes along a straight line to effectively represent the input in this design example. Also, a transparent acrylic plate is placed over the prototype to alleviate potential out-of-plane deformation. For each mode, we apply three load levels and measure the nodal displacements at the targeted DOF. The results are listed in Table~\ref{tab:LobA-P1-measurement}, and the structural responses at which measurements are taken are shown in Fig.~\ref{fig:LobA-P1-measures}.

As expected, due to non-negligible rotating stiffness of joints, bending of ligaments, and friction between the prototype and the base platform, the magnitudes of structural responses at outputs are lower than the numerical results. Note that the entire truss network system is treated as linear elastic in simulations, so the nodal displacements at all free DOF scale linearly with the magnitude of input in FEA results. In reality, there should be a limited range within which the linear elastic assumption is justifiable due to potential material and structural non-linearity under large structural deformations. Therefore, the numerical results should not be considered as the correct reference values here. 

\begin{table*}[htb]
\small
\centering
\caption{Measured output responses of Lobster-A, Prototype-1}\label{tab:LobA-P1-measurement}
\begin{threeparttable}
\begin{tabular}{c c c c c c}
    \toprule
    & \multirow{3}{*}{Input (mm)} & \multicolumn{4}{c}{Output (mm)} \\
    \cmidrule(lr){3-6} 
    & & \multicolumn{2}{c}{Top claw} & \multicolumn{2}{c}{Bottom claw} \\
    \cmidrule(lr){3-4}
    \cmidrule(lr){5-6}
    & & FEA & Prototype & FEA & Prototype \\ [1ex]
    \hline
    \multirow{3}{*}{Mode-0} & -7.19 & -9.56 & -6.42 & 9.52 & 6.37 \\
     & -14.82 & -19.71 & -15.19 & 19.63 & 10.37 \\
     & -24.93 & -33.15 & -21.12 & 33.03 & 13.76 \\
     \hline
     \multirow{3}{*}{Mode-1} & 4.51 & 5.98 & 4.68 & -5.96 & -4.08 \\
     & 8.56 & 11.38 & 8.63 & -11.34 & -10.79 \\
     & 11.12 & 14.77 & 11.09 & -14.72 & -12.74 \\
    \bottomrule
\end{tabular}
\begin{tablenotes}\footnotesize
\item[*] Input is the amount of pinch or spread applied on the two input nodes. Output is the change in distance between the two end nodes of the top or bottom claw. Details are shown in Fig.~\ref{fig:LobA-example} and Fig.~\ref{fig:LobA-P1-measures}. 
\end{tablenotes}
\end{threeparttable}
\end{table*}

\begin{figure*}[htb]
\centering
\includegraphics[width= 0.98\textwidth]{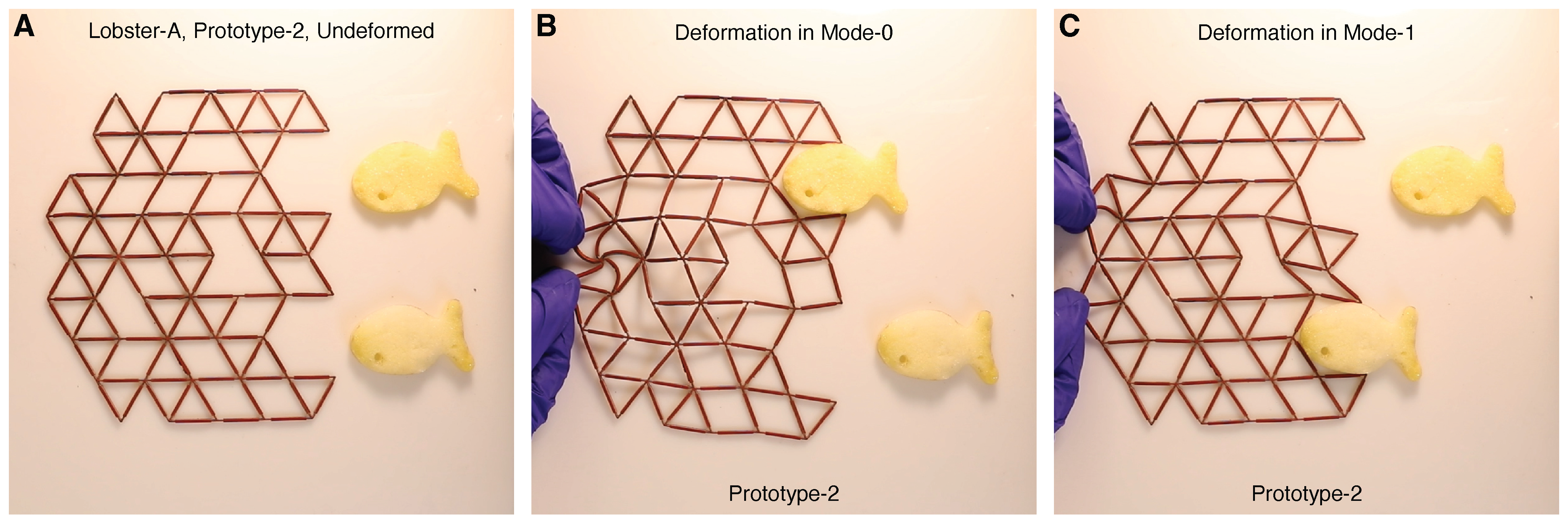}
\caption{Additional design and prototype example from Lobster-A (different topology from Fig.~\ref{fig:LobA-example}).
(A) Fabricated Prototype-2. (B) In Mode-0, the top claw closes to seize an object, while the bottom claw remains open. (C) In Mode-1, the bottom claw closes to seize the other object, while the top claw remains open.
}
\label{fig:LobA-P2-snaps}
\end{figure*}

In addition to Prototype-1, Fig.~\ref{fig:LobA-P2-snaps} shows prototype of another design (i.e., different network topology) from the Lobster-A design example. As expected, this design also realizes the prescribed functionalities under corresponding inputs.

\end{document}